\documentclass[preprint,12pt]{elsarticle}
\journal{Journal of Molecular Liquids}

\usepackage{amsmath}
\usepackage{amssymb}
\usepackage{graphicx}
\usepackage{natbib}
\biboptions{numbers,sort&compress}
\usepackage{csquotes}
\usepackage{siunitx}
\usepackage{hyperref}
\usepackage{enumitem}
\usepackage{physics}
\usepackage{caption}
\usepackage{subcaption}
\usepackage{booktabs}
\usepackage{multirow}
\usepackage{cancel}
\usepackage{bm}
\usepackage[capitalise]{cleveref}
\usepackage{color}
\usepackage[toc,page]{appendix}

\DeclareSIUnit\Molar{\textsc{M}}
\DeclareSIUnit\ergs{ergs}
\DeclareSIUnit\esu{esu}
\DeclareSIUnit\statV{stat\textsc{V}}
\DeclareSIUnit\cm{\centi\metre}

\begin{document}
	\begin{frontmatter}
	
		%\ead{marcelolcmx@ier.unam.mx}
	
		\title{Topological impact of nanopore electrodes on the structure of the electrical double layer and the  differential capacitance}
		\author{A. Silva-Caballero}
		\author{A. Lozada-Hidalgo}
		\author{M. Lozada-Cassou\corref{cor1}}
		\ead{marcelolcmx@gmail.com}
		\address{Instituto de Energías Renovables, Universidad Nacional Autónoma de México, Priv. Xochicalco S/N Temixco, Morelos 62580 México.}
		\cortext[cor1]{Corresponding author:}
		\begin{abstract}
			The electrical double layer for three different topologies of nanopore electrodes is studied, i.e., the interior and exterior electrical double layers of planar, cylindrical and spherical nanopores immersed into a point-ions electrolyte, and \textit{not connected to a power source}, are analytically attained through the linearized Poisson-Boltzmann equation. Thus, analytical formulas for the mean electrostatic potential, electrolyte's reduced concentration, and electrical field profiles, are exhibited. Their corresponding analytical expressions for their  differential capacitances are presented. All the nanopores are treated as permeable, so the electrolyte outside and inside the electrodes are at the same chemical potential. Analogous analytical formulas for solid nano-electrodes are obtained as a corollary of those for nanopores. In particular, their analytical expressions for the  differential capacitance here derived are shown to be consistent with the capacitive compactness proposed in the past by one of us. Numerical results of all of the above functions are analyzed as a function of the nanopores geometrical parameters and the electrolyte’s temperature and molar concentration. It is found that the spherical topology, at lower temperatures, has the higher differential capacitance. It is demonstrated that for the three nanopore topologies here considered their capacitances reduce to that of a single planar electrode, in the limit of infinitely wide nanopores.The electrical double layer  and mean electrostatic potential of the three topologies are in qualitatively agreement with those from the non-linearized Poisson-Boltzmann, hypernetted chain/mean-spherical approximation (HNC/MSA) equations and computer simulations results presented in the past, within the low mean electrostatic potential assumption. 
		\end{abstract}
\end{frontmatter}

\section{Introduction}

In the development of highly efficient energy storage systems, electric double layer capacitors (EDLC) play an important role.
In recent years, there has been several experimental~\cite{Chmiola_2006,Kim_2013,Supercapacitors-Book-2013,Beguin_2014,ElKady_2014,Ke_2016} and theoretical~\cite{Lozada_1984,Mier_1988,Vlachy1989,Yeomans1993,Yu_1997,Vlachy2001,Grosse-2002,Henderson2005,Aguilar_2007,Peng2009,Henderson2012,Pizio2012,Lamperski-2014,Henderson2015,Yang2019,biagooi-Nature2020,Enrique-Henry2021,Keshavarzi_2022,Feng-nanopores-topology-2023} interest on developing porous electrodes for EDLC. On the other hand, the electrical double layer structure have been long time recognized as a very important factor in the study of colloids~\cite{Verwey_TheoryStabilityLyophobicColloids_1948,Israelachvili-book}, biological systems~\cite{Evans-Wennerstrom-1999}, biomaterials~\cite{Bohinc_2008,Bohinc-2018}, medicine~\cite{Coffey-Biology-2023}, and oil industry~\cite{HUANG-oil-1996,Oil-Recovery-book-2019}, among others.
Hence, analytical solutions of simple models can help to understand the electrical double layer (EDL) and its impact in colloids, micelles and porous materials, and be useful for the comprehension and design of chemicals for oil recovery, polymers, medical procedures and pore-electrodes.
Here, we propose a solution for nanopores, \textit{immersed into a bulk electrolyte}, that takes in consideration the inner and outer pore's EDL.
Our theoretical approach is both, simple and easily reproducible, and it takes into account the different parameters of the system.

The electrical double layer is formed in interfacial systems, such as those mentioned above, where one of the phases is an electrolyte solution.
For such systems, the EDL is formed by a layer of adsorbed ions to the solid phase, plus a diffuse layer of ions~\cite{Grahame_1947}.
As such, the EDL is the result of an uneven distribution of negative and positive ions that are next to the solid phase, i.e., the electrode, colloidal particle or membrane.

Throughout the years there have been many models that have tried to explain the EDL behavior, beginning with the pioneer works of Helmholtz~\cite{Helmholtz_1879}, and Gouy-Chapman~\cite{Gouy_1910,Chapman_1913}, up to well founded statistical mechanical theories, such as integral equation~\cite{Greberg_1998,Henderson_1992_FIF, Lozada_1992_FIF,Attard_1996,Croxton_1981,Henderson_1982}, density functional~\cite{Shi_1996,Goel_2008,Huang_2008,Lian_2016,Hartel_2017,Patra_1994,Patra_2020,Gillespie_2005,Stevens_1990}, and modified Poisson-Boltzmann~\cite{Bhuiyan_1994,Outhwaite_1986,Bhuiyan_1993_CMT}.

The first model of the EDL was proposed by Helmholtz~\cite{Helmholtz_1879}.
He considered that the arrangement of charges was of a rigid fashion on both sides of the interface, hence, neglecting the interactions far from the electrode, and the dependence of the electrolyte's concentration.
Later,  Gouy~\cite{Gouy_1910} and Chapman~\cite{Chapman_1913} models originated from the assumption that the electrode's charge is not rigid and that the ions in the EDL are subjected to an electrical and thermal field.
However, this model also fails as it does not consider the ions' size on its formulation, regarding them as point charges.
As a consequence, Stern~\cite{Stern_1924} proposed a hybrid model; combining the Helmholtz model and the Gouy-Chapman model.
He took into consideration the ion size by recognizing two regions of ions distributions; a rigid one, next to the electrode, and, then, a diffused ions cloud.
The first region is known as the Stern or  Helmholtz layer, where the distance of maximum approach of an ion to the electrode is restricted by its ionic size.
In the second region, the ions are considered to be point charges. Thence, in this model the electrostatics in the first region is ruled by the Laplace equation, whereas in the second region it is governed by the Poisson-Boltzmann (PB) equation. The PB equation is a non-linear, inhomogeneous, second order differential equation, and has been analytically solved  for a planar~\cite{Verwey_TheoryStabilityLyophobicColloids_1948} electrode. Analytical solutions of the linearized Poisson-Boltzmann (LPB) equation have been given in the past for cylindrical~\cite{Lyklema_FundIntCollSci}, and  spherical~\cite{Dyachkov_2005} solid electrodes. The  integral equation version of the PB equation, has been numerically solved for planar~\cite{Lozada_1982}, cylindrical~\cite{Gonzalez_1985}, and spherical~\cite{Gonzalez_1989}, solid electrodes geometries, including the Stern correction. 

A more complex system is that of nanopores immersed into an electrolyte solution, as for example in vesicles into a bulk electrolyte solution~\cite{Evans-Wennerstrom-1999,Coffey-Biology-2023}, or nanopores in a large array of nanocapacitors closely packed as components of a  supercapacitor device~\cite{Supercapacitors-Book-2013}, or self-assembled, according with several recent self-assembly techniques~\cite{Limin-self-assembly-2012,Chai-Advance-Materials-self-assembly-2020,vialetto-JACS-self-assembly-2021,jia-RSC-adv-2023}. In this case, although numerical solutions of the PB equation, for point ions, and integral equations and computer simulations for the restricted primitive model (RPM) electrolyte have been published in the past for planar~\cite{Lozada1996}, cylindrical~\cite{Yeomans1993,Aguilar_2007} and spherical~\cite{Yu_1997} nanopores, immersed into an electrolyte, to the best of our knowledge there are no analytical solutions of the LPB equation for spherical or cylindrical nanopores \textit{immersed into an electrolyte}, although there are molecular dynamics~\cite{biagooi-Nature2020} and density functional~\cite{Lian_2016,Wu-nano-letters2011,Pizio2012}, and LPB~\cite{Yang2019} results for an electrolyte confined in a planar, cylindrical and spherical cavities, at a fixed constant potential,  or at constant surface charge~\cite{Peng2009}. However, in all cases the EDL outside the nanopore is not considered. However, in the past it has been shown that there is an strong correlation between the fluids inside and outside a nanopore immersed into an electrolyte~\cite{Lozada-Cassou-PRL1996,Lozada-Cassou-PRE1997}.
Here, we analytically solve the LPB equation for three nanopore electrode topologies planar, cylindrical and spherical, \textit{immersed into an electrolyte and not connected to a power source}, \textit{at infinite dilution}; and analyze our results in terms of the different topologies, with focus on their  differential capacitance. While, of course, a topology goes far beyond a geometry and, hence, we could refer simply to the geometry of the nanopores, here we emphasize their topological differences since we think this is a relevant characteristic for the electrical double structure and  differential capacitance of nanopores which go beyond the simple geometries that are revised in this article~\cite{Feng-nanopores-topology-2023}.

We wish to emphasized that in this paper we deal with nanopores, immersed into an electrolyte, at infinite dilution. Hence, here we will not address to systems of self-assembled or designed arrays of nanocapacitors, which can have relevant application in medicine~\cite{jia-RSC-adv-2023} and supercapacitors~\cite{Supercapacitors-Book-2013} or consider charge regulation effects~\cite{ninham-JTB-1971,Bohinc-2018}. 

This paper is organized as follows.
In \cref{Theory}, we depict the models for planar, cylindrical and spherical nanopores, immersed into a point ion electrolyte, and their corresponding analytical solutions of the LPB equation.
In \cref{Res_Disc}, we numerically compare the EDL structure and induced electrical field for several system parameters of the nanopores, and their impact on their  differential capacitance. Finally, in  \cref{Conclusions}, some conclusions are presented.

\section{Theory}\label{Theory}

\begin{figure}[!htb]
	\centering
	\begin{subfigure}{.495\textwidth}
		\centering
		\includegraphics[width=.95\linewidth]{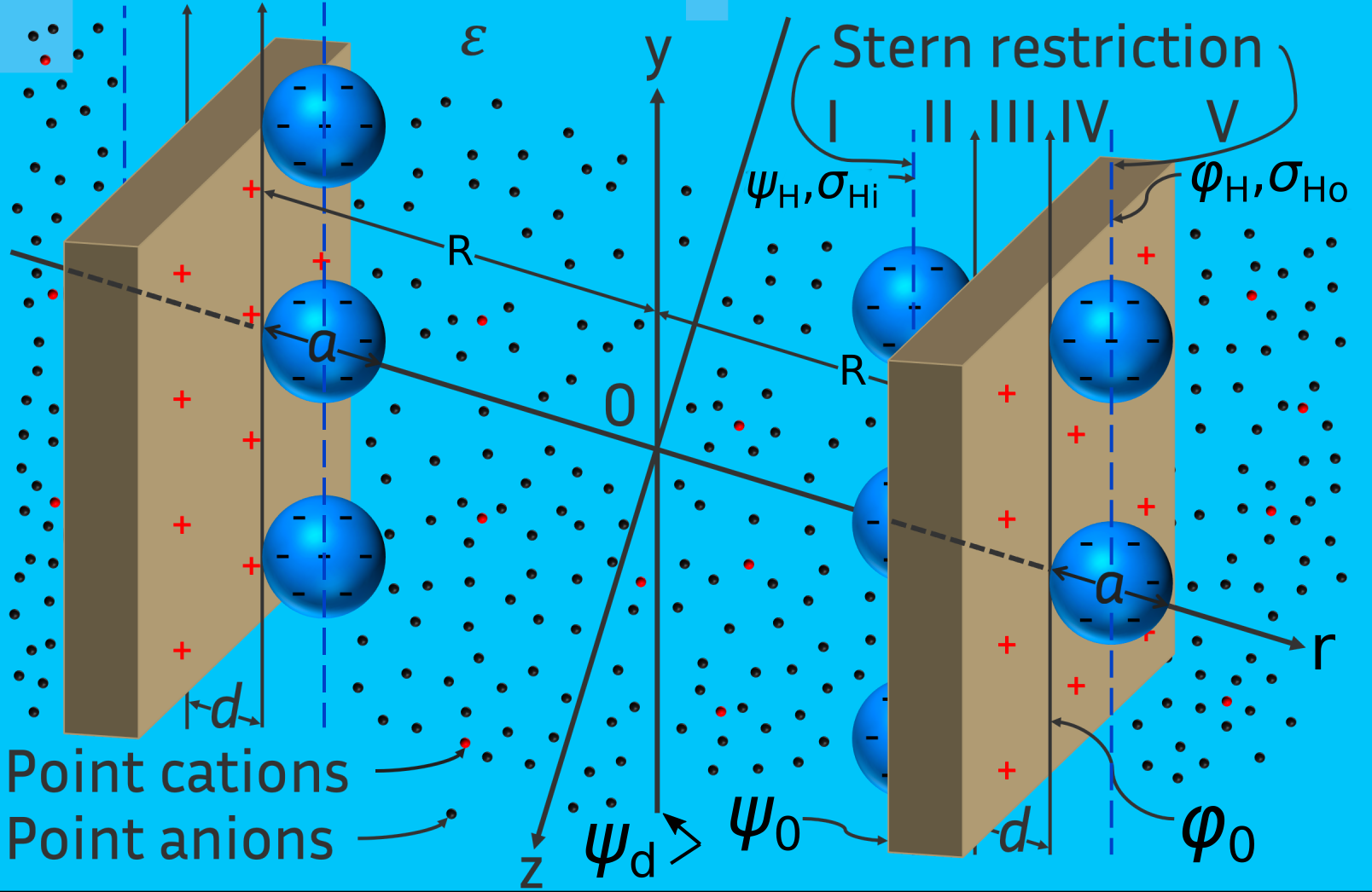}
		\caption{Slit nanopore electrode}	\label{Geometry_twoplates}
	\end{subfigure}
	\begin{subfigure}{.495\textwidth}
		\centering
		\includegraphics[width=.98
		\linewidth]{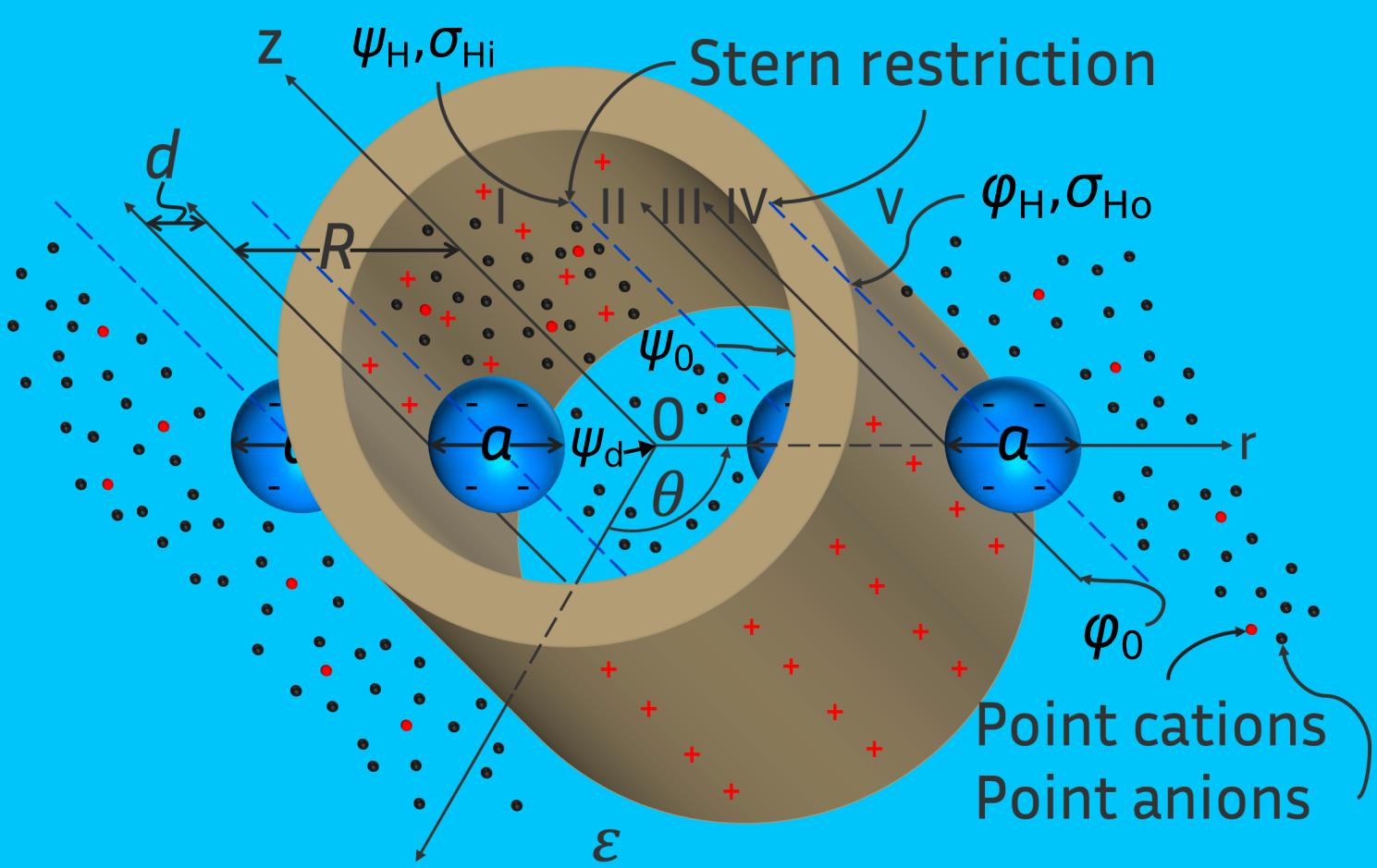}
		\caption{Cylindrical nanopore electrode}
		\label{Geometry_cylinder}
	\end{subfigure}\\
	\begin{subfigure}{.425\textwidth}
		\centering
		\includegraphics[width=\linewidth]{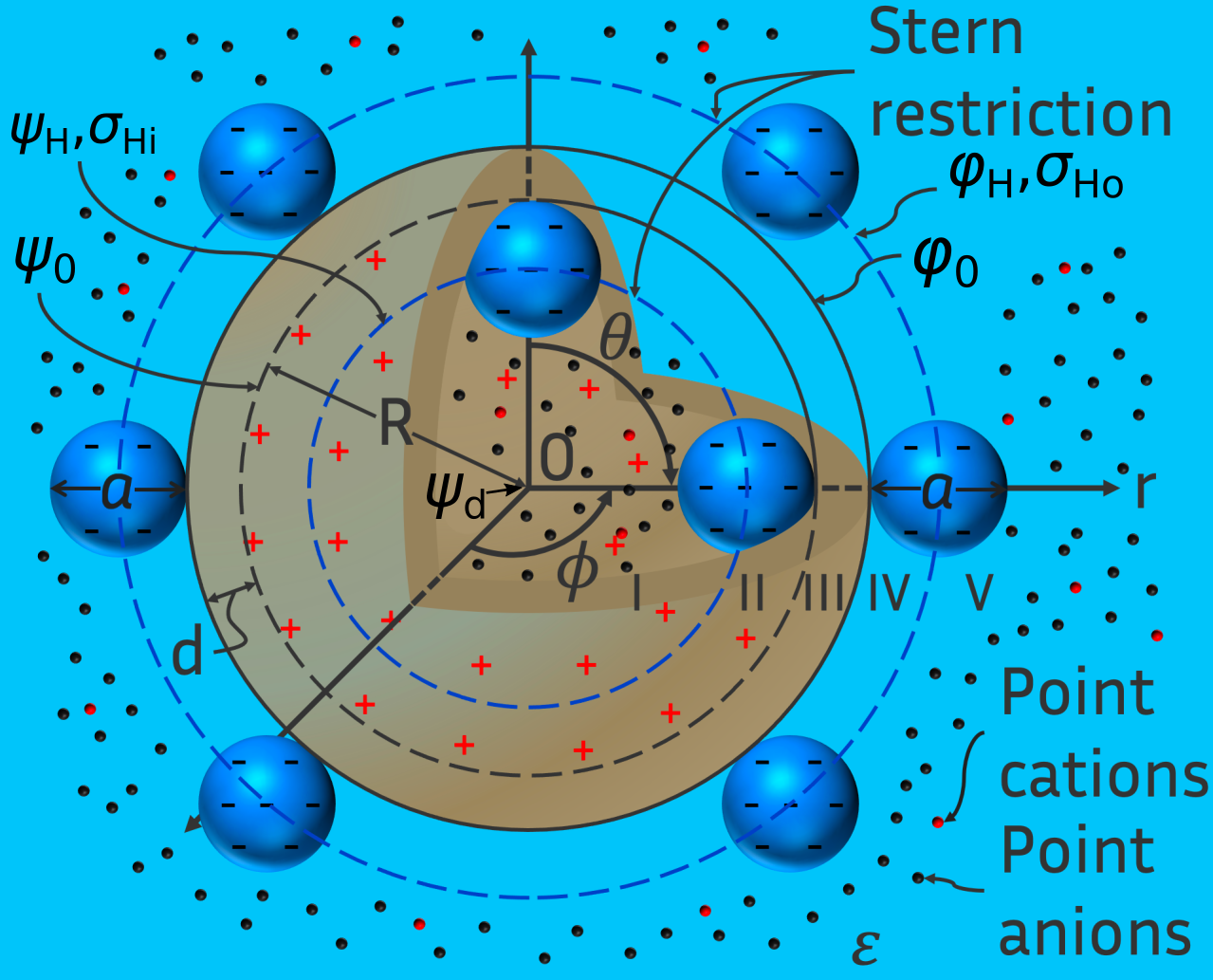}
		\caption{Spherical nanopore electrode}
		\label{Geometry_sphere}
	\end{subfigure}
	\caption{Topologies of nanopore electrodes.}
	\label{Geometry_electrodes}
\end{figure}

The EDL for nanopore electrodes is obtained from the analytical solution of the LPB equation.
The models for the different nanopores are shown in \cref{Geometry_electrodes}, where five regions of ions distributions are shown.
Due to the symmetry of the planar nanopore, the left and right sides of the pore's center are symmetrical.
Hence, to solve the electrodes' electrostatics for the planar electrode, we consider only its five right-hand regions.

The models for two interacting charged plates (slit nanopore), a cylindrical nanopore (open at both extremes), and a spherical nanopore are depicted in \cref{Geometry_electrodes}.
In \cref{Geometry_twoplates} both plates are positively charged and they are separated by a distance of $2\,R$. In \cref{Geometry_cylinder,Geometry_sphere} the cylindrical and spherical pores are also positively charged, and their inner radius is $R$. The walls of the three nanopores have the same thickness $d$. The surface charge density on the nanopores walls, say $\sigma_{\scriptscriptstyle{I}}$, for the interior wall and $\sigma_{\scriptscriptstyle{II}}$, for the external wall, are assumed to be equal to $\sigma_{\scriptscriptstyle{0}}$. Let us point out that the charge on the two faces of the wall of the  nanopores need not to be equal or we could have assumed a fixed constant potential. However, in this article we will solve the case in which the nanopores walls have a fixed, symmetrical electrical charge. The nanopores are taken to be immersed into an electrolyte.
The electrolyte is modeled as a hard spheres' fluid of diameter $a$, with charge in their centers. However, the ionic diameter is considered only when interacting with the pore's walls, while the ion-ion interaction is taken as that of point charges, i.e., the Stern correction~\cite{Verwey_TheoryStabilityLyophobicColloids_1948}.

Both, the electrodes and the electrolyte are assumed to have the same dielectric constant (relative permittivity) $\varepsilon$, to avoid image potentials. The electrodes are permeable, so \textit{the electrolyte outside and inside the nanopores are at the same chemical potential}. Because the nanopores are immersed into an electrolyte, an effective charge density is induced inside, $\sigma_{\scriptscriptstyle{Hi}}$, and outside, $\sigma_{\scriptscriptstyle{Ho}}$, the nanopores, located at $r=R-a/2$ and $r=R+d+a/2$, respectively.

In regions II, III and IV, electrical charge of the ions is excluded since their centers cannot penetrate in these regions due to their hard-sphere diameter. Hence, in these regions the electrostatics is ruled by the Laplace equation, whereas for regions I and V, the electrostatics obeys the Poisson equation~\cite{Jackson_2001}. The Laplace equation is given by,

\begin{equation}
	\laplacian{\psi(r)}=0\label{Ec.Laplace},
\end{equation}

\noindent whereas in MKS units the Poisson equation is

\begin{equation}
	\laplacian{\psi(r)}=-\frac{1}{\varepsilon_{\scriptscriptstyle{0}} \varepsilon}\,\rho_{\scriptscriptstyle{el}}(r) \label{Ec.Poisson}.
\end{equation}

 In \cref{Ec.Poisson}, $\rho_{\scriptscriptstyle{el}}(r)$ and $\psi(r)$ are the local charge concentration distribution function, and the mean electrostatic potential profile (MEP), respectively; both induced in the electrolyte by the charge on the nanopores walls. $\varepsilon_{\scriptscriptstyle{0}}$ is the vacuum permittivity. The distance $r$ is measured from the geometrical center of the nanopores.

 For a n-species electrolyte solution,
 
\begin{equation}
	\rho_{\scriptscriptstyle{el}}(r)=\sum_{i=1}^{n}e\,z_{i}\,\rho_{\scriptscriptstyle{i}}(r)=\sum_{i=1}^{n}e\,z_{i}\,\rho_{\scriptscriptstyle{i0}}g_{\scriptscriptstyle{i}}(r) \label{Rho_elx},
\end{equation}
where $\rho_{\scriptscriptstyle{i}}(r)$ is the ions concentration profile of species $i$, whereas $g_{\scriptscriptstyle{i}}(r)$ is its corresponding ions reduced concentration profile. $z_{\scriptscriptstyle{i}}$ and $\rho_{\scriptscriptstyle{i0}}$ are the valence and bulk concentration of species $i$, respectively. Thus, $	\rho_{\scriptscriptstyle{el}}(r)$ is the charge concentration profile, at the distance $r$ from the nanopore's geometrical center. In general, according with the canonical Boltzmann distribution function, $g_{\scriptscriptstyle{i}}(r)=exp(e\,z_{i}\,\beta\, W_i (r))$~\cite{McQuarrie_StatMech}, where $W_i (r))$ is the potential of mean force, which for a point ions electrolyte reduces to $\psi (r)$. Thenceforth, for a point ions electrolyte, in the mean field approximation \cref{Rho_elx} becomes 

\begin{equation}
	\rho_{\scriptscriptstyle{el}}(r)=\sum_{i=1}^{n}e\,z_{i}\,\rho_{\scriptscriptstyle{i0}}\,g_{\scriptscriptstyle{i}}(r)=\sum_{i=1}^{n}e\,z_{i}\,\rho_{\scriptscriptstyle{i0}}\exp(-e\,z_{i}\,\beta\, \psi (r)) \label{Rho_elx2},
\end{equation}

\noindent where $\beta=\frac{1}{k\,T}$, being $k$, the Boltzmann constant, and $T$ the temperature. Accordingly, \cref{Ec.Poisson} becomes the Poisson-Boltzmann equation,

\begin{equation}
	\laplacian{\psi(r)}=-\frac{1}{\varepsilon_{\scriptscriptstyle{0}} \varepsilon}\sum_{i=1}^{n}e\,z_{i}\,\rho_{\scriptscriptstyle{i0}}\exp(-e\,z_{i}\,\beta \,\psi (r)) \label{Ec.Poisson.2}.
\end{equation}

\noindent To obtain the EDL for the different nanopores, \cref{Ec.Poisson.2} will be linearized, and analytically solved for a two-species symmetrical electrolyte. Thence, for small enough mean electrostatic potentials, such that $z\,e\,\psi(r)$ $/k\,T$ $\ll1$, the PB linearization (\cref{Ec.Poisson.2}), leads to
\begin{equation}
	\laplacian{\psi(r)}=\kappa^2\,\psi(r) \label{Ec.ALS_Poisson}
\end{equation} 
where $\kappa$ [\si{\per\m}] is the inverse of the Debye length, $\lambda_{\scriptscriptstyle_{D}}$, and it is given by
\begin{equation}
	\kappa=\sqrt{ \frac{2\,\rho_{\scriptscriptstyle{i0}}\,e^2\, z^2}{\varepsilon_{\scriptscriptstyle{0}}\varepsilon\,k\,T}}\label{Ec.kappa}
\end{equation}

The general solutions of the Laplace and LPB equations for $\psi(r)$, and their respective derivatives, for the planar geometry are

\begin{gather}
	\psi(r)=C+D\,r\quad\quad \dv{\psi}{x}=D\label{Ec.ALS_Laplace_GENsol}\\
	\psi(r)=A\,e^{\kappa\,r}+B\,e^{-\kappa\,r}\quad\quad \dv{\psi}{x}=A\,\kappa\,e^{\kappa\,r}-B\,\kappa\,e^{-\kappa\,r}\label{Ec.ALS_Poisson_GENsol}
\end{gather}
Here $\kappa r$ is the distance to the geometrical center plane of the nanopore, located at $r=0$, and normalized with the Debye length, $\kappa^{-1}$, (see \cref{Geometry_twoplates}). As pointed out above, we will focus only on the right-hand side of the nanopore since the left side is, by symmetry, identical to the right-hand side.

For the cylindrical geometry, the linearization of the PB  leads to

\begin{equation}
	\dv[2]{\psi}{r}+\frac{1}{r}\dv{\psi}{r}=\kappa^2\,\psi(r)\label{Ec.ACS_Poisson}
\end{equation}

\noindent Correspondingly, the general solutions of the Laplace and Poisson-Boltzmann equations for the cylindrical geometry are, with their respective derivatives,

\begin{gather}
	\psi(r)=C+D\,\ln(r)\quad\quad\dv{\psi}{r}=\frac{D}{r}\label{Ec.ACS_Laplace_GENsol}\\
	\psi(r)=A\,I_{\scriptscriptstyle{0}}(\kappa\,r)+B\,K_{\scriptscriptstyle{0}}(\kappa\,r)\quad\quad\dv{\psi}{r}=A\,\kappa\,I_{\scriptscriptstyle{1}}(\kappa\,r)-B\,\kappa\,K_{\scriptscriptstyle{1}}(\kappa\,r)\label{Ec.ACS_Poisson_GENsol}
\end{gather}
\noindent where $I_{\scriptscriptstyle{0}}(\kappa\,r)$, $K_{\scriptscriptstyle{0}}(\kappa\,r)$,  $I_{\scriptscriptstyle{1}}(\kappa\,r)$ and $K_{\scriptscriptstyle{1}}(\kappa\,r)$ are the modified Bessel functions of first and second kind of order $0$ and $1$, respectively; as a function of the scaled distance to the axial center of the cylindrical nanopore, $\kappa r$ (see \cref{Geometry_cylinder}).

Lastly, for the spherical geometry the linearization of the PB  leads to

\begin{equation}
	\dv[2]{\psi}{r}+\frac{2}{r}\,\dv{\psi}{r}=\kappa^2\,\psi(r)\label{Ec.ASS_Poisson}
\end{equation}

\noindent Likewise, the general solutions of the Laplace and Poisson-Boltzmann equations for the spherical geometry are, with their respective derivatives,
\begin{gather}
	\psi(r)=C+\frac{D}{r}\quad\quad\dv{\psi}{r}=-\frac{D}{r^2}\label{Ec.ASS_Laplace_GENsol}\\
	\psi(r)=A\,\frac{e^{\kappa\, r}}{r}+B\,\frac{e^{-\kappa\, r}}{r}\quad\quad\dv{\psi}{r}=\frac{A\,e^{\kappa\, r}(\kappa\,r+1)-B\,e^{-\kappa\, r}(\kappa\,r+1)}{r^2}\label{Ec.ASS_Poisson_GENsol}
\end{gather}

\noindent As before, in \cref{Ec.ASS_Poisson_GENsol} $\kappa r$ is the scaled distance to the center of the nanopore (see \ref{Geometry_sphere}).

For each of the three geometries we need to find the expressions for the six different constants $A$ and $B$, for regions I, II, III, and the four different constants $C$ and $D$, in regions I and V. These constants will be determined from the boundary conditions (BCs) of the electrodynamics theory~\cite{Jackson_2001}. The BCs are exactly the same for the three nanopore electrodes, when properly considering the different curvatures of the nanopores. Thence, we detail the solution of the slit nanopore, and present the main results for the cylindrical and spherical nanopores in the main text. However, the complete solutions for the cylindrical and spherical nanopores are outlined in the appendixes.

\subsection{Slit Nanopore}\label{Slit-pore}

For the slit nanopore, the right-hand side five regions defined in \cref{Geometry_electrodes} will be treated individually with their respective BCs.
In each region the  electrical field, $E(r)$, can be obtained from the mean electrostatic potential, $\psi (r)$,  through the equation

\begin{equation}
	\abs{\overrightarrow{\mathbf{E(r)}}}\equiv E(r)=-\abs{\grad \psi(r)}.\label{elecrical-field}
\end{equation}

\noindent Thus, for each of the five regions, from \cref{Ec.ALS_Laplace_GENsol,Ec.ALS_Poisson_GENsol}, the mean electrostatic potential is

\begin{equation}
	\psi(r) = \begin{cases}
		Le^{-\kappa r} + M e^{\kappa r} & 0 \leq r \leq R - \frac{a}{2} \\
		Hr+I & R - \frac{a}{2} \leq r \leq R \\
		Fr+G & R \leq r \leq R+d  \\
		Cr+D & R + d \leq r \leq R+d+a/2\\
		Ae^{-\kappa r} + B e^{\kappa r} & R+d+a/2 \leq r .\label{Plates_MEP_five-regions}
	\end{cases}
\end{equation}

\noindent To obtain the values of the constants in \cref{Plates_MEP_five-regions}, in each of the five regions two BC will be introduced.

\textbf{1. Region V:} $\bm{r\ge R+d+a/2}$\vspace{5pt}

\noindent Here, the solution of the LPB equation is $\psi_{\scriptscriptstyle{5}}(r)=Ae^{\kappa\,r}+Be^{-\kappa\,r}$. A convenient first BC is that, at $r\rightarrow \infty$, the MEP is taken to be equal to zero. Thus, if
the $\lim_{r \to \infty} \psi_{\scriptscriptstyle{5}}(r) = 0$, then $A=0$. Hence, $\psi_{\scriptscriptstyle{5}}(r) = Be^{-\kappa r}$. Thence, the electric field in region $V$ is
\begin{equation}
	E_5(r) = -\nabla \psi_{\scriptscriptstyle{5}}(r) = B\kappa e^{-\kappa \left(r- R_{\scriptscriptstyle_{H}} \right)}= \frac{\sigma_{\scriptscriptstyle{5}}(r)}{\varepsilon_{\scriptscriptstyle{0}} \varepsilon \label{Plates_EF_RV}},
\end{equation}

\noindent where, according with the Gauss law of the electrostatics, $\sigma_{\scriptscriptstyle{5}}(r)$ is the induced surface charge density on the electrolyte, in region V. Defining $R_{\scriptscriptstyle_{H}}\equiv R+d+a/2$ and $\sigma_{\scriptscriptstyle{5}}(r=R_{\scriptscriptstyle_{H}})\equiv \sigma_{\scriptscriptstyle{Ho}}$, \cref{Plates_EF_RV} becomes

\begin{equation}
	E_5(R_{\scriptscriptstyle_{H}}) = B\kappa e^{-\kappa \left(r- R_{\scriptscriptstyle_{H}}\right)}= \frac{\sigma_{\scriptscriptstyle{Ho}}}{\varepsilon_{\scriptscriptstyle{0}} \varepsilon \label{Plates_EF2_RV}}.
\end{equation}
Hence, $B=\sigma_{\scriptscriptstyle{Ho}}e^{\kappa \left(r- R_{\scriptscriptstyle_{H}}\right)}/(\varepsilon_{\scriptscriptstyle{0}} \varepsilon \kappa)$, i.e.,

\begin{equation}
	E_5(r) = \frac{\sigma_{\scriptscriptstyle{Ho}}}{\varepsilon_{\scriptscriptstyle{0}}\varepsilon}e^{-\kappa \left(r- R_{\scriptscriptstyle_{H}}\right)} \label{Plates_EF3_RV}.
\end{equation}

\noindent Similarly, defining $\varphi_{\scriptscriptstyle H}\equiv\psi (r=R_{\scriptscriptstyle_{H}})=Be^{-\kappa R_{\scriptscriptstyle_{H}}}$, we find 

\begin{equation}
	\psi_{\scriptscriptstyle{5}}(r) = \varphi_{\scriptscriptstyle H}e^{-\kappa \left(r- R_{\scriptscriptstyle_{H}}\right)} \label{Plates_MEP1_RV},
\end{equation}

\noindent or
\begin{equation}
	\psi_{\scriptscriptstyle{5}}(r) = \frac{\sigma_{\scriptscriptstyle{Ho}}}{\varepsilon_{\scriptscriptstyle{0}}\varepsilon\kappa}e^{-\kappa \left(r- R_{\scriptscriptstyle_{H}}\right)} \label{Plates_MEP2_RV}.
\end{equation}

\noindent Therefore, from \cref{Plates_MEP1_RV}, the electric field in region $V$ can also be expressed in terms of $\varphi_{\scriptscriptstyle{H}}$, as

\begin{equation}
	E_5(r) = \frac{\sigma_{\scriptscriptstyle{5}}(r)}{\varepsilon_{\scriptscriptstyle{0}}\varepsilon}=\kappa \varphi_{\scriptscriptstyle_{H}}e^{-\kappa \left(r- R_{\scriptscriptstyle_{H}}\right)} \label{Plates_EF4_RV}.
\end{equation}
\noindent Consequently, 

\begin{equation}
	\varphi_{\scriptscriptstyle_{H}}=\frac{\sigma_{\scriptscriptstyle{Ho}}}{\varepsilon_{\scriptscriptstyle{0}}\varepsilon\kappa} \label{Plates_phiH-sigmaH_RV}.
\end{equation}

\textbf{2. Region IV:} $\bm{R+d\le r\le R+d+a/2}$\vspace{5pt}

\noindent
Here, $\psi_{\scriptscriptstyle{4}}(r)=Cr+D$, and $E_4(r)=-\bigtriangledown\psi_{\scriptscriptstyle{4}}(r)=-C$. In this region, since there is not a fixed charge at $r=R_{\scriptscriptstyle_{H}}$, the BC on the electric field given by $\varepsilon_{\scriptscriptstyle{0}}\varepsilon E_5(r=R_{\scriptscriptstyle_{H}})-\varepsilon_{\scriptscriptstyle{0}}\varepsilon E_4(r=R_{\scriptscriptstyle_{H}})=0$, implies the continuity of the electric field across the boundary at $r=R_{\scriptscriptstyle_{H}}$, i.e., $E_4(R_{\scriptscriptstyle_{H}})=E_5(R_{\scriptscriptstyle_{H}})$. Thus $-C=\sigma_{\scriptstyle{Ho}}/(\varepsilon_{\scriptscriptstyle{0}}\varepsilon\kappa)$. Also, the MEP must be continuous across the boundary at $r=R+d$. Thence, $\psi_{\scriptscriptstyle{4}}(R_{\scriptscriptstyle_{H}}) =CR_{\scriptscriptstyle_{H}}+D=\varphi_{\scriptscriptstyle_{H}}$. Defining the MEP at $r=R+d$ as $\varphi_{\scriptscriptstyle_{0}}\equiv\psi_{\scriptscriptstyle{4}}(R+d)=C(R+d)+D$, eliminating D from the above equations for $\psi_{\scriptscriptstyle{4}}(R_{\scriptscriptstyle_{H}})$ and $\psi_{\scriptscriptstyle{4}}(R+d)$, and using the expression for $-C$, above, and \cref{Plates_phiH-sigmaH_RV}, we find the useful relation

\begin{equation}
	\varphi_{\scriptscriptstyle_{0}} = \varphi_{\scriptscriptstyle_{H}}+\frac{\sigma_{\scriptscriptstyle{Ho}}}{\varepsilon_{\scriptscriptstyle{0}}\varepsilon}\left(\frac{a}{2}\right)=\left[1+\frac{\kappa a}{2}\right]\varphi_{\scriptscriptstyle_{H}} \label{Plates_phi0},
\end{equation}

\noindent while

\begin{equation}
	\psi_{\scriptscriptstyle{4}}(r)=\frac{\sigma_{\scriptscriptstyle{Ho}}}{\varepsilon_{\scriptscriptstyle{0}}\varepsilon\kappa}[1-\kappa(r-R_{\scriptscriptstyle_{H}})]  \label{Plates_phi-r-sectionIV}.
\end{equation}

\noindent In the slit nanopore geometry, the electrical field in region $IV$ is constant and

\begin{equation}
	 E_4(r)=\frac{\sigma_{\scriptscriptstyle{4}}(r)}{\varepsilon_{\scriptscriptstyle{0}}\varepsilon}=\frac{\sigma_{\scriptscriptstyle{Ho}}}{\varepsilon_{\scriptscriptstyle{0}}\varepsilon}=\kappa\varphi_{\scriptscriptstyle_{H}}\label{Plates_E4r}.
\end{equation}

\textbf{3. Region III:} $\bm{R\le r\le R+d}$	\vspace{5pt}

\noindent
Here, $\psi_{\scriptscriptstyle{3}}(r)=Fr+G$, and the BC for the electric field, at $r=R+d$, is $\varepsilon_{\scriptscriptstyle{0}}\varepsilon E_4(r=R+d)-\varepsilon_{\scriptscriptstyle{0}}\varepsilon E_3(r=R+d)=\sigma_{\scriptscriptstyle{0}}$. Hence, $\varepsilon_{\scriptscriptstyle{0}}\varepsilon\frac{\sigma_{\scriptscriptstyle{Ho}}}{\varepsilon_{\scriptscriptstyle{0}}\varepsilon}-\varepsilon_{\scriptscriptstyle{0}}\varepsilon E_3(r=R+d)=\sigma_{\scriptscriptstyle{0}}$. Thus

\begin{equation}
	E_3(r=R+d)=\frac{\sigma_{\scriptscriptstyle{Ho}}}{\varepsilon_{\scriptscriptstyle{0}}\varepsilon}-\frac{\sigma_{\scriptscriptstyle{0}}}{\varepsilon_{\scriptscriptstyle{0}}\varepsilon}=\frac{1}{\varepsilon_{\scriptscriptstyle{0}}\varepsilon}[\sigma_{\scriptscriptstyle{Ho}}-\sigma_{\scriptscriptstyle{0}}]\label{Plates_E3(R+d)}.
\end{equation}

\noindent Now, defining $\psi_{\scriptscriptstyle{0}}\equiv\psi_{\scriptscriptstyle{3}}(R)$, and since the MEP must be continuous at $r=R+d$, we find that $\psi_{\scriptscriptstyle{3}}(R+d)=\psi_{\scriptscriptstyle{4}}(R+d)=\varphi_{\scriptscriptstyle{0}}=F(R+d)+G$ and $\psi_{\scriptscriptstyle{3}}(R)=\psi_{\scriptscriptstyle{0}}=FR+G$. Eliminating the constant G, from these two expressions, we find $\psi_{\scriptscriptstyle{0}}-\varphi_{\scriptscriptstyle{0}}=F(-d)$. But $E_3 (r)=-\bigtriangledown\psi_{\scriptscriptstyle{3}}(r)=-F$, and using \cref{Plates_E3(R+d)}, we obtain 
\begin{equation}
	E_3(r)=\frac{\sigma_{\scriptscriptstyle{3}}(r)}{\varepsilon_{\scriptscriptstyle{0}}\varepsilon}=\frac{1}{\varepsilon_{\scriptscriptstyle{0}}\varepsilon}[\sigma_{\scriptscriptstyle{Ho}}-\sigma_{\scriptscriptstyle{0}}]\label{Plates_E3r}
\end{equation}

\noindent and
\begin{equation}
	\psi_{\scriptscriptstyle{0}}=\varphi_{\scriptscriptstyle{0}}+\frac{1}{\varepsilon_{\scriptscriptstyle{0}}\varepsilon}[\sigma_{\scriptscriptstyle{Ho}}-\sigma_{\scriptscriptstyle{0}}]d.\label{Plates_MEP3(R)}
\end{equation}

\noindent Finally, from the above equations, straightforwardly one finds

\begin{equation}
\psi_{\scriptscriptstyle{3}}(r)=\varphi_{\scriptscriptstyle{0}}+\frac{1}{\varepsilon_{\scriptscriptstyle{0}}\varepsilon}[\sigma_{\scriptscriptstyle{Ho}}-\sigma_{\scriptscriptstyle{0}}](r-R-d).\label{Plates_MEP3(r)}
\end{equation}

\textbf{4. Region II:} $\bm{R-\frac{a}{2}\le r\le R}$\vspace{5pt}

\noindent
\noindent
Here, $\psi_{\scriptscriptstyle{2}}(r)=Hr+I$, and the BC for the electric field at $r=R$ is $\varepsilon_{\scriptscriptstyle{0}}\varepsilon E_3(r=R)-\varepsilon_{\scriptscriptstyle{0}}\varepsilon E_2(r=R)=\sigma_{\scriptscriptstyle{0}}$. Hence, from \cref{Plates_E3r} $(\sigma_{\scriptscriptstyle{Ho}}-\sigma_{\scriptscriptstyle{0}})-\varepsilon_{\scriptscriptstyle{0}}\varepsilon E_2(r=R)=\sigma_{\scriptscriptstyle{0}}$. Thus

\begin{equation}
	E_2(R)=\frac{\sigma_{\scriptscriptstyle{Ho}}}{\varepsilon_{\scriptscriptstyle{0}}\varepsilon}-\frac{2\sigma_{\scriptscriptstyle{0}}}{\varepsilon_{\scriptscriptstyle{0}}\varepsilon}=\frac{1}{\varepsilon_{\scriptscriptstyle{0}}\varepsilon}[\sigma_{\scriptscriptstyle{Ho}}-2\sigma_{\scriptscriptstyle{0}}]\label{Plates_E2(R)}.
\end{equation}

\noindent \textit{The electroneutrality condition for a planar, charged nanopore, immersed into an electrolyte}, is given by

\begin{equation}
	\sigma_{\scriptscriptstyle{Ho}}+\sigma_{\scriptscriptstyle{Hi}}=2\sigma_{\scriptscriptstyle{0}},\label{Plates_electroneutrality}
\end{equation}
\noindent where, by definition $\sigma_{\scriptscriptstyle{Hi}}\equiv\sigma_{\scriptscriptstyle{2}}(r=[R-a/2])$. However, since within the region $II$, $E_2(r)$ is constant, and

 \begin{equation}
 	E_2(r)=\frac{\sigma_{\scriptscriptstyle{2}}(r)}{\varepsilon_{\scriptscriptstyle{0}}\varepsilon}=E_2(R)=\frac{1}{\varepsilon_{\scriptscriptstyle{0}}\varepsilon}[\sigma_{\scriptscriptstyle{Ho}}-2\sigma_{\scriptscriptstyle{0}}]=-\frac{\sigma_{\scriptscriptstyle{Hi}}}{\varepsilon_{\scriptscriptstyle{0}}\varepsilon},\label{Plates_E2(r)}
 \end{equation}

\noindent the value of $\sigma_{\scriptscriptstyle{2}}(r=[R-a/2])=-\sigma_{\scriptscriptstyle{Hi}}$. On the other hand, $E_2(r)=-\bigtriangledown\psi_{\scriptscriptstyle{2}}(r)=-H$. Thus, from \cref{Plates_E2(r)}, $H=\sigma_{\scriptscriptstyle{Hi}}/(\varepsilon_{\scriptscriptstyle{0}}\varepsilon)$. Hence, the MEP  at $r=R-a/2$ is $\psi_{\scriptscriptstyle_{H}}\equiv\psi_{\scriptscriptstyle{2}}(R-a/2)=(\sigma_{\scriptscriptstyle{Hi}}/(\varepsilon_{\scriptscriptstyle{0}}\varepsilon))[R-a/2]+I$, and at $r=R$,  $\psi_{\scriptscriptstyle{0}}\equiv\psi_{\scriptscriptstyle{2}}(R)=(\sigma_{\scriptscriptstyle{Hi}}/(\varepsilon_{\scriptscriptstyle{0}}\varepsilon))[R]+I$, where we have defined the MEP at $r=R-a/2$ as $\psi_{\scriptscriptstyle_{H}}\equiv\psi_{\scriptscriptstyle{2}}(R-a/2)$. Thus, eliminating the constant I, from these two expressions, we find

\begin{equation}
	\psi_{\scriptscriptstyle_{0}}=\psi_{\scriptscriptstyle_{H}}+\frac{\sigma_{\scriptscriptstyle{Hi}}}{\varepsilon_{\scriptscriptstyle{0}}\varepsilon}\left(\frac{a}{2}\right)=\psi_{\scriptscriptstyle_{H}}+\frac{[2\sigma_{\scriptscriptstyle{0}}-\sigma_{\scriptscriptstyle{Ho}}]}{\varepsilon_{\scriptscriptstyle{0}}\varepsilon}\left(\frac{a}{2}\right)\label{Plates_MEP2(R)}
\end{equation}

Finally, from the above equations, straightforwardly one finds

\begin{equation}
	\psi_{\scriptscriptstyle{2}}(r)=\psi_{\scriptscriptstyle{0}}+\frac{\sigma_{\scriptscriptstyle{Hi}}}{\varepsilon_{\scriptscriptstyle{0}}\varepsilon}(r-R)\label{Plates_MEP2(r)}
\end{equation}
\noindent or

\begin{equation}
	\psi_{\scriptscriptstyle{2}}(r)=\psi_{\scriptscriptstyle{H}}+\frac{\sigma_{\scriptscriptstyle{Hi}}}{\varepsilon_{\scriptscriptstyle{0}}\varepsilon}(r-(R-a/2))\label{Plates_MEP2(r)bis}
\end{equation}

\textbf{5. Region I:} $\bm{0\le r\le R-\frac{a}{2}}$\vspace{5pt}

\noindent
Lastly, here, $\psi_{\scriptscriptstyle{1}}(r)=Le^{\kappa\,r}+Me^{-\kappa\,r}$, and one of the final BCs is that the electric field at the center of the plates, $r=0$, is equal to zero, due to the  symmetry of the nanopore, and in accordance to Gauss' law. Therefore, if $\lim\limits_{r\to0}E_1(r)=0$

\begin{gather}
 \abs{\overrightarrow{\mathbf{E_1(r)}}}\equiv E_1(r)=-\abs{\grad \psi_{\scriptscriptstyle{1}}(r)}=-L\kappa\,e^{\kappa\,r}+M\kappa\,e^{-\kappa\,r}\nonumber\\
	\lim\limits_{r\to0}E_1(r)=-L\kappa\,\cancelto{1}{e^{\kappa\,r}}+M\kappa\,\cancelto{1}{e^{-\kappa\,r}}=0\quad\quad\quad\quad\therefore\quad L=M\label{Ec.ALS_c9}
\end{gather}
Hence, the MEP in this interval and its electric field are

\begin{equation}
\psi_{\scriptscriptstyle{1}}(r)=2L\cosh(\kappa\,r)\label{Plates_MEP-r-region1}
\end{equation}

\begin{equation}
	 E_1(r)=-\dv{\psi_{\scriptscriptstyle{1}}(r)}{r}=-2L\kappa\sinh(\kappa\,r)\label{Plates_EF-r-region1}
\end{equation}

\noindent From the electrostatic theory, at $r=R-a/2$, the electric field must be continuous, i.e., $\varepsilon_{\scriptscriptstyle{0}}\varepsilon E_2(r=R-a/2)-\varepsilon_{\scriptscriptstyle{0}}\varepsilon E_1(r=R-a/2)=0$. Hence, from this relation and \cref{Plates_E2(r)}, $E_1(r=R-a/2)=-\sigma_{\scriptscriptstyle{Hi}}/(\varepsilon_{\scriptscriptstyle{0}}\varepsilon)$. Thus, from this expression for $E_1(R-a/2)$ and \cref{Plates_EF-r-region1}, the constant $L$ is obtained, and therefore

\begin{equation}
	E_1(r)=\frac{\sigma_{\scriptscriptstyle{1}}(r)}{\varepsilon_{\scriptscriptstyle{0}}\varepsilon}=-\frac{\sigma_{\scriptscriptstyle{Hi}}}{\varepsilon_{\scriptscriptstyle{0}}\varepsilon}\frac{sinh[\kappa r]}{sinh[\kappa(R-a/2)]}\label{Plates_EF-final-r-region1}
\end{equation}

\noindent and

\begin{equation}
	\psi_{\scriptscriptstyle{1}}(r)=\frac{\sigma_{\scriptscriptstyle{Hi}}}{\varepsilon_{\scriptscriptstyle{0}}\varepsilon\kappa}\frac{cosh[\kappa r]}{sinh[\kappa(R-a/2)]}\label{Plates_MEP-final-r-region1}
\end{equation}

\noindent In particular,

\begin{equation}
	\psi_{\scriptscriptstyle{d}}\equiv\psi_{\scriptscriptstyle{1}}(r=0)=-\frac{\sigma_{\scriptscriptstyle{Hi}}}{\varepsilon_{\scriptscriptstyle{0}}\varepsilon\kappa}csch[\kappa (R-a/2)],\label{Plates_MEP-r0-region1}
\end{equation}

\noindent and

\begin{equation}
	\psi_{\scriptscriptstyle{d}}-\psi_{\scriptscriptstyle{H}}=\frac{\sigma_{\scriptscriptstyle{Hi}}}{\varepsilon_{\scriptscriptstyle{0}}\varepsilon\kappa}\frac{\left(cosh[\kappa (R-a/2)]-1\right)}{sinh[\kappa (R-a/2)]}.\label{Plates_MEP-(rd-rH)-region1}
\end{equation}

\noindent Therefore the mean electrostatic potential for the slit nanopore is
\begin{equation}
	\psi(r) = 
	\begin{cases}
		\frac{\sigma_{\scriptscriptstyle{Hi}}}{\varepsilon_{\scriptscriptstyle{0}}\varepsilon\kappa}\frac{cosh[\kappa r]}{sinh[\kappa(R-a/2)]}&0\le r\le R-\frac{a}{2}\\
		
		\psi_{\scriptscriptstyle{H}}+\frac{\sigma_{\scriptscriptstyle{Hi}}}{\varepsilon_{\scriptscriptstyle{0}}\varepsilon}(r-(R-a/2))&R-\frac{a}{2}\le r\le R\\
		
		\varphi_{\scriptscriptstyle{0}}-\frac{1}{\varepsilon_{\scriptscriptstyle{0}}\varepsilon}[\sigma_{\scriptscriptstyle{Ho}}-\sigma_{\scriptscriptstyle{0}}](r-R-d)&R\le r\le R+d\\
		
		\frac{\sigma_{\scriptscriptstyle{Ho}}}{\varepsilon_{\scriptscriptstyle{0}}\varepsilon\kappa}[1-\kappa(r-R_{\scriptscriptstyle_{H}})] &R+d\le r\le R_{\scriptscriptstyle_{H}}\\
		
		\frac{\sigma_{\scriptscriptstyle{Ho}}}{\varepsilon_{\scriptscriptstyle{0}}\varepsilon\kappa}e^{-\kappa \left(r- R_{\scriptscriptstyle_{H}}\right)} & R_{\scriptscriptstyle_{H}}\leq r,
	\end{cases}
	\label{Plates-MEP(r)}
\end{equation}

\noindent and its electrical field is

\begin{equation}
	E(r) =\frac{\sigma(r)}{\varepsilon_{\scriptscriptstyle{0}}\varepsilon}= \begin{cases}
	-\frac{\sigma_{\scriptscriptstyle{Hi}}}{\varepsilon_{\scriptscriptstyle{0}}\varepsilon}\frac{sinh[\kappa r]}{sinh[\kappa(R-a/2)]}& 0 \leq r \leq R - \frac{a}{2} \\
		\frac{1}{\varepsilon_{\scriptscriptstyle{0}}\varepsilon}(\sigma_{\scriptscriptstyle{Ho}}-2\sigma_{\scriptscriptstyle{0}}) & R - \frac{a}{2} \leq r \leq R \\
		\frac{1}{\varepsilon_{\scriptscriptstyle{0}}\varepsilon}(\sigma_{\scriptscriptstyle{Ho}}-\sigma_{\scriptscriptstyle{0}}) & R < r < R+d  \\
		\frac{\sigma_{\scriptscriptstyle{Ho}}}{\varepsilon_{\scriptscriptstyle{0}}\varepsilon} & R + d 
		\leq r \leq R_{\scriptscriptstyle_{H}}\\
		\frac{\sigma_{\scriptscriptstyle{Ho}}}{\varepsilon_{\scriptscriptstyle{0}}\varepsilon}e^{-\kappa \left(r- R_{\scriptscriptstyle_{H}} \right)} & R_{\scriptscriptstyle_{H}} \leq r .
	\end{cases}\label{Plates-E(r)}
\end{equation}
%%%%%%%%%%%%%%%%%%%%%%%%%%%%%%%%%%%%%%%%%%

Finally, we need to find expressions for $\sigma_{\scriptscriptstyle{Ho}}$ and $\sigma_{\scriptscriptstyle{Hi}}$ in terms of $\sigma_{\scriptscriptstyle{0}}$, which is the given parameter of the system. From \cref{Plates_phiH-sigmaH_RV,Plates_electroneutrality,Plates_MEP2(R),Plates_MEP3(R),Plates_phi0} we find

\begin{equation}
	\psi_{\scriptscriptstyle_{H}}=\frac{\sigma_{\scriptscriptstyle{Ho}}}{\varepsilon_{\scriptscriptstyle{0}}\varepsilon\kappa}[1+\kappa(a+d)]-\frac{\sigma_{\scriptscriptstyle{0}}}{\varepsilon_{\scriptscriptstyle{0}}\varepsilon\kappa}[\kappa(a+d)].\label{Plates_psiH1-region1}
\end{equation}

However, from \cref{Plates_MEP-final-r-region1}, also

\begin{equation}
	\psi_{\scriptscriptstyle{1}}(R-a/2)=\psi_{\scriptscriptstyle_{H}}=\frac{\sigma_{\scriptscriptstyle{Hi}}}{\varepsilon_{\scriptscriptstyle{0}}\varepsilon\kappa}coth[\kappa(R-a/2)].\label{Plates_psiH2-region1}
\end{equation}

Thus,
\begin{equation}
	\sigma_{\scriptscriptstyle{Ho}}[1+\kappa(a+d)]-\sigma_{\scriptscriptstyle{Hi}}coth[\kappa(R-a/2)]=\sigma_{\scriptscriptstyle{0}}[\kappa(a+d)].\label{Plates_sigma1-region1}
\end{equation}

\noindent Solving the system of \cref{Plates_electroneutrality,Plates_sigma1-region1}, we obtain

\begin{equation}
	\sigma_{\scriptscriptstyle{Ho}}=\left[\frac{\kappa(a+d)+2coth[\kappa(R-a/2)]}{[1+\kappa(a+d)]+coth[\kappa(R-a/2)]}\right]\sigma_{\scriptscriptstyle{0}}.\label{Plates_sigmaHO}
\end{equation}

and
\begin{equation}
	\sigma_{\scriptscriptstyle{Hi}}=\left[\frac{2+\kappa(a+d)}{[1+\kappa(a+d)]+coth[\kappa(R-a/2)]}\right]\sigma_{\scriptscriptstyle{0}}.\label{Plates_sigmaHI}
\end{equation}

%%%%%%%%%%%%%%%%%%%%%%%%%%%%%%%%%%%%%%%%%%%%%%%%%%%%%%%%%%%%%%%%%%%%%%%%%%%%%%%%%%%%%%%%%%%

\subsection{Cylindrical Nanopore}\label{Cylindrical-pore}
Following the same procedure for the cylindrical nanopore as that for the slit nanopore, we obtain the analytical expressions for the MEP, $\psi(r)$, and its associated electric field, $E(r)$, profiles, as a function of $r$:

\begin{equation}
	\psi(r) = \begin{cases}
		-\left( \frac{\left(R_{\scriptscriptstyle_{H}}\right)\sigma_{\scriptscriptstyle{Ho}}}{\varepsilon_{\scriptscriptstyle{0}}\varepsilon} - \frac{\left( 2R+d \right)\sigma_{o} }{\varepsilon_{\scriptscriptstyle{0}} \varepsilon} \right) \frac{I_{\scriptscriptstyle{0}}(\kappa r)}{\kappa \left( R -\frac{a}{2} \right) I_{\scriptscriptstyle{1}}\left[\kappa \left( R -\frac{a}{2} \right) \right]} & 0 \leq r \leq R - \frac{a}{2} \\
		\psi_{\scriptscriptstyle{H}} + \left( \frac{\left(R_{\scriptscriptstyle_{H}}\right)\sigma_{\scriptscriptstyle{Ho}}}{\varepsilon_{\scriptscriptstyle{0}}\varepsilon} - \frac{\left( 2R+d \right)\sigma_{o} }{\varepsilon_{\scriptscriptstyle{0}} \varepsilon} \right)ln\left( \frac{R - \frac{a}{2}}{r} \right) & R - \frac{a}{2} \leq r \leq R \\
		\psi_{\scriptscriptstyle{0}} + \left( \frac{ \left(R_{\scriptscriptstyle_{H}}\right)\sigma_{\scriptscriptstyle{Ho}}}{\varepsilon_{\scriptscriptstyle{0}}\varepsilon} - \frac{(R+d)\sigma_{o}}{\varepsilon_{\scriptscriptstyle{0}} \varepsilon} \right)ln\left( \frac{R}{r} \right) & R \leq r \leq R+d  \\
		\varphi_{\scriptscriptstyle{0}} + \frac{\sigma_{\scriptscriptstyle{Ho}} \left(R_{\scriptscriptstyle_{H}}\right)}{\varepsilon_{\scriptscriptstyle{0}}\varepsilon} ln \left( \frac{R+d}{r} \right) & R + d \leq r \leq R_{\scriptscriptstyle_{H}}\\
		\frac{\varphi_{\scriptscriptstyle{H}}}{K_{\scriptscriptstyle{0}} \left[\kappa R_{\scriptscriptstyle_{H}} \right]} K_{\scriptscriptstyle{0}}[\kappa r] & R_{\scriptscriptstyle_{H}} \leq r .
	\end{cases}\label{Cylinder-MEP(r)}
\end{equation}

\begin{equation}
	E(r) = \frac{\sigma(r)}{\varepsilon_{\scriptscriptstyle{0}}\varepsilon}=\begin{cases}
		\left( \frac{\left(R_{\scriptscriptstyle_{H}}\right)\sigma_{\scriptscriptstyle{Ho}}}{\varepsilon_{\scriptscriptstyle{0}}\varepsilon} - \frac{\left( 2R+d \right)\sigma_{o} }{\varepsilon_{\scriptscriptstyle{0}} \varepsilon} \right) \frac{I_{\scriptscriptstyle{1}}(\kappa r) }{\left( R -\frac{a}{2} \right) I_{\scriptscriptstyle{1}}\left[\kappa \left( R -\frac{a}{2} \right) \right]} & 0 \leq r \leq R - \frac{a}{2} \\
		\left( \frac{\left(R_{\scriptscriptstyle_{H}}\right)\sigma_{\scriptscriptstyle{Ho}}}{\varepsilon_{\scriptscriptstyle{0}}\varepsilon} - \frac{\left( 2R+d \right)\sigma_{o} }{\varepsilon_{\scriptscriptstyle{0}} \varepsilon} \right) \frac{1}{r} & R - \frac{a}{2} \leq r \leq R \\
		\left( \frac{ \left(R_{\scriptscriptstyle_{H}}\right)\sigma_{\scriptscriptstyle{Ho}}}{\varepsilon_{\scriptscriptstyle{0}}\varepsilon} - \frac{(R+d)\sigma_{o}}{\varepsilon_{\scriptscriptstyle{0}} \varepsilon} \right)\frac{1}{r} & R < r < R+d  \\
		\frac{\sigma_{\scriptscriptstyle{Ho}} \left(R_{\scriptscriptstyle_{H}}\right)}{\varepsilon_{\scriptscriptstyle{0}}\varepsilon}\frac{1}{r} & R + d \leq r \leq R_{\scriptscriptstyle_{H}}\\
		\frac{\sigma_{\scriptscriptstyle{Ho}}}{\varepsilon_{\scriptscriptstyle{0}}\varepsilon K_{\scriptscriptstyle{1}} [\kappa R_{\scriptscriptstyle_{H}} ]} K_{\scriptscriptstyle{1}}\left(\kappa r \right) & R_{\scriptscriptstyle_{H}} \leq r .
	\end{cases}\label{Cylinder-E(r)}
\end{equation}

\noindent Where the induced surface charge density $\sigma_{\scriptscriptstyle{Ho}}$  at $r=R_{\scriptscriptstyle_{H}}$ is given by

\begin{equation}
	\sigma_{\scriptscriptstyle{Ho}} = \frac{L_2}{L_1}\sigma_{o}\label{Cylinder-sigmaHO},
\end{equation}

\noindent and since \textit{the electroneutrality condition for the cylindrical nanopore} is given by

\begin{equation}
	(R-a/2)\sigma_{\scriptscriptstyle{Hi}}+(R_{\scriptscriptstyle_{H}})\sigma_{\scriptscriptstyle{Ho}}=R\sigma_{\scriptscriptstyle{0}}+(R+d)\sigma_{\scriptscriptstyle{0}},\label{Cylinder-electroneutrality}
\end{equation}

\noindent an expression for the induced surface charge density at $R-a/2$, $\sigma_{\scriptscriptstyle{Hi}}$, can be directly obtained from \cref{Cylinder-sigmaHO,Cylinder-electroneutrality}.

\noindent with

\begin{eqnarray*}
	L_1 &=& \frac{R_{\scriptscriptstyle_{H}}I_{\scriptscriptstyle{0}} \left[\kappa \left( R -\frac{a}{2} \right) \right]}{\left( R -\frac{a}{2} \right) \kappa I_{\scriptscriptstyle{1}}\left[\kappa \left( R -\frac{a}{2} \right) \right]} + \left(R_{\scriptscriptstyle_{H}}\right)ln \left( \frac{R_{\scriptscriptstyle_{H}}}{R - \frac{a}{2}} \right) + \frac{K_{\scriptscriptstyle{0}} \left[\kappa  R_{\scriptscriptstyle_{H}}  \right]}{\kappa K_{\scriptscriptstyle{1}}\left[\kappa R_{\scriptscriptstyle_{H}}  \right]}\\
	L_2 &=& \frac{\left( 2R+d \right)I_{\scriptscriptstyle{0}} \left[\kappa \left( R -\frac{a}{2} \right) \right]}{\left( R -\frac{a}{2} \right) \kappa I_{\scriptscriptstyle{1}}\left[\kappa \left( R -\frac{a}{2} \right) \right]} + \left[ 2R+d \right]ln\left( \frac{R}{R - \frac{a}{2}} \right)\\ & & + (R+d)ln\left( \frac{R+d}{R} \right)
\end{eqnarray*}

Below we give the analytical expressions for some specific, relevant values of the MEP for the cylindrical nanopore:

\begin{gather}
	\varphi_{\scriptscriptstyle_{H}}\equiv\psi \left( R_{\scriptscriptstyle_{H}} \right) = \frac{K_{\scriptscriptstyle{0}} \left[\kappa R_{\scriptscriptstyle_{H}} \right] \sigma_{\scriptscriptstyle{Ho}}}{\varepsilon_{\scriptscriptstyle{0}}\varepsilon \kappa K_{\scriptscriptstyle{1}} \left[\kappa R_{\scriptscriptstyle_{H}} \right]}\label{Cyl-spe-values-MEP1} \\
	\varphi_{\scriptscriptstyle{0}}\equiv\psi(R+d) =\varphi_{\scriptscriptstyle{H}} + \frac{R_{\scriptscriptstyle_{H}}\sigma_{\scriptscriptstyle{Ho}} }{\varepsilon_{\scriptscriptstyle{0}}\varepsilon} ln \left( \frac{R_{\scriptscriptstyle_{H}}}{R+d} \right)\label{Cyl-spe-values-MEP2} \\
	\psi_{\scriptscriptstyle{0}}\equiv\psi(R) = \varphi_{\scriptscriptstyle{0}} + \left( \frac{ R_{\scriptscriptstyle_{H}}\sigma_{\scriptscriptstyle{Ho}}}{\varepsilon_{\scriptscriptstyle{0}}\varepsilon} - \frac{(R+d)\sigma_{o}}{\varepsilon_{\scriptscriptstyle{0}} \varepsilon} \right)ln\left( \frac{R+d}{R} \right)\label{Cyl-spe-values-MEP3} \\
	\psi_{\scriptscriptstyle{H}}\equiv\psi\left( R-\frac{a}{2} \right) = \psi_{\scriptscriptstyle{0}} + \left( \frac{R_{\scriptscriptstyle_{H}}\sigma_{\scriptscriptstyle{Ho}}}{\varepsilon_{\scriptscriptstyle{0}}\varepsilon} - \frac{(2R+d)\sigma_{o} }{\varepsilon_{\scriptscriptstyle{0}} \varepsilon} \right)ln\left( \frac{R}{R - \frac{a}{2}} \right)\label{Cyl-spe-values-MEP4} \\ 	\psi_{\scriptscriptstyle{d}}\equiv\psi\left(r=0 \right) = \psi_{\scriptscriptstyle{H}} + \frac{\left[R_{\scriptscriptstyle_{H}}\sigma_{\scriptscriptstyle{Ho}}-(2R+d)\sigma_{o}\right](I_{\scriptscriptstyle{0}}[\kappa(R-a/2)]-1)}{\varepsilon_{\scriptscriptstyle{0}}\varepsilon\kappa(R-a/2)I_{\scriptscriptstyle{1}}[\kappa(R-a/2)]} \label{Cyl-spe-values-MEP5}
\end{gather}
\noindent As in the case of the slit nanopore $R_{\scriptscriptstyle_{H}}\equiv R+d+a/2$. A derivation of all the equations for the cylindrical nanopore, and in particular for \crefrange{Cyl-spe-values-MEP1}{Cyl-spe-values-MEP5}, is given in appendix A.

%%%%%%%%%%%%%%%%%%%%%%%%%%%%%%%%%%%%%%%%%%%%%%%%%%%%%%%%%%%%%%%%%%%%%%%%%%%%%%%%%%%%%%%
\subsection{Spherical Nanopore}\label{Spherical-pore}
Similarly for the spherical nanopore we obtain the analytical expressions for the MEP, $\psi(r)$, and its associated electric field, $E(r)$, profiles, as a function of $r$:

\begin{equation}
	\psi(r) = \begin{cases}
		\frac{  \left(R_{\scriptscriptstyle_{H}}^2\sigma_{\scriptscriptstyle{Ho}} - \left[ (R+d)^2 + R^2 \right] \sigma_{o} \right)}{\varepsilon_{\scriptscriptstyle{0}}\varepsilon\left( sinh \left[\kappa \left( R -\frac{a}{2} \right) \right] -\kappa \left[ R -\frac{a}{2} \right] cosh[\kappa \left( R -\frac{a}{2} \right) ] \right)} \frac{sinh\left[ \kappa r \right]}{r} & 0 \leq r \leq R - \frac{a}{2} \\[15pt]
		\psi_{\scriptscriptstyle{H}} + \frac{\left(R_{\scriptscriptstyle_{H}}^2\sigma_{\scriptscriptstyle{Ho}} - \left[ (R+d)^2 + R^2 \right] \sigma_{o}\right)}{\varepsilon_{\scriptscriptstyle{0}}\varepsilon (R-\frac{a}{2})} \left( \frac{R-\frac{a}{2}}{r} - 1 \right) & R - \frac{a}{2} \leq r \leq R \\[15pt]
		\psi_{\scriptscriptstyle{0}} + \frac{\left(  R_{\scriptscriptstyle_{H}}^2\sigma_{\scriptscriptstyle{Ho}} - (R+d)^2 \sigma_{o} \right)}{\varepsilon_{\scriptscriptstyle{0}}\varepsilon R} \left( \frac{R}{r} - 1 \right) & R \leq r \leq R+d  \\[15pt]
		\varphi_{\scriptscriptstyle{0}} +  \frac{R_{\scriptscriptstyle_{H}}^2\sigma_{\scriptscriptstyle{Ho}}}{\varepsilon_{\scriptscriptstyle{0}}\varepsilon(R+d)} \left( \frac{(R+d)}{r} - 1 \right) & R + d \leq r \leq R_{\scriptscriptstyle_{H}}\\[15pt]
		R_{\scriptscriptstyle_{H}} \varphi_{\scriptscriptstyle{H}} \frac{e^{-\kappa \left(r-\left[ R_{\scriptscriptstyle_{H}} \right] \right)}}{r} & R_{\scriptscriptstyle_{H}} \leq r .
	\end{cases}\label{Sphere-MEP(r)}
\end{equation}

\begin{equation}
	E(r) = \begin{cases}
		\frac{ \left( R_{\scriptscriptstyle_{H}}^2\sigma_{\scriptscriptstyle{Ho}} - \left[ (R+d)^2 + R^2 \right] \sigma_{o}\right) }{\varepsilon_{\scriptscriptstyle{0}}\varepsilon\left( sinh \left[\kappa \left( R -\frac{a}{2} \right) \right] -\kappa \left[ R -\frac{a}{2} \right] cosh[\kappa \left( R -\frac{a}{2} \right) ] \right)} \frac{sinh[\kappa r] -\kappa r cosh[\kappa r] }{r^2}  & 0 \leq r \leq R - \frac{a}{2} \\[15pt]
		\left(\frac{R_{\scriptscriptstyle_{H}}^2\sigma_{\scriptscriptstyle{Ho}}-\left[(R+d)^2 + R^2\right]\sigma_{\scriptscriptstyle{0}}}{\varepsilon_{\scriptscriptstyle{0}}\varepsilon}\right) \frac{1}{r^2}  & R - \frac{a}{2} \leq r \leq R \\[15pt]
		\left(\frac{R_{\scriptscriptstyle_{H}}^2\sigma_{\scriptscriptstyle{Ho}}-(R+d)^2\sigma_{\scriptscriptstyle{0}}}{\varepsilon_{\scriptscriptstyle{0}}\varepsilon}\right)\frac{1}{r^2} & R < r < R+d  \\[15pt]
		\frac{R_{\scriptscriptstyle_{H}}^2\sigma_{\scriptscriptstyle{Ho}}}{\varepsilon_{\scriptscriptstyle{0}}\varepsilon}\frac{1}{r^2} & R + d \leq r \leq R_{\scriptscriptstyle_{H}}\\[15pt]
		\frac{R_{\scriptscriptstyle_{H}}^2 \left( 1+\kappa r \right) \sigma_{\scriptscriptstyle{Ho}}}{\varepsilon_{\scriptscriptstyle{0}}\varepsilon \left( 1+\kappa \left[ R_{\scriptscriptstyle_{H}} \right] \right)} \left[ \frac{e^{-\kappa\left(r -  \left[ R_{\scriptscriptstyle_{H}} \right] \right )}}{r^2} \right] & R_{\scriptscriptstyle_{H}} \leq r .
	\end{cases}\label{Sphere-E(r)}
\end{equation}

Where the induced surface charge density $\sigma_{\scriptscriptstyle{Ho}}$  on the surfaces at $r=R+d+a/2=R_{\scriptscriptstyle_{H}}$ is given by

\begin{equation}
	\sigma_{\scriptscriptstyle{Ho}} = \frac{L_2}{L_1}\sigma_{o}\label{Sphere-sigmaHO},
\end{equation}

\noindent and since \textit{the electroneutrality condition for the spherical nanopore} is given by

\begin{equation}
	(R-a/2)^2\sigma_{\scriptscriptstyle{Hi}}+(R_{\scriptscriptstyle_{H}})^2\sigma_{\scriptscriptstyle{Ho}}=R^2\sigma_{\scriptscriptstyle{0}}+(R+d)^2\sigma_{\scriptscriptstyle{0}},\label{Sphere-electroneutrality}
\end{equation}

\noindent an expression for for the surface charge density at $r=R-a/2$, $\sigma_{\scriptscriptstyle{Hi}}$, can be directly obtained from \cref{Sphere-sigmaHO,Sphere-electroneutrality}.

\noindent with
\begin{eqnarray*}
	L_1 &=& \frac{R_{\scriptscriptstyle_{H}}}{\left( 1+\kappa R_{\scriptscriptstyle_{H}} \right)} + \frac{\frac{a}{2} R_{\scriptscriptstyle_{H}}}{(R+d)} + \frac{dR_{\scriptscriptstyle_{H}}^2}{R(R+d)} + \frac{\frac{a}{2}R_{\scriptscriptstyle_{H}}^2}{R\left(R-\frac{a}{2} \right)}\\
	& &- \frac{sinh\left[ \kappa \left(R- \frac{a}{2}\right) \right] R_{\scriptscriptstyle_{H}}^2}{\left( R- \frac{a}{2} \right) \left[ sinh \left[ \kappa \left(R- \frac{a}{2}\right) \right] -\kappa \left(R- \frac{a}{2}\right) cosh\left[ \kappa\left(R- \frac{a}{2}\right) \right] \right]} \\[15pt]
	L_2 &=& \frac{d(R+d)}{R} + \frac{\frac{a}{2}\left[ (R+d)^2 + R^2 \right]}{R\left(R-\frac{a}{2} \right)}\\
	& & - \frac{sinh\left[ \kappa \left(R- \frac{a}{2}\right) \right]\left( R^2+ (R+d)^2 \right)}{\left( R- \frac{a}{2} \right) \left[ sinh \left[ \kappa \left(R- \frac{a}{2}\right) \right] -\kappa \left(R- \frac{a}{2}\right) cosh\left[ \kappa\left(R- \frac{a}{2}\right) \right] \right]}
\end{eqnarray*}

Specific, relevant relations of the MEP for the spherical nanopore are provided below:

\begin{gather}
	\varphi_{\scriptscriptstyle_{H}}\equiv\psi (R_{\scriptscriptstyle_{H}}) = \frac{R_{\scriptscriptstyle_{H}} \sigma_{\scriptscriptstyle{Ho}}}{\varepsilon_{\scriptscriptstyle{0}}\varepsilon \left( 1+\kappa R_{\scriptscriptstyle_{H}} \right)}\label{Sph-spe-values-MEP1} \\
	\varphi_{\scriptscriptstyle_{0}}\equiv\psi(R+d) = \varphi_{\scriptscriptstyle_{H}} + \frac{ R_{\scriptscriptstyle_{H}}}{\varepsilon_{\scriptscriptstyle{0}}\varepsilon(R+d)}\frac{a}{2}\sigma_{\scriptscriptstyle{Ho}}\label{Sph-spe-values-MEP2}\\
	\psi_{\scriptscriptstyle_{0}}\equiv\psi(R) =  \varphi_{\scriptscriptstyle_{0}} + \frac{\left[\left(R_{\scriptscriptstyle_{H}}\right)^2\sigma_{\scriptscriptstyle{Ho}}-(R+d)^2\sigma_{\scriptscriptstyle{0}}\right]d}{\varepsilon_{\scriptscriptstyle{0}}\varepsilon R(R+d)}\label{Sph-spe-values-MEP3}\\
	\psi_{\scriptscriptstyle_{H}}\equiv \psi\left( R-\frac{a}{2} \right) = \psi_{\scriptscriptstyle{0}} +  \frac{[\left(R_{\scriptscriptstyle_{H}}\right)^2\sigma_{\scriptscriptstyle{Ho}}-\left[ (R+d)^2 + R^2 \right]\sigma_{\scriptscriptstyle{0}}]\frac{a}{2}}{\varepsilon_{\scriptscriptstyle{0}}\varepsilon R \left(R-\frac{a}{2} \right)}\label{Sph-spe-values-MEP4}\\
	\psi_{\scriptscriptstyle_{d}}\equiv \psi( r=0 ) = \psi_{\scriptscriptstyle{H}} +  \frac{[\left(R_{\scriptscriptstyle_{H}}\right)^2\sigma_{\scriptscriptstyle{Ho}}-\left[ (R+d)^2 + R^2 \right]\sigma_{\scriptscriptstyle{0}}]}{\varepsilon_{\scriptscriptstyle{0}}\varepsilon  \left(R-\frac{a}{2} \right)}F_s(\kappa(R-a/2)),\label{Sph-spe-values-MEP5}
\end{gather}

\noindent where

\begin{equation*}
	F_s(\kappa(R-a/2)) \equiv  \frac{(\kappa(R-a/2)-sinh[\kappa(R-a/2)])}{(sinh[\kappa(R-a/2)]-\kappa(R-a/2)cosh[\kappa(R-a/2)])}
\end{equation*}
\noindent A derivation of all the equations for the spherical nanopore, and in particular for \crefrange{Sph-spe-values-MEP1}{Sph-spe-values-MEP5},  is outlined in appendix B.
\subsection{Nanopores capacitances equations}\label{Nanopores-capacitances}

As shown above, a charged nanopore immersed into an electrolyte induces an electrical double layer inside and outside its walls. The self-capacitance of this nanopore electrode is given by the charge delivered from a reference point, say infinity, to their walls, thus producing a potential difference. The voltage differences through the five electrostatic regions are depicted in \cref{Geometry_electrodes}.
Hence, the capacitance of the nanopore is that of five capacitors, corresponding to each of these five regions, connected in series, i.e.,

\begin{equation}
	C_d=\left(\frac{1}{C_1}+\frac{1}{C_2}+\frac{1}{C_3}+\frac{1}{C_4}+\frac{1}{C_5}\right)^{-1}\label{Ec.CT}
\end{equation}

\noindent The capacitance \textit{per unit area} in each of these regions is given by

\begin{gather}
	c_j=\frac{C_j}{A_j(r_i)}=\frac{(Q_j(r_i)/A_j(r_i))}{\Delta\psi_j(r)}=\sigma_j(r_i)/\Delta\psi_j(r_i)
	\label{Ec.Cs.ind-definition}
\end{gather}

\noindent where the charge, $Q_j(r_i)$, surface area, $A_j(r_i)$, and voltage difference, $\Delta\psi_j(r_i)$, are specific for each region of the nanopore, i.e., $j$ refers to the region of the nanopore, and $i$ to the specific location of the charge, area and potential difference, where the capacitance $c_j$ is calculated. Correspondingly, the total capacitance, \textit{per unit of area}, i.e., the \textit{specific capacitance}, is given by 

\begin{equation}
	c_{\scriptscriptstyle{d}}= \left(\sum_{j=1} ^{5}\frac{1}{c_j}\right)^{-1}.\label{Ec.cT}
\end{equation}

 \noindent Analytical expressions for $\sigma_j(r_i)$ and $\Delta \psi_j(r_i)$ have been given above for each of the five regions, of the three nanopore topologies. Thus, from \cref{Cylinder-E(r)} and \crefrange{Cyl-spe-values-MEP1}{Cyl-spe-values-MEP5} for the cylindrical nanopore, its five regions capacitances are
\begin{gather}	
	c_1=\frac{\sigma_{\scriptscriptstyle{1}}(R-a/2)}{\psi_{\scriptscriptstyle{d}}-\psi_{\scriptscriptstyle_{H}}}
	=\frac{\varepsilon_{\scriptscriptstyle{0}}\varepsilon\kappa I_{\scriptscriptstyle_{1}}[\kappa(R-a/2)]}{(I_{\scriptscriptstyle_{0}}[\kappa (R-a/2)]-1)}	\label{Equ.Cap-Cyl1}\\
	c_2=\frac{\sigma_{\scriptscriptstyle{2}}(R)}{\psi_{\scriptscriptstyle{H}}-\psi_{\scriptscriptstyle_{0}}}=\frac{\varepsilon_{\scriptscriptstyle{0}}\varepsilon}{Rln\left[\frac{R}{R-a/2}\right]}	\label{Equ.Cap-Cyl2}\\c_3=\frac{\sigma_{\scriptscriptstyle{3}}(R)}{\psi_{\scriptscriptstyle{0}}-\varphi_{\scriptscriptstyle{0}}}=\frac{\varepsilon_{\scriptscriptstyle{0}}\varepsilon}{Rln\left[\frac{R+d}{R}\right]} 
		\label{Equ.Cap-Cyl3}\\c_4=\frac{\sigma_4(R+d)}{\varphi_0-\varphi_H}
	= \frac{\varepsilon_{\scriptscriptstyle{0}}\varepsilon}{(R+d)ln\left[ \frac{R_{\scriptscriptstyle_{H}}}{R+d} \right]} 	\label{Equ.Cap-Cyl4}\\c_5= \frac{\sigma_{\scriptscriptstyle{Ho}}}{\varphi_H} =\frac{\varepsilon_{\scriptscriptstyle{0}}\varepsilon\kappa K_{\scriptscriptstyle_{1}}\left[\kappa R_{\scriptscriptstyle_{H}}\right]}{K_{\scriptscriptstyle_{0}}\left[\kappa R_{\scriptscriptstyle_{H}}\right]}  \label{Equ.Cap-Cyl5}
\end{gather}
\noindent The derivation of \crefrange{Equ.Cap-Cyl1}{Equ.Cap-Cyl5} is given in appendix A.
 
\noindent Similarly, for the spherical nanopore,

\begin{gather}	
	c_1=\frac{\sigma_{\scriptscriptstyle{1}}(R-a/2)}{\psi_{\scriptscriptstyle{d}}-\psi_{\scriptscriptstyle_{H}}}
	=\frac{\varepsilon_{\scriptscriptstyle{0}}\varepsilon\left[sinh[\kappa(R-a/2)]-\kappa(R-a/2)cosh[\kappa(R-a/2)]\right]}{(R-a/2)\left[\kappa(R-a/2)-sinh[\kappa(R-a/2)]\right]}  \label{Equ.Cap-Sphe1} \\c_2=\frac{\sigma_{\scriptscriptstyle{2}}(R)}{\psi_{\scriptscriptstyle{H}}-\psi_{\scriptscriptstyle_{0}}}=\frac{\varepsilon_{\scriptscriptstyle{0}}\varepsilon(R-a/2)}{R(a/2)}  \label{Equ.Cap-Sphe2} \\c_3=\frac{\sigma_{\scriptscriptstyle{3}}(R)}{\psi_{\scriptscriptstyle{0}}-\varphi_{\scriptscriptstyle{0}}}=\frac{\varepsilon_{\scriptscriptstyle{0}}\varepsilon(R+d)}{Rd} \label{Equ.Cap-Sphe3}
	\\c_4=\frac{\sigma_4(R+d)}{\varphi_0-\varphi_H}
	= \frac{\varepsilon_{\scriptscriptstyle{0}}\varepsilon R_{\scriptscriptstyle_{H}}}{(R+d)(a/2)}  \label{Equ.Cap-Sphe4} \\c_5= \frac{\sigma_{\scriptscriptstyle{Ho}}}{\varphi_H} =\frac{\varepsilon_{\scriptscriptstyle{0}}\varepsilon (1+\kappa R_{\scriptscriptstyle_{H}})}{R_{\scriptscriptstyle_{H}}}  \label{Equ.Cap-Sphe5}
\end{gather}
\noindent The derivation of \crefrange{Equ.Cap-Sphe1}{Equ.Cap-Sphe5} is given in appendix B.
Lastly, for the right-hand side of the slit nanopore, we have

\begin{gather}	
	c_1=\frac{\sigma_{\scriptscriptstyle{1}}(R-a/2)}{\psi_{\scriptscriptstyle{d}}-\psi_{\scriptscriptstyle_{H}}}
	=\frac{\varepsilon_{\scriptscriptstyle{0}}\varepsilon\kappa sinh[\kappa(R-a/2)]}{\left[cosh[\kappa(R-a/2)]-1\right]}	\label{Equ.Cap-Plate1}\\c_2=\frac{\sigma_{\scriptscriptstyle{2}}(R)}{\psi_{\scriptscriptstyle{H}}-\psi_{\scriptscriptstyle_{0}}}=\frac{\varepsilon_{\scriptscriptstyle{0}}\varepsilon}{(a/2)}	\label{Equ.Cap-Plate2}\\c_3=\frac{\sigma_{\scriptscriptstyle{3}}(R)}{\psi_{\scriptscriptstyle{0}}-\varepsilon_{\scriptscriptstyle{0}}}=\frac{\varepsilon_{\scriptscriptstyle{0}}\varepsilon}{d} 	\label{Equ.Cap-Plate3}
	\\c_4=\frac{\sigma_4(R+d)}{\varphi_0-\varphi_H}
	= \frac{\varepsilon_{\scriptscriptstyle{0}}\varepsilon}{(a/2)} 	\label{Equ.Cap-Plate4}\\c_5= \frac{\sigma_{\scriptscriptstyle{Ho}}}{\varphi_H} =\varepsilon_{\scriptscriptstyle{0}}\varepsilon\kappa	\label{Equ.Cap-Plate5} 
\end{gather}
\noindent Therefore, an analytical expression for the \textit{specific} capacitance, $c_{\scriptscriptstyle{d}}$, for the planar, cylindrical and spherical nanopores can be straightforwardly found from \cref{Ec.cT} and their corresponding $c_i$, given above for each geometry. In \cref{Equ.Cap-Plate1,Equ.Cap-Plate2,Equ.Cap-Plate3,Equ.Cap-Plate4,Equ.Cap-Plate5} for the slit nanopore capacitance, the left side of the slit is implicitly taken into account through the right side of the EDL inside and outside the nanopore, since they are in all respects correlated with those associated with the left plate. In fact, the two plates plus the associated EDLs inside and outside the slit form a single device.

It should be point out that in general a capacitance or, sometimes referred to as integral capacitance, is defined as $C=Q/V$ and the differential capacitance as $C_d=dQ/dV$. However, because we are using a linear theory, in all geometries the capacitance does not directly depends of the charge densities. Hence, in this approximation, the capacitance is equal to its differential capacitance. However, hereinafter we will use only the terms capacitance, $C$, and specific capacitance, $c$.

Our study, still within the linear Poisson-Boltzmann, can be extended to non-symmetrical electrolytes in charge and ionic size. However, we will not include these results here. Additionally, although perhaps not relevant for nanocapacitors, but important for micelles and cellular biology is the net pressure on the nanopore walls. For this purpose, the use of contact theorems, based on the balance of steric and electrostatic forces (Maxwell stress tensor) on a given surface, are convenient. Contact theorems for planar~\cite{Henderson-Contact-JCP-1978}, cylindrical and spherical electrodes~\cite{Bari-Contact-Mol-Phys-2015,Holovko-Contact-2023}, as well as for planar~\cite{Lozada_1984,Lozada_1990-I}, cylindrical and spherical nanopores have published in the past~\cite{Yu_1997,Aguilar_2007}. The nanopores contact conditions, of course, reduces to those for nano-electrodes, making the internal radius equal to zero. The application of these contact theorems, together with the concentration profiles and electrical field analytical expressions here derived, to calculate the aforementioned pressure is possible, but we will leave this study for a future publication.

%%%%%%%%%%%%%%%%%%%%%%%%%%%%%%%%%%%%%%%%%%%%%%%%%%%%%%%%%%%%%%%%%%%%%%%%%%%%%%%%%%%%%%%%%%%%

\section{Results and discussion}\label{Res_Disc}

 From \cref{Rho_elx2}, for the mean electrostatic potential of the different nanopores topologies, their corresponding co-ion, $g_{\scriptscriptstyle{+}}(r)$, and counter-ion, $g_{\scriptscriptstyle{-}}(r)$, distribution functions, can be obtained, i.e.,

\begin{equation}
	g_{\scriptscriptstyle{+}}(r) = 
	\begin{cases}
		\exp(-\frac{z\,e\,\psi_{\scriptscriptstyle_{1}}(r)}{k\,T})&0\le r\le R-\frac{a}{2}\\
		
		0&R-\frac{a}{2}< r< R_{\scriptscriptstyle_{H}}\\
		
		\exp(-\frac{z\,e\,\psi_{\scriptscriptstyle_{5}}(r)}{k\,T})&r\ge R_{\scriptscriptstyle_{H}}
	\end{cases}
	\label{Ec.gmas}
\end{equation}
\begin{equation}
	g_{\scriptscriptstyle{-}}(r) = 
	\begin{cases}
		\exp(\frac{z\,e\,\psi_{\scriptscriptstyle_{1}}(r)}{k\,T})&0\le r\le R-\frac{a}{2}\\
		
		0&R-\frac{a}{2}< r< R_{\scriptscriptstyle_{H}}\\
		
		\exp(\frac{z\,e\,\psi_{\scriptscriptstyle_{5}}(r)}{k\,T})&r\ge R_{\scriptscriptstyle_{H}}
	\end{cases}
	\label{Ec.gmenos}
\end{equation}

\noindent Although \cref{Ec.gmas,Ec.gmenos} are symmetrically valid for positively or negatively charged nanopores, hereinafter we will assume a positive charge on the nanopores, and, hence, the electrolyte's cations and anions become the nanopore's co-ion and counter-ion, respectively. As pointed in \cref{Theory}, we will refer to $r$, as the distance to the nanopore's geometrical center, for all nanopore's topologies. \textit{In all of our calculations, we have taken the first-order Taylor expansion of \cref{Ec.gmas,Ec.gmenos}, to be consistent with \cref{Ec.ALS_Poisson}}.

Unless stated otherwise, for all our results, we consider a positively charged electrode immersed into an aqueous symmetric electrolyte (1:1), with an electrical relative permittivity, $\varepsilon$, of $78.5$, and an ion's size, $a$, of $\SI{4.25}{\angstrom}$, and pore's walls width, $d$, of one ionic diameter, i.e., $d$=$\SI{4.25}{\angstrom}$, at room temperature (\SI{298}{\kelvin}).
\subsection{The ions concentration profiles}

\begin{figure}[!htb]
	\begin{subfigure}{.5\textwidth}
		\centering
		\includegraphics[width=0.95\linewidth]{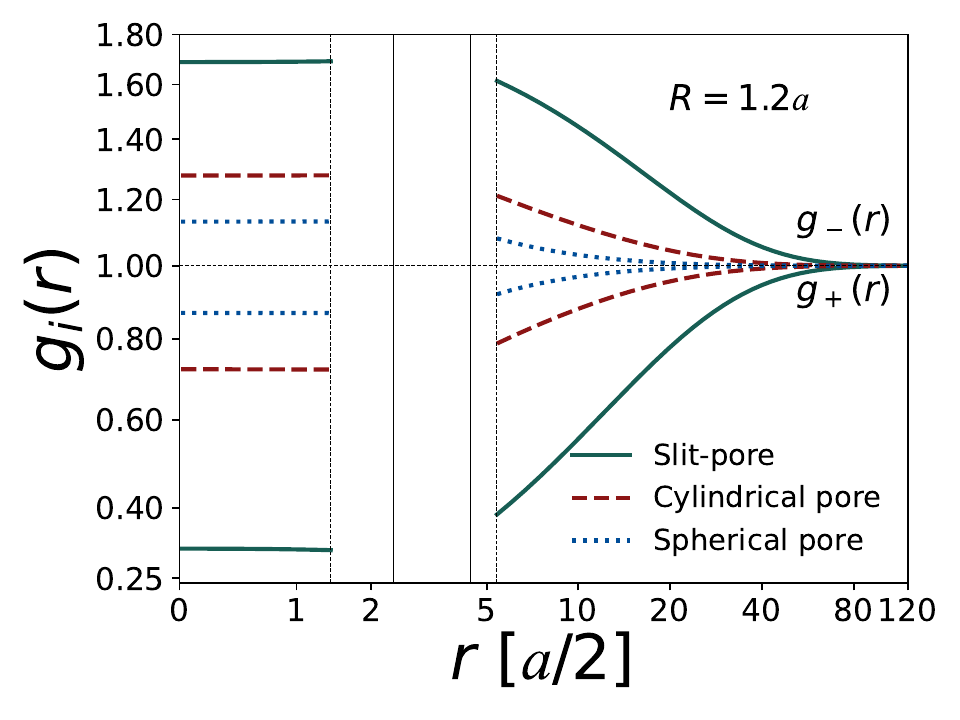}
		\caption{The concentration profiles for $R=\SI{1.2} {\textit{a}}$.}
		\label{Fig.PCS.g(r).R1.2}
	\end{subfigure}
	\begin{subfigure}{.5\textwidth}
		\centering
		\includegraphics[width=0.95\linewidth]{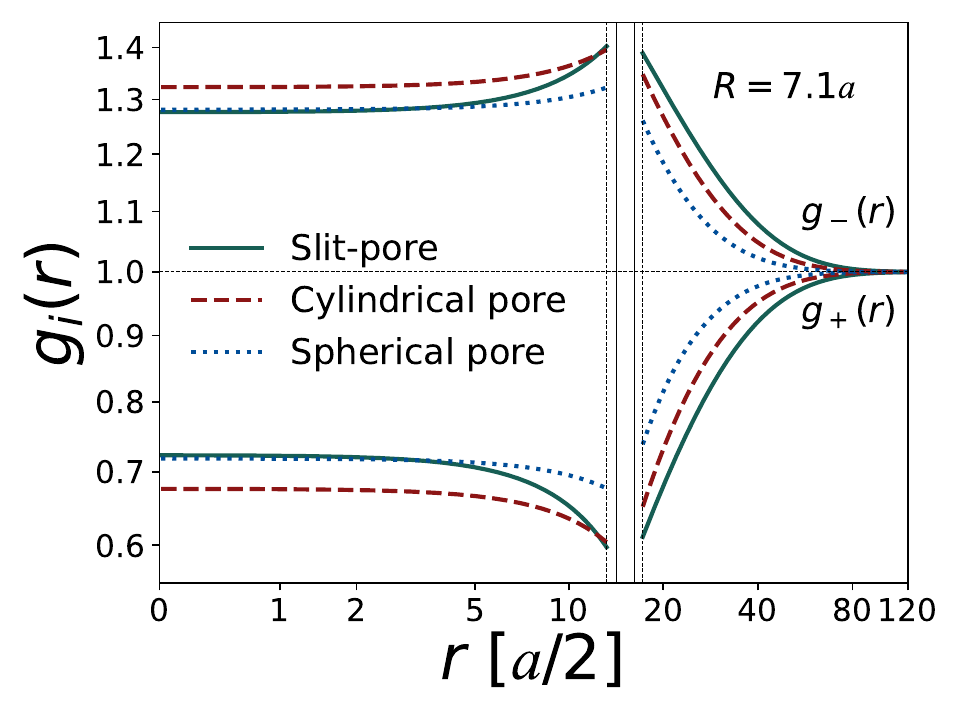}
		\caption{The concentration profiles for $R=\SI{7,1}{\textit{a}}$.}
		\label{Fig.PCS.g(r).R7.1}
	\end{subfigure}\\
	\begin{subfigure}{.5\textwidth}
		\centering
		\includegraphics[width=0.95\linewidth]{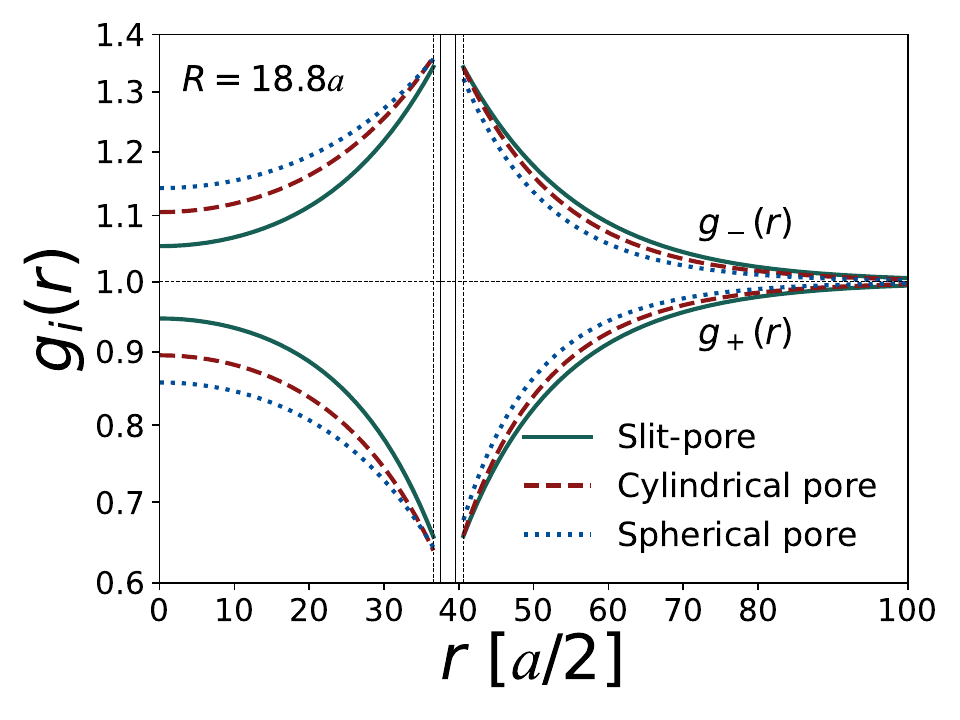}
		\caption{The concentration profiles for $R=\SI{18.8}{\textit{a}}$.}
		\label{Fig.PCS.g(r).R18p8}
	\end{subfigure}
	\begin{subfigure}{.5\textwidth}
		\centering
		\includegraphics[width=0.95\linewidth]{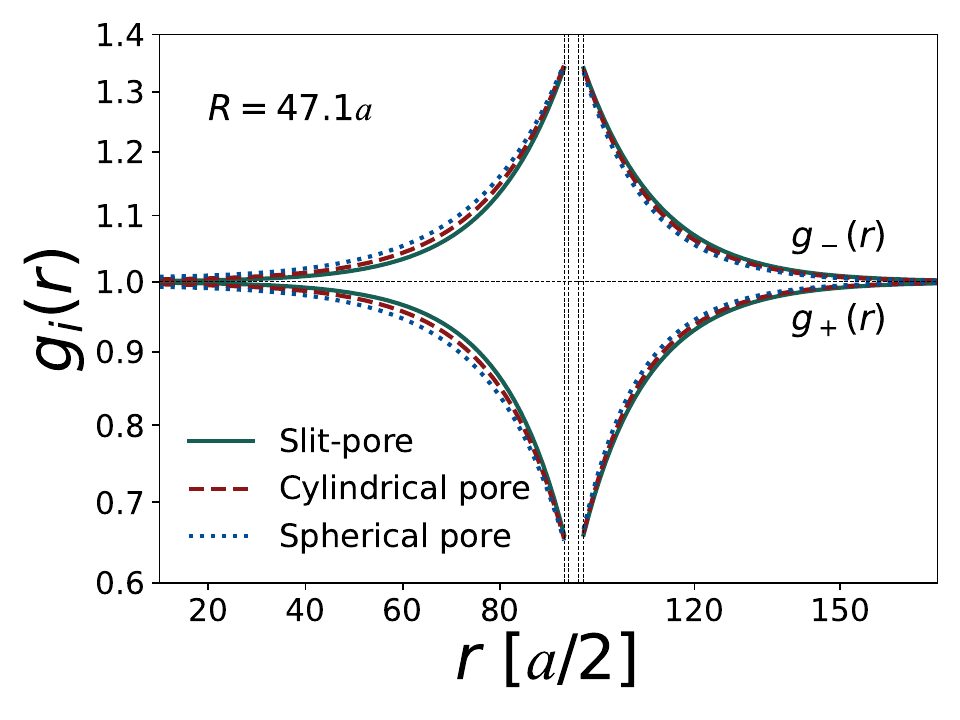}
		\caption{The concentration profiles for $R=\SI{47,1}{\textit{a}}$.} 		\label{Fig.PCS.g(r).R47p1}
	\end{subfigure}
	\caption{The co-ion, $g_{\scriptscriptstyle{+}}(r)$, and counter-ion, $g_{\scriptscriptstyle{-}}(r)$, distribution functions of different nanopore geometries, with distinct radii values and a constant pore's width ($d=a$).
	The vertical dashed and straight lines represent the outer Helmholtz plane (OHP), and inner Helmholtz plane (IHP).
	In \cref{Fig.PCS.g(r).R1.2,Fig.PCS.g(r).R7.1,Fig.PCS.g(r).R18p8,Fig.PCS.g(r).R47p1}, the electrode is immersed in an aqueous symmetric electrolyte (1:1, $\varepsilon=78.5$, $a=\SI{4.25}{\angstrom}$, $\rho_{\scriptscriptstyle{0}}=\SI{0.01}{\Molar}$) at room temperature (\SI{298}{\kelvin}), and $\sigma_{\scriptscriptstyle{0}}$=$\,\SI{0.002}{\si[per-mode=symbol]{\coulomb\per\square\metre}}$. In (a) and (b) a $sinh^-1(r)$ scale is used.}
	\label{Fig.PCS.g(r)_s0.002}
\end{figure}

In relation to the surface charge density on the nanopore's walls, in this paper we will limit ourselves to the case in which the inner and outer surface charges are equal, i.e.,  $\sigma_{\scriptscriptstyle{I}}$=$\,\sigma_{\scriptscriptstyle{II}}$=$\sigma_{\scriptscriptstyle{0}}$.
In \cref{Fig.PCS.g(r)_s0.002} we show the ionic distribution functions for a constant surface charge density, $\,\sigma_{\scriptscriptstyle{0}}$=$\,\SI{0.002}{\si[per-mode=symbol]{\coulomb\per\square\metre}}$, and bulk electrolyte concentration, $\rho_{\scriptscriptstyle{0}}=\SI{0.01}{\Molar}$, for the three nanopore's topologies considered above; and for different radii sizes for the slit, cylindrical and spherical nanopores. 
However, notice that the total charge on the slit pore is $4\,\sigma_{\scriptscriptstyle{0}}\times A_{\scriptscriptstyle {M}}$ and for the cylindrical and spherical nanopores are $\sigma_{\scriptscriptstyle{0}}\times 2\,\pi\,[R+(R+d)]\,L$, and  $\sigma_{\scriptscriptstyle{0}}\times4\,\pi[R^2+(R+d)^2]$, respectively.  However, considering the symmetry of the slit nanopore, we can focus only on half of it, and thus the total surface charge on half of the nanopores is $2\,\sigma_{\scriptscriptstyle{0}}\times A_{\scriptscriptstyle {M}}$.
$A_{\scriptscriptstyle {M}}$ is the area of the slit surfaces, and $L$ is the length of the cylinder.
In our models these quantities are infinite for the slit and cylindrical nanopores, for any value of $R$, and also for the spherical nanopore, when $R\rightarrow \infty$, however in our results they are in units of $\mathrm{area}$.

From \cref{Fig.PCS.g(r)_s0.002}, it can be seen that the higher contact values of the counter-ions reduced concentration profiles, $g_{\scriptscriptstyle{-}}(r)$, are obtained for the smaller slit nanopores, followed by the cylindrical and spherical nanopores.
The opposite is true for the contact values of the co-ions reduced concentration profiles, $g_{\scriptscriptstyle{+}}(r)$, implying the expulsion of the co-ions.
However, as expected, when the radius increases the inner and outer EDL tend to become symmetrically equal. Additionally, the difference among the EDL of the three topologies become significantly smaller, such that at a pore's radius of \SI{47.1}{\textit{a}}, they almost overlap, although, they become equal only for an infinite radius. 

For a radius of \SI{1.2}{\textit{a}}, it is observed that inside the pores, i.e., for $0\le r \le(R-a/2)\,$=$\,(1.2-0.5)\,a\approxeq \SI{2.98}{\angstrom}$, the adsorption of co-ions and counter-ions is nearly constant, and it is highest for the slit nanopore, followed by the cylindrical and spherical pores, see \cref{Fig.PCS.g(r).R1.2}.
At this small radius, only a little more than two ions in straight line can be fitted inside the pore.
When the pore's radius is increased, the inner reduced counter-ion concentration profiles, $g_{\scriptscriptstyle{-}}(r)$, of all the nanopores, decay in the entire inner interval.
The opposite is true for the co-ions concentration profiles, $g_{\scriptscriptstyle{+}}(r)$. However, the absorption of counter-ions, inside the pores, always surmount that of the co-ions, driven mainly, but not only, by the imposed electrical field produced by the positively charged walls of the nanopores. Additionally, because in our model the ion-ion interaction potential is neglecting their ionic size, no oscillations of the reduced concentration profiles can be present, as has been widely recognized in the past~\cite{ Attard_1996}. For a radius of \SI{7.1}{\textit{a}}, inside the pores, the counter-ion concentration profile of the cylindrical nanopore, $g_{\scriptscriptstyle{-}}^{\scriptstyle{c}}(r)$, becomes higher than those for the other two pore's topologies.
However, the spherical $g_{\scriptscriptstyle{-}}^{\scriptstyle{s}}(r)$ surpass that of the slit nanopore, $g_{\scriptscriptstyle{-}}^{\scriptstyle{p}}(r)$, at least in part of the interval inside the pores. For a larger radius, $R=\SI{18.8}{\textit{a}}$, $g_{\scriptscriptstyle{-}}^{\scriptstyle{s}}(r) > g_{\scriptscriptstyle{-}}^{\scriptstyle{c}}(r) > g_{\scriptscriptstyle{-}}^{\scriptstyle{p}}(r), \ \forall\, r \  \cdot\ni\cdot \ (0 \le r \le R-a/2)$. Notwithstanding, outside the nanopores $g_{\scriptscriptstyle{-}}^{\scriptstyle{s}}(r) < g_{\scriptscriptstyle{-}}^{\scriptstyle{c}}(r) < g_{\scriptscriptstyle{-}}^{\scriptstyle{p}}(r), \ \forall\, r \geq R_{\scriptscriptstyle_{H}}$ and $R\geq 0$, since the negative charge induced inside the nanopores decreases their effective surface charge at $r=R+d$. If the radius is increased enough, i.e., $R=\SI{47.1}{\textit{a}}$ (see \cref{Fig.PCS.g(r).R47p1}), the concentration profiles of all the pores become nearly equal $\forall\, r \  \cdot \ni \cdot\ (0 \le r \le R-a/2) \ \bigcup \ (R_{\scriptscriptstyle_{H}} \le r \le \infty)$. Nevertheless, even at this large pore radius, for $0 \le r \le R-a/2$, the counter-ion profile of the spherical pore slightly overcomes that of the cylindrical pore, and this last surpass that of the slit nanopore.
Increasing $R$, implies augmenting the charge on the spherical and cylindrical nanopores, to keep the same surface charge density, while that of the slit nanopore is independent of inter-plate distance $R$; with more charge on the spherical pore than in the cylindrical pore, for a given value of $R$, of course.
This explains the observed higher counter-ion contact values inside of the spherical and cylindrical nanopores, over that of the slit nanopore; and, partially, the inverse order in the outside reduced concentration profiles of the nanopores.
However, this effect can not explain the lowering of the counter-ion contact values in all the pores, with increasing $R$, neither the lower counter-ion profiles of all the nanopores, for all $r$, inside and outside of the pores.
\textit{This is a confinement effect, due to the need of the system to maintain the inner chemical potential equal to that outside the pore}, or, in other words, the larger $R$, the thicker the inner and outer EDL must be, at the expenses of their contact value, to satisfy overall electroneutrality around the nanopore.
External to the pores the chemical potential is that of the bulk electrolyte, which is the same, of course, for all the pore geometries here considered, since the bulk concentration profile is taken to be same for all the pore topologies, i.e., $\rho_{\scriptscriptstyle{0}}=\SI{0.01}{\Molar}$.
\textit{The Poisson-Boltzmann equation is the statistical mechanics version of the Gauss' law (in differential form), where the charge density is taken from the canonical partition function of the system. Hence this equation forces the conservation of charge, energy and probability in the system where it is applied, and, therefore, guaranties a constant chemical potential throughout the system~\cite{Odriozola_2017}}. We will come back to this point later in this section, when discussing the charge concentration profile.

%%%%%%%%%%%%%%%%%%%%%%%%%%%%%%%%%%%%%%%%%%%%%%%%%%%%%%%%%%%%%%%%%%%%%%55

Moreover, the counter-ion concentration profiles decrease, as a function of the nanopores topology, according with their corresponding effective electrical fields, $E_{\scriptscriptstyle{N}}(r)$.
Quite generally, by Gauss' law, the \textit{effective or net} electrical field for a charged nanopore of any geometry,  $E_{\scriptscriptstyle{N}}(r)$, at a distance $r$ from the center of the nanopore is 

\begin{equation}
	E_{\scriptscriptstyle N}(r)=\frac{1}{\varepsilon_{\scriptscriptstyle{0}}\varepsilon}\,\sigma_{\scriptscriptstyle{N}}(r) \label{Ec.elec_a},
\end{equation}

\noindent where $\sigma_{\scriptscriptstyle{N}}(r)$=$Q_{\scriptscriptstyle {N}} (r)/A_{\scriptscriptstyle {N}} (r)$, is the effective charge density profile inside and outside the nanopores. $A_{\scriptscriptstyle {N}}(r)$ is the area around the geometrical center of the nanopore, at the distance r, and $Q_{\scriptscriptstyle {N}} (r)$ is the total charge inside the volume $V_{\scriptscriptstyle {N}}(r)$. For example, $Q_{\scriptscriptstyle {N}} (r)$ is the charge  induced on the electrolyte inside the nanopore, plus the charge on the nanopore's walls, plus the induced charge, outside the nanopore, up to the distance $r\equiv \abs{\overrightarrow{\mathbf{r}}}$, for $r\geq (R+d+a/2)$.
Thus, $V_{\scriptscriptstyle {N}}(r)$ is the volume defined between the distance $r=0$, i.e., from the coordinates' origin, to $r$, such that $r$ can be inside or outside the nanopore. $E_{\scriptscriptstyle_{N}}(r)$ induces a charge density $\sigma_{\scriptscriptstyle{M}}(r)$ in the electrolyte, such that inside the nanopore, $\sigma_{\scriptscriptstyle{M}}(r)=\sigma_{\scriptscriptstyle{N}}(r)$, but outside $\sigma_{\scriptscriptstyle{M}}(r)=-\sigma_{\scriptscriptstyle{N}}(r)$. Hereinafter, we will focus on the induced surface charge density, $\sigma_{\scriptscriptstyle{M}}(r)$. Thence, from Gauss' law, \textit{the induced surface charge density, $\sigma_{\scriptscriptstyle{M}}(r)$}, inside and outside the nanopores are given by

\begin{equation}
	\sigma_{\scriptscriptstyle{M}}(r)=\frac{1}{A_{\scriptscriptstyle{M}}(r)}\int_{0}^{V_{\scriptscriptstyle {M}}^{^{\scriptscriptstyle{in}}}(r)}\rho_{\scriptscriptstyle{el}}(r)\,\dd^3{r}, \quad \forall r \ni 0\leq r \eqslantless (R-a/2) \label{Ec.esf.SIGMA2_r}
\end{equation}

\noindent and

\begin{equation}
	\sigma_{\scriptscriptstyle{M}}(r)=\frac{1}{A_{\scriptscriptstyle{M}}(r)}\int_{{V_{\scriptscriptstyle {M}}^{^{\scriptscriptstyle{out}}}(r)}}^{V_{\scriptscriptstyle {M}}^{^{\scriptscriptstyle{out}}}(r\,\rightarrow\,\infty)}\rho_{\scriptscriptstyle{el}}(r)\,\dd^3{r}, \quad \forall r \ni (R+d+a/2)\leq r \eqslantless \infty\label{Ec.esf.SIGMA3_r},
\end{equation}

\noindent respectively. $\rho_{\scriptscriptstyle{el}}(r)$ is the induced electrolyte charge concentration profile inside and outside, around the nanopore, given by \cref{Rho_elx}, $V_{\scriptscriptstyle {M}}^{^{\scriptscriptstyle{in}}}(r)$ is the volume defined between the center of the pore, $r=0$, to distance $ r \ni 0\leq r \eqslantless (R-a/2)$, and $V_{\scriptscriptstyle {M}}^{^{\scriptscriptstyle{out}}}(r)$ is the volume defined between the distance $r \ni (R+d+a/2)\leq r \eqslantless \infty$.
The electroneutrality condition guaranties that the total charge induced in the electrolyte sourrounding the nanopore, must be equal to that on their walls. Notice that the integrals in \cref{Ec.esf.SIGMA2_r,Ec.esf.SIGMA3_r} give an induced negative charge, which gives the correct effective electrical field for $r\leq (R-a/2)$. However, outside the nanopore the effective surface charge density at $r\geq R+d+a/2$ is positive, hence, for this quantity  \cref{Ec.esf.SIGMA3_r} must have have a minus sign, i.e., by the electroneutrality condition, the induced charge from $(R+d+a/2)\leq r$ to $r\rightarrow \infty$ is the negative of the effective positive charge of the nanopore at $r$.

\begin{figure}[!htb]
	\begin{subfigure}{.5\textwidth}
		\centering
		\includegraphics[width=0.95\linewidth]{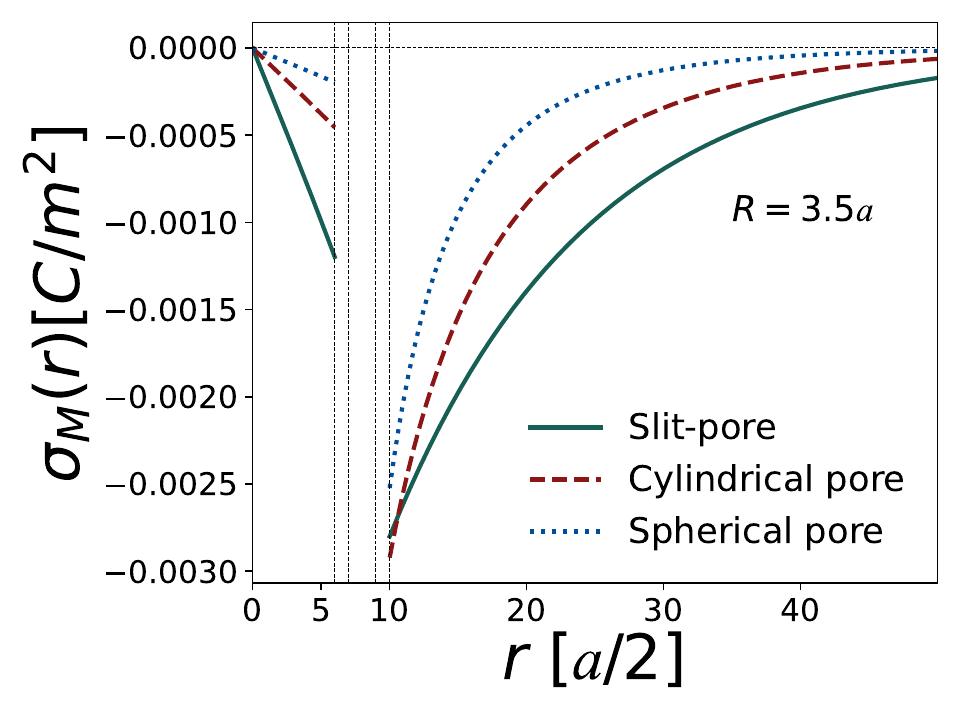}
		\caption{Charge density profile, with $d=a$.}
		\label{Fig.PCS_sigma[r]_R3.5}
	\end{subfigure}
	\begin{subfigure}{.5\textwidth}
		\centering
		\includegraphics[width=0.95\linewidth]{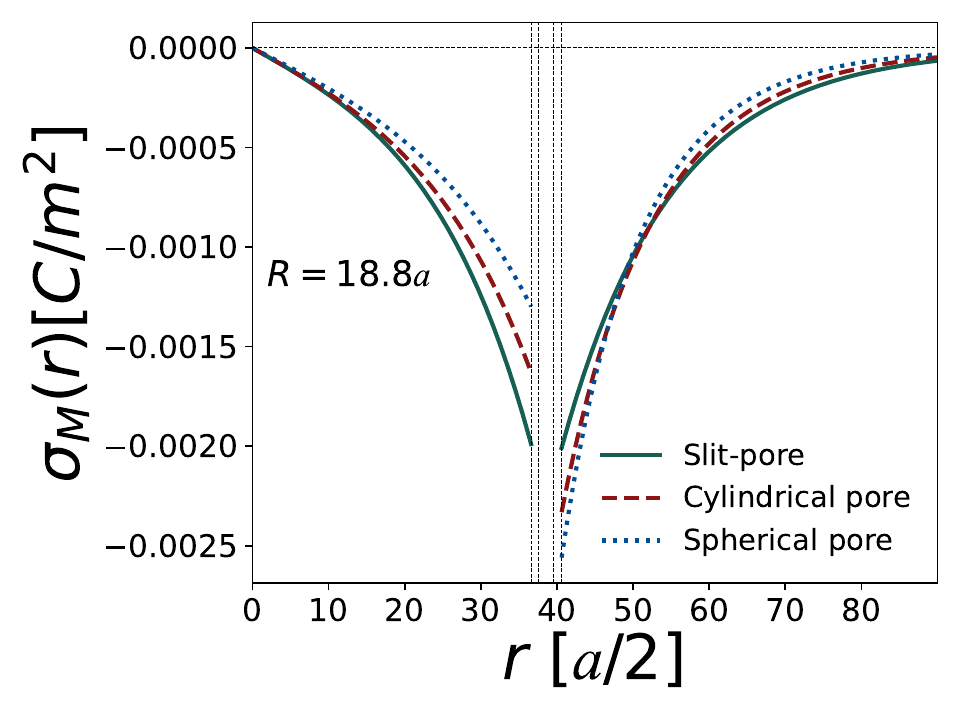}
		\caption{Charge density profile. with $d=a$.}
		\label{Fig.PCS_sigma[r]_R18.8}
	\end{subfigure}
\begin{subfigure}{.5\textwidth}
	\centering
	\includegraphics[width=0.95\linewidth]{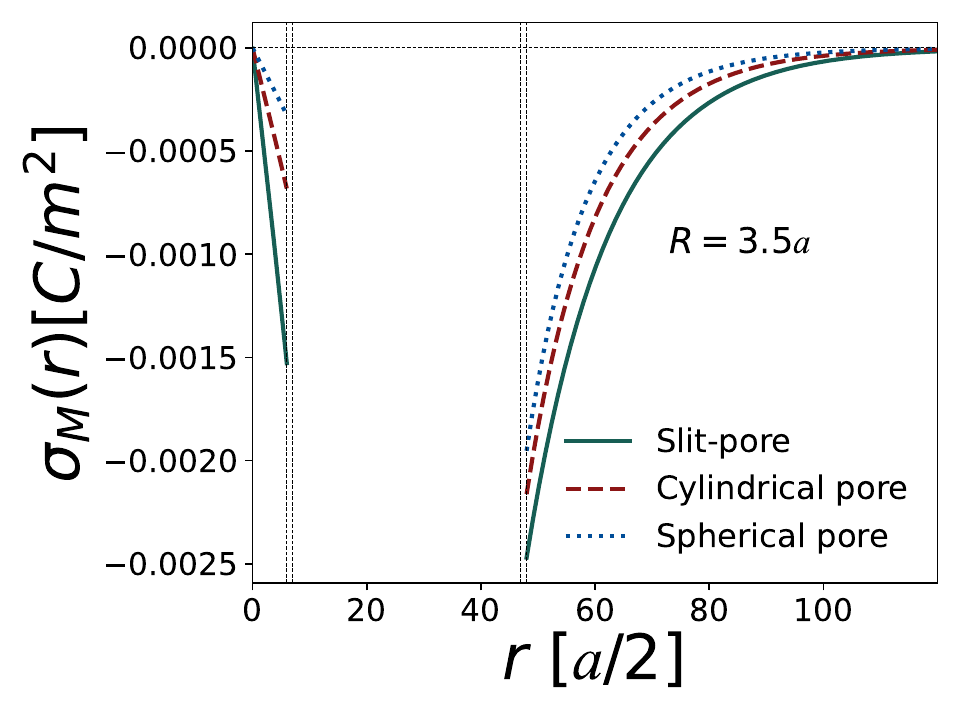}
	\caption{Charge density profile, with $d=20a$.}
	\label{Fig.PCS_sigma[r]_R3.5_d20}
\end{subfigure}
\begin{subfigure}{.5\textwidth}
	\centering
	\includegraphics[width=0.95\linewidth]{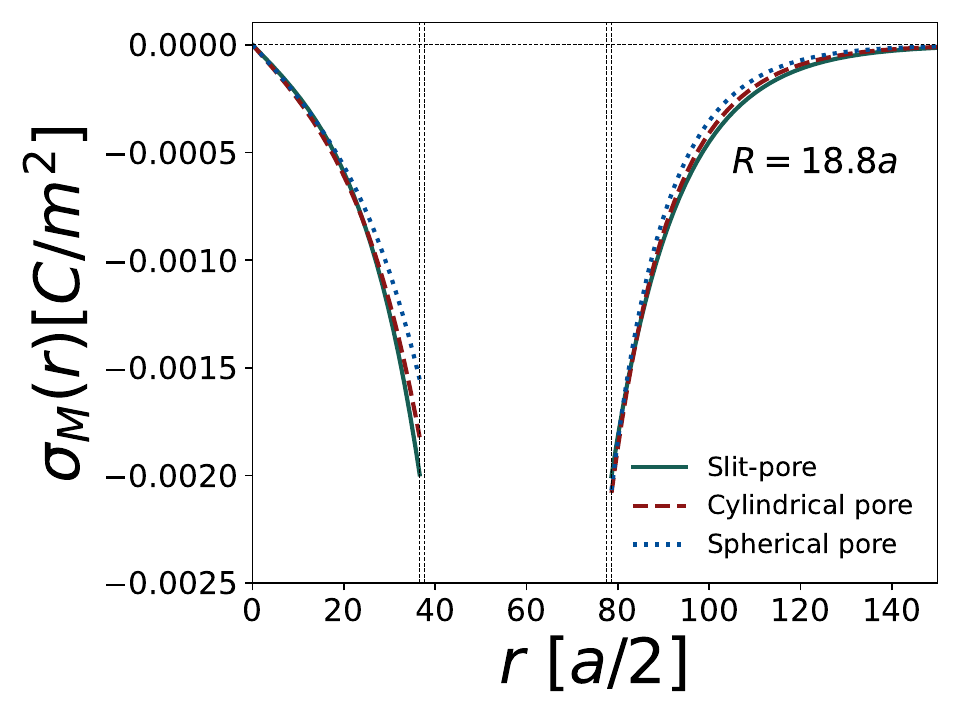}
	\caption{Charge density profile, with $d=20a$.}
	\label{Fig.PCS_sigma[r]_R18.8_d20}
\end{subfigure}\\
	\caption{Induced charge density profile, $\sigma_{\scriptscriptstyle{M}}(r)$, as a function of $r$, for three different nanopore topologies. The solid and dot-dash vertical lines represent the nanopore's walls, and the Stern layer, respectively. Two different nanopore's radius, $R$, and wall thickness, $d$, are shown, with walls' surface charge densities, $\sigma_{\scriptscriptstyle{I}}=\sigma_{\scriptscriptstyle{II}}=\sigma_{\scriptscriptstyle{0}}=\SI{0.002}{\si[per-mode=symbol]{\coulomb\per\square\metre}}$. As in \cref{Fig.PCS.g(r)_s0.002}  the solution is a 1:1 electrolyte, with $\varepsilon=78.5$, $a=\SI{4.25}{\angstrom}$, $\rho_{\scriptscriptstyle{0}}=\SI{0.01}{\Molar}$, $T=\SI{298}{\kelvin}$.}
	\label{Fig.PCS_sigma[r]_R3.5-18.8}
\end{figure}

In particular, for the slit nanopore \cref{Ec.esf.SIGMA2_r,Ec.esf.SIGMA3_r} become

\begin{equation}
	\sigma_{\scriptscriptstyle{P}}(r)=\int_{0}^{r}\rho_{\scriptscriptstyle{el}}(r)\,\dd^3{r}\label{Ec.esf.SIGMA2_r_slit},
\end{equation}

\begin{equation}
	\sigma_{\scriptscriptstyle{P}}(r)=\int_{r}^{\infty}\rho_{\scriptscriptstyle{el}}(r)\,\dd^3{r}\label{Ec.esf.SIGMA3_r_slit},
\end{equation}

\noindent and for the cylindrical and spherical pores, they become

\begin{equation}
	\sigma_{\scriptscriptstyle{C}}(r)=\frac{1}{r}\int_{0}^{r}\rho_{\scriptscriptstyle{el}}(r)\,r\,\dd^3{r}\label{Ec.esf.SIGMA2_r_cyl},
\end{equation}

\begin{equation}
	\sigma_{\scriptscriptstyle{C}}(r)=\frac{1}{r}\int_{r}^{\infty}\rho_{\scriptscriptstyle{el}}(r)\,r\,\dd^3{r}\label{Ec.esf.SIGMA3_r_cyl},
\end{equation}
\noindent and

\begin{equation}
	\sigma_{\scriptscriptstyle{S}}(r)=\frac{1}{r^2}\int_{0}^{r}\rho_{\scriptscriptstyle{el}}(r)\,r^2\,\dd^3{r}\label{Ec.esf.SIGMA2_r_spher},
\end{equation}

\begin{equation}
	\sigma_{\scriptscriptstyle{S}}(r)=\frac{1}{r^2}\int_{r}^{\infty}\rho_{\scriptscriptstyle{el}}(r)\,r^2\,\dd^3{r}\label{Ec.esf.SIGMA3_r_spher},
\end{equation}

\noindent respectively.
\cref{Ec.esf.SIGMA2_r_slit,Ec.esf.SIGMA3_r_slit,Ec.esf.SIGMA2_r_cyl,Ec.esf.SIGMA3_r_cyl,Ec.esf.SIGMA2_r_spher,Ec.esf.SIGMA3_r_spher}  explain the decreasing counter-ion concentration profiles, as a function of the increasing electrolyte confinement, due to the nanopore's topology, at least in the region outside the nanopores. Inside the nanopores, the electrolyte confinement modifies this behavior, i.e., for very narrow pores, see \cref{Fig.PCS.g(r).R1.2}, very few counter-ions can get inside the slit nanopore, even much less inside the cylindrical and spherical pores, and such that they basically arrange in two layers of counter-ions, next to the inner walls of the nanopores. The very high counter-ions \textit{reduced} concentration profile, $g_{\scriptscriptstyle{-}}(r)$, observed for the slit nanopore is a consequence of its geometry. This comportment is accentuated in the co-ions \textit{reduced} concentration profile, $g_{\scriptscriptstyle{+}}(r)$, although now, in the inverted order shown by the counter-ions. For larger values of $R$, the order of $g_{\scriptscriptstyle{-}}(r)$, in relation to the nanopore's topology, gradually, changes: First, the counter-ion concentration profile of the cylindrical nanopore overcomes that of the slit nanopore, see  \cref{Fig.PCS.g(r).R7.1}, and then, that for the spherical nanopore becomes larger than those for the cylindrical and slit nanopores (see  \cref{Fig.PCS.g(r).R18p8}). This is, of course, a result of the different degree of confinement on the electrolyte. The amount of electrolyte that can squeeze into the nanopores is ruled by the chemical potential of the bulk, which is equal inside and outside the nanopores; the nanopores size and surface charge, and the ions size. The constant chemical potential condition is imposed by the construction of the PB equation. In our model, the ions size is only considered by the Stern layer. Lastly, the nanopore's radius, surface charge, and their walls thickness determines the electrical field profile, $E_{\scriptscriptstyle N}(r)$, both, inside and outside the nanopores. Within the internal walls of the nanopores, $E_{\scriptscriptstyle N}(r)$, $\forall \quad r \ni 0\leq r \eqslantless (R+a/2)$, is proportional to the induced charge density on the electrolyte, whereas outside of the nanopores $E_{\scriptscriptstyle N}(r)$,  $\forall \quad r \ni (R+d+a/2)\leq r \eqslantless \infty $, is the sum of the induced electrical field inside the nanopore, plus that due to the surface charge on their walls, plus the electrical field induced up to the distance $r$.  Hence, outside the nanopores this electrical electrical field is $E_{\scriptscriptstyle_{N}}(r)=\sigma_{\scriptscriptstyle{N}}/(\varepsilon_{\scriptscriptstyle{0}}\varepsilon)=-\sigma_{\scriptscriptstyle{M}}(r)/(\varepsilon_{\scriptscriptstyle{0}}\varepsilon)$ (see \cref{Ec.elec_a}). For the different nanopore's topologies, whether inside or outside the nanopores, 	$\sigma_{\scriptscriptstyle{M}}(r)$ is given by \cref{Ec.esf.SIGMA2_r_slit,Ec.esf.SIGMA3_r_slit,Ec.esf.SIGMA2_r_cyl,Ec.esf.SIGMA3_r_cyl,Ec.esf.SIGMA2_r_spher,Ec.esf.SIGMA3_r_spher}. In \cref{Fig.PCS_sigma[r]_R3.5-18.8}, we plot the \textit{induced }charge density profile, $\sigma_{\scriptscriptstyle{M}}(r)$, inside and outside the nanopores, for the slit, cylindrical and spherical nanopores. In consistence with our above discussion, we see that the lower amount of ions inside the nanopores generates \textit{negative} weaker electrical field inside the nanopores (see \cref{Fig.PCS.g(r)_s0.002}). This electrical field is, of course, negative because of the larger amount of counter-ions inside the nanopore (see \cref{Fig.PCS.g(r)_s0.002}). Increasing the internal radius of the nanopore, increases the negative electrical field inside (see \cref{Fig.PCS_sigma[r]_R18.8}).

For $r=(R-a/2)$, $\sigma_{\scriptscriptstyle{M}}(R-a/2)$ becomes the total induced charge inside the nanopore, while for $r=(R+d+a/2)$, $\sigma_{\scriptscriptstyle{M}}(R+d+a/2)$ becomes the total induced charge outside the nanopore. Defining $\bigtriangleup \sigma_{\scriptscriptstyle{in}}$ as the charge on the inside wall of the nanopore, $\sigma_{\scriptscriptstyle{I}}=\sigma_{\scriptscriptstyle{0}}$, plus the total induced charge inside the nanopore, and $\bigtriangleup \sigma_{\scriptscriptstyle{out}}$ as the charge on the outside wall of the nanopore, $\sigma_{\scriptscriptstyle{II}}=\sigma_{\scriptscriptstyle{0}}$ plus the total induced charge outside the nanopore, i.e.,

\begin{equation}
	\bigtriangleup \sigma_{\scriptscriptstyle{in}}\equiv \sigma_{\scriptscriptstyle{0}}+\sigma_{\scriptscriptstyle{M}}(R-a/2)=\sigma_{\scriptscriptstyle{0}}+\sigma_{\scriptscriptstyle{Hi}},
	\label{deltain}
\end{equation}

 and 
 
 \begin{equation}
 \bigtriangleup \sigma_{\scriptscriptstyle{out}}\equiv \sigma_{\scriptscriptstyle{0}}+\sigma_{\scriptscriptstyle{M}}(R+d+a/2)=\sigma_{\scriptscriptstyle{0}}+\sigma_{\scriptscriptstyle{Ho}},
 \label{deltaout}
 \end{equation}

\noindent in \cref{Fig.PCS_sigma[r]_R3.5} we notice that $\bigtriangleup \sigma_{\scriptscriptstyle{in}}>0$, for the three nanopore's topologies, implying a violation of the local electroneutrality condition (LEC). This violation is greater for the spherical nanopore and lower for the slit nanopore. On the other hand, $\bigtriangleup \sigma_{\scriptscriptstyle{out}}<0$, and such that $\bigtriangleup \sigma_{\scriptscriptstyle{in}}+\bigtriangleup \sigma_{\scriptscriptstyle{out}}=0$, in all cases, and, thence, satisfying the total electroneutrality condition. We signal again the the induced charge is negative. For larger nanopore's radius the violation of the LEC decreases in the cylindrical and spherical nanopores, while almost disappear in the slit nanopore (see \cref{Fig.PCS_sigma[r]_R18.8}). In fact, for the cylindrical and spherical nanopores the LEC is rigorously satisfied only for $R\rightarrow \infty $. These findings are in qualitative agreement with more elaborated theories~\cite{Lozada1996,Aguilar_2007}.

Coming  back to the electrolyte distribution functions of \cref{Fig.PCS.g(r)_s0.002}, we see that for very large values of $R$ the counter-ion concentration profiles of the three nanopores tend to become equal among them, and to those outside the nanopores (see  \cref{Fig.PCS.g(r).R47p1}).  In general, as $R$ increases, $g_{\scriptscriptstyle{-}}(r)$ decreases, $\forall \ r$. A similar, but inverted, behavior have the co-ions concentration profiles. For $R\rightarrow\infty$, the EDL inside and outside the three nanopores become symmetrically equal and identical to that of a single solid, planar electrode, with surface charge density equal to $2\sigma_{\scriptscriptstyle{0}}$. We will come back to this point later below.

In general, the EDLs inside and outside the pores are correlated~\cite{Lozada-Cassou-PRL1996,Lozada-Cassou-PRE1997}; this correlation depends on the radius, $R$, and the pore's thickness wall, $d$~\cite{Yu_1997}. In \cref{Fig.PCS_sigma[r]_R3.5,Fig.PCS_sigma[r]_R18.8} we see that this correlation decreases, as $R$ increases. In \cref{Fig.PCS_sigma[r]_R3.5_d20,Fig.PCS_sigma[r]_R18.8_d20} the induced charge density profiles, inside and outside the pores, are depicted for thicker nanopore's walls, i.e. $d=20a$. In \cref{Fig.PCS_sigma[r]_R3.5_d20}, for $R=3.5a$ and $d=20a$, it is observed that the violation of the LEC inside the pores decreases, compared to that in \cref{Fig.PCS_sigma[r]_R3.5}, for thinner walls ($R=3.5a$ and $d=a$). This is also the case, when comparing the induced charge densities inside and outside the pores for $R=18.8a$ and $d=20a$, shown in \cref{Fig.PCS_sigma[r]_R18.8_d20}, with those for $R=18.8a$ and $d=a$, in \cref{Fig.PCS_sigma[r]_R18.8}. Comparison of \cref{Fig.PCS_sigma[r]_R18.8_d20} with \cref{Fig.PCS_sigma[r]_R3.5_d20} also shows a decrease of the violation of the LEC, when increasing $R$. This comportment of the induced profiles inside and outside the nanopores is necessary to keep the chemical potential inside equal to that outside. However, we detected a contrasting behavior for the charge concentration profiles inside and outside the nanopores, for a decrease of $\rho_{\scriptscriptstyle{0}}$/$T$, i.e., a sufficient decrease of $\kappa$, and at sufficiently large nanopore internal radius. We will come back to this point further below.

%%%%%%%%%%%%%%%%%%%%%%%%%%%%%%%%%%%%%%%%%%%%55
%%%%%%%%%%%%%%%%%%%%%%%55
\begin{figure}[!htb]
	\begin{subfigure}{.5\textwidth}
		\centering
		\includegraphics[width=0.99\linewidth]{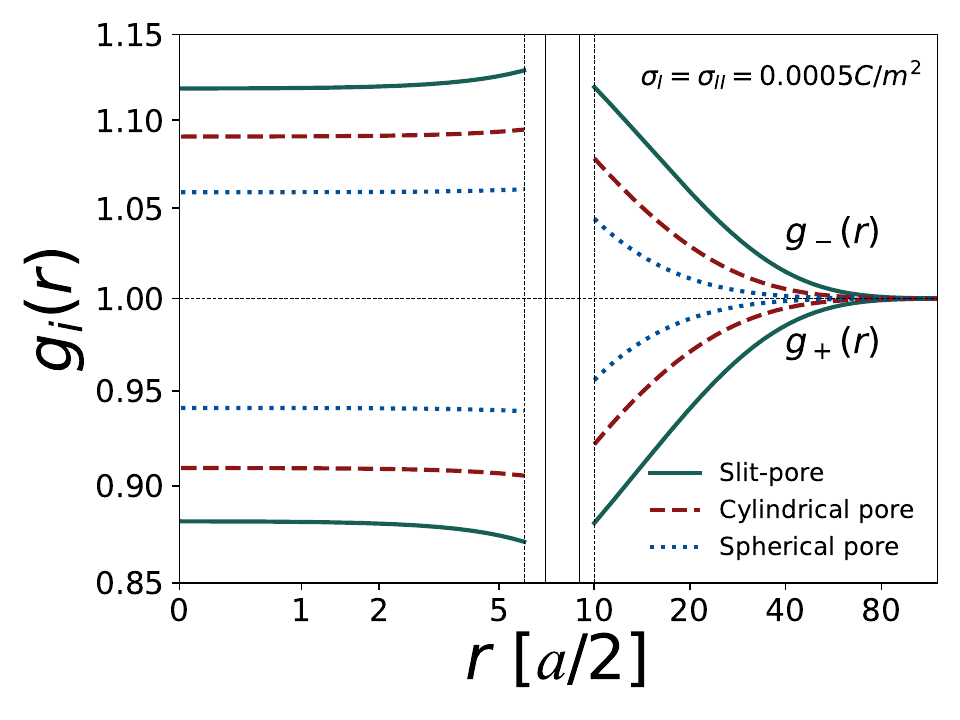}
		\caption{The concentration profiles for $\sigma_{\scriptscriptstyle{0}}=\SI{0.0005}{\si[per-mode=symbol]{\coulomb\per\square\metre}}$.}
		\label{Fig.PCS.g(r).s0.005R3.5}
	\end{subfigure}
	\begin{subfigure}{.5\textwidth}
		\centering
		\includegraphics[width=0.99\linewidth]{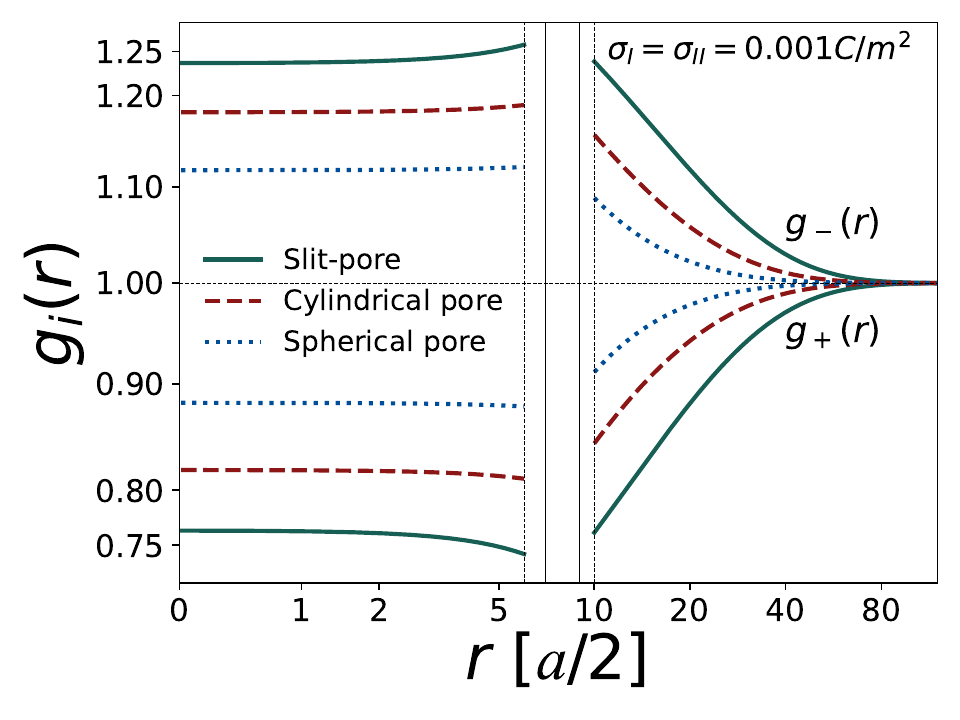}
		\caption{The concentration profiles for $\sigma_{\scriptscriptstyle{0}}=\SI{0.001}{\si[per-mode=symbol]{\coulomb\per\square\metre}}$.}
		\label{PCS_g[r]_s_0.01_R3.5}
	\end{subfigure}\\
	\begin{subfigure}{.5\textwidth}
		\centering
		\includegraphics[width=0.99\linewidth]{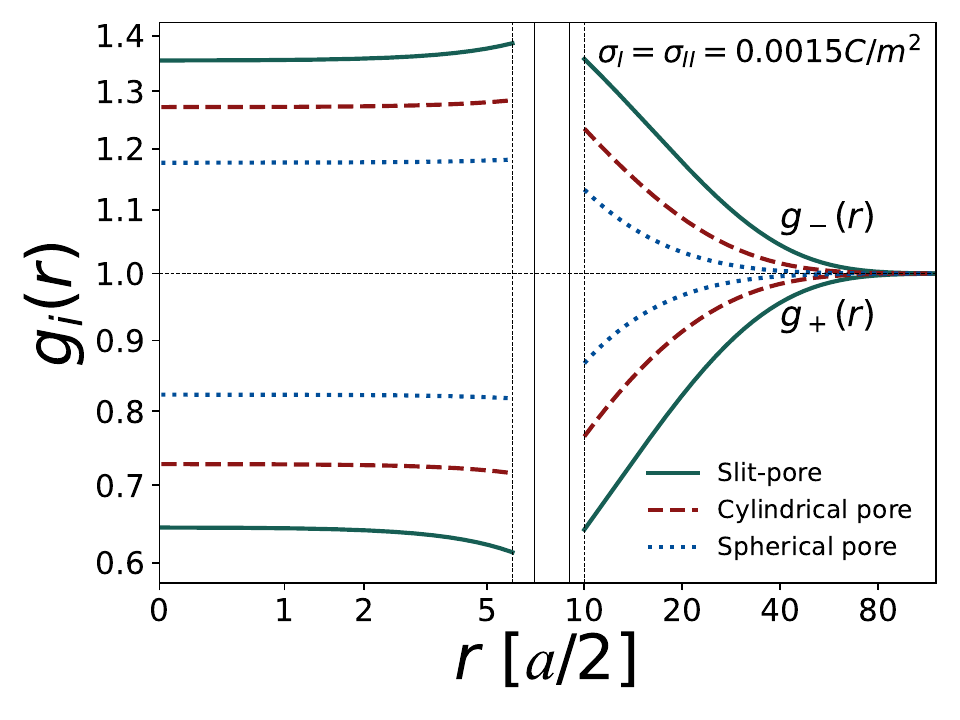}
		\caption{The concentration profiles for $\sigma_{\scriptscriptstyle{0}}=\SI{0.0015}{\si[per-mode=symbol]{\coulomb\per\square\metre}}$.}
		\label{Fig.PCS.g(r).s0.015R3.5}
	\end{subfigure}
	\begin{subfigure}{.5\textwidth}
		\centering
		\includegraphics[width=0.99\linewidth]{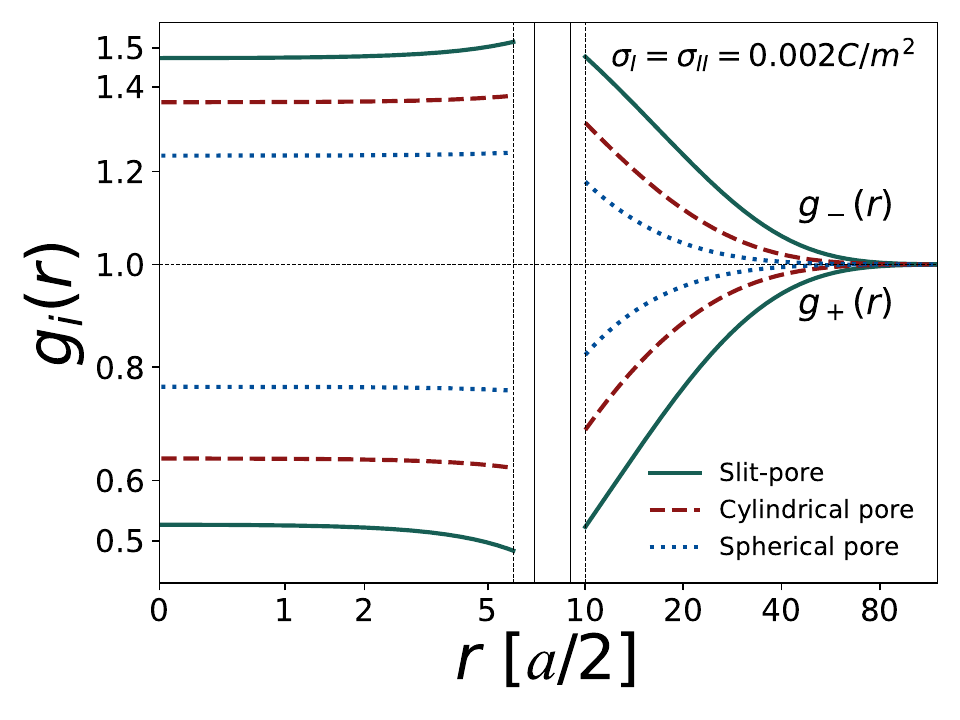}
		\caption{The concentration profiles for $\sigma_{\scriptscriptstyle{0}}=\SI{0.002}{\si[per-mode=symbol]{\coulomb\per\square\metre}}$.}
		\label{Fig.PCS.g(r).s0.02R3.5}
	\end{subfigure}
	\caption{LPB reduced concentration profile, $g_{i}(r)$, for different electrode topologies.
		A radius of $\SI{3,5}{\textit{a}}$ is used, while the electrode's surface charge, $\sigma_{\scriptscriptstyle{I}}=\sigma_{\scriptscriptstyle{II}}=\sigma_{\scriptscriptstyle{0}}$, is varied, and a $sinh^-1(r)$ scale is used.
		As in \cref{Fig.PCS.g(r)_s0.002}  the solution is a 1:1 electrolyte, with $\varepsilon=78.5$, $a=\SI{4.25}{\angstrom}$, $\rho_{\scriptscriptstyle{0}}=\SI{0.01}{\Molar}$, $T=\SI{298}{\kelvin}$. The vertical lines have the same meaning as in \cref{Fig.PCS.g(r)_s0.002}}
	\label{Fig.PCS.g(r).sigmaMR3.5}
\end{figure}
%%%%%%%%%%%%%%%%%%%%%%%%%
%%%%%%%%%%%%%%%%%%%%%%%%%%%%%%%%%%%%%%%%%%%%55
In \cref{Fig.PCS.g(r).sigmaMR3.5} the concentration profiles are shown for the different pore topologies, for a pores' radius of \SI{3.5}{\textit{a}}, while the electrode's surface charge is varied.
As expected, it is observed that for all the electrodes' surface charges, the absorption of ions is higher for the slit nanopore, followed by the cylindrical and spherical pores.
Furthermore, it is observed that the concentration profiles of the slit nanopore growth quicker with higher values of $\sigma_{\scriptscriptstyle{0}}$, compared to that of the cylindrical-pore, and that this one, correspondingly, growths faster than that for the spherical-pore.
As such, due to the packed topology of the systems, the slit nanopore outranks the cylindrical and spherical pores, regardless of the electrode's surface charge.
This behavior translates that due to the small pore's radii, the pores are filled up by counter-ions, with a very strong attraction to the counter-ions, and a very strong repulsion to the co-ions.
If the ionic size had been considered in the ion-ion interaction potential, the ionic size would be an additional obstacle for the co-ions to squeeze into the pores.
\begin{figure}[!htb]
	\begin{subfigure}{.5\textwidth}
		\centering
		\includegraphics[width=0.99\linewidth]{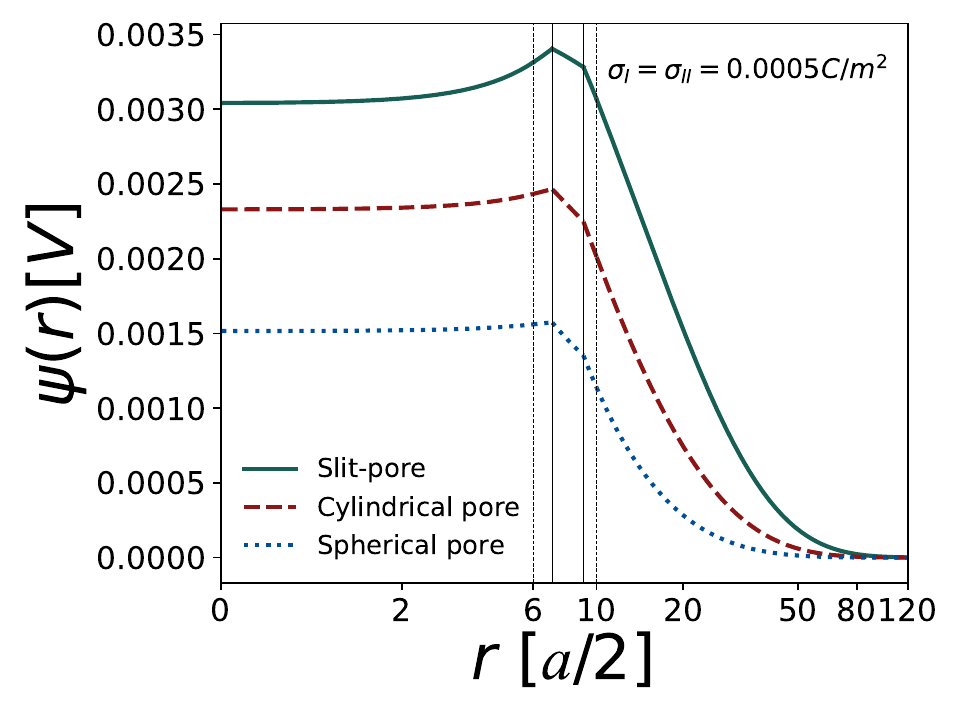}
		\caption{The mean electrostatic potential profiles for $\sigma_{\scriptscriptstyle{0}}=\SI{0.005}{\si[per-mode=symbol]{\coulomb\per\square\metre}}$.}
		\label{Fig.PCS.psi(r).sigmaM0.005R3.5}
	\end{subfigure}
	\begin{subfigure}{.5\textwidth}
		\centering
		\includegraphics[width=0.99\linewidth]{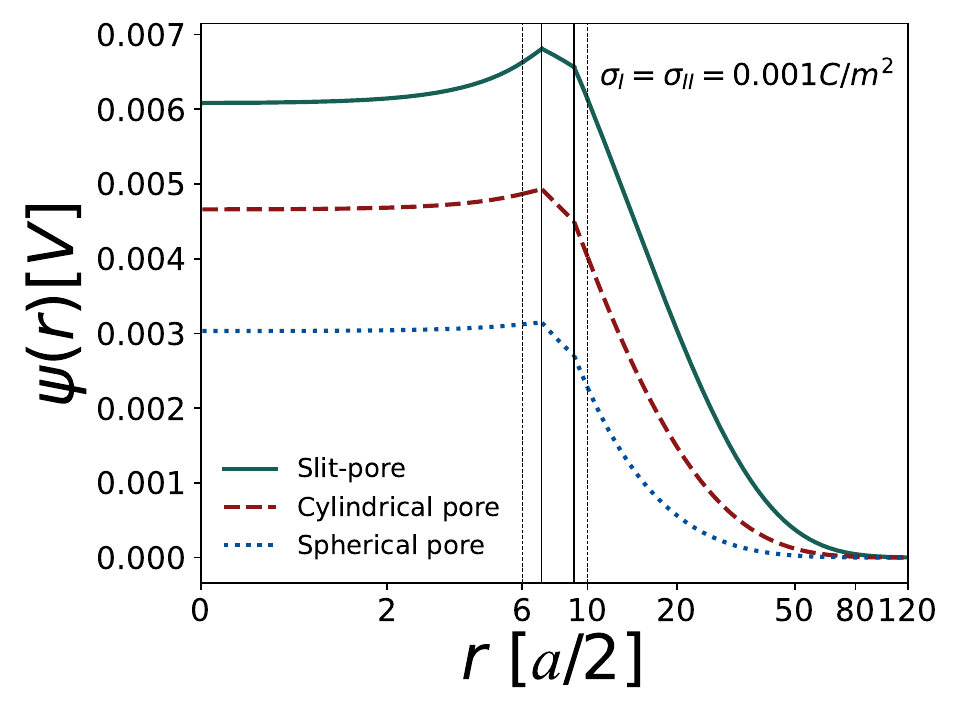}
		\caption{The mean electrostatic potential profiles for $\sigma_{\scriptscriptstyle{0}}=\SI{0.01}{\si[per-mode=symbol]{\coulomb\per\square\metre}}$.}
		\label{Fig.PCS.psi(r).sigmaM0.01R3.5}
	\end{subfigure}\\
	\begin{subfigure}{.5\textwidth}
		\centering
		\includegraphics[width=0.99\linewidth]{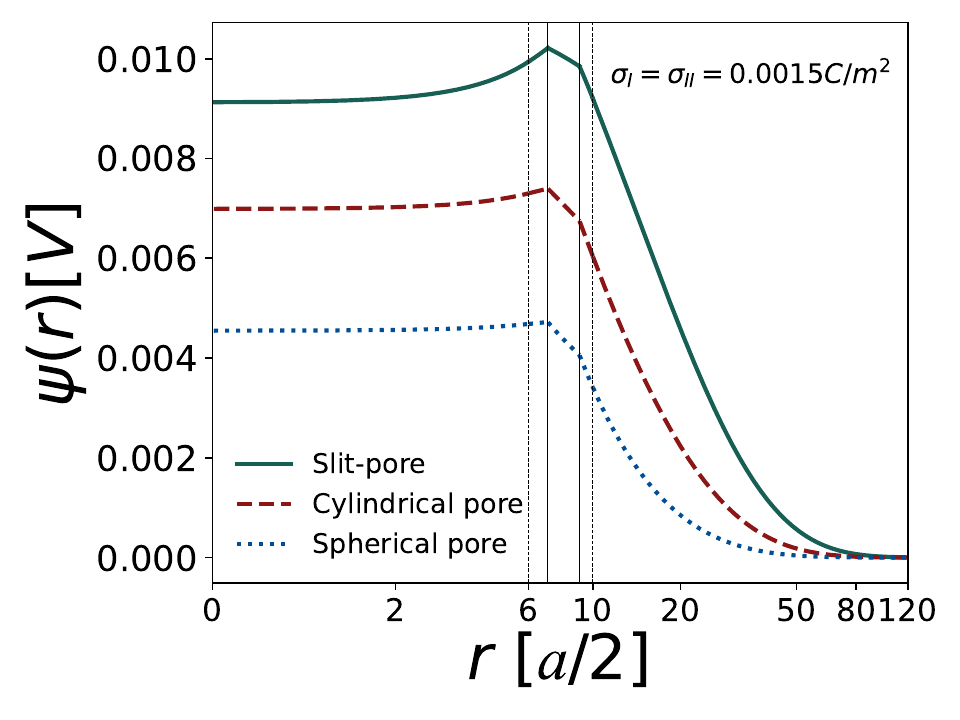}
		\caption{The mean electrostatic potential profiles for $\sigma_{\scriptscriptstyle{0}}=\SI{0.0015}{\si[per-mode=symbol]{\coulomb\per\square\metre}}$.}
		\label{Fig.PCS.psi(r).sigmaM0.015R3.5}
	\end{subfigure}
	\begin{subfigure}{.5\textwidth}
		\centering
		\includegraphics[width=0.99\linewidth]{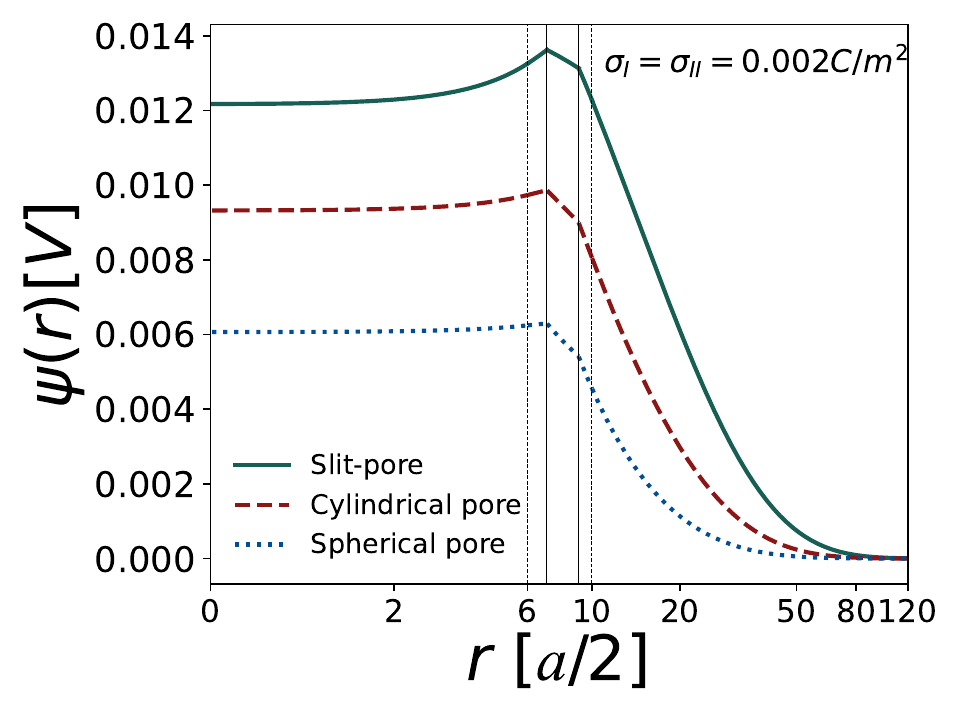}
		\caption{The mean electrostatic potential profiles for $\sigma_{\scriptscriptstyle{0}}=\SI{0.002}{\si[per-mode=symbol]{\coulomb\per\square\metre}}$.}
		\label{Fig.PCS.psi(r).sigmaM0.02R3.5}
	\end{subfigure}
	\caption{The mean electrostatic potential, $\psi(r)$, obtained from the Laplace and LPB equation for different electrode topologies.
		A radius of $\SI{3,5}{\textit{a}}$ is used, while the electrode's surface charge, $\sigma_{\scriptscriptstyle{I}}=\sigma_{\scriptscriptstyle{II}}=\sigma_{\scriptscriptstyle{0}}$, is varied, and a $sinh^-1(r)$ scale is used.
		As in \cref{Fig.PCS.g(r)_s0.002}  the solution is a 1:1 electrolyte, with $\varepsilon=78.5$, $a=\SI{4.25}{\angstrom}$, $\rho_{\scriptscriptstyle{0}}=\SI{0.01}{\Molar}$, $T=\SI{298}{\kelvin}$.
		The vertical lines have the same meaning as in \cref{Fig.PCS.g(r)_s0.002}.}
	\label{Fig.PCS.psi(r).sigmaMR3.5}
\end{figure}
Comparatively, when the pore's radii and the electrode's surface charge are increased the difference between the reduced concentration profiles among the different nanopore's geometries, will decrease as the cylindrical and spherical nanopores concentration profiles become closer to those of the slit nanopore, as seen in \cref{Fig.PCS.g(r)_s0.002,Fig.PCS.g(r).sigmaMR3.5}.
Furthermore, in \cref{Fig.PCS.g(r)_s0.002} it is observed that for higher radii, the inner concentration profiles, do not stay constant through all the pore.
It is found that this value softly decays as it approaches the pore's center, showing that the counter-ions are not as highly compacted toward the pore's center.

In \cref{Fig.PCS.psi(r).sigmaMR3.5}, the mean electrostatic potential profile, $\psi(r)$, obtained from the Laplace and LPB equation is plotted for different electrode topologies. A radius of $\SI{3,5}{\textit{a}}$ is used, while the electrode's surface charge, $\sigma_{\scriptscriptstyle{I}}=\sigma_{\scriptscriptstyle{II}}=\sigma_{\scriptscriptstyle{0}}$, is varied. The MEP plotted in \cref{Fig.PCS.psi(r).sigmaMR3.5}, correspond to the reduced concentration profiles, $g_i(r)$, depicted in \cref{Fig.PCS.g(r).sigmaMR3.5}. The MEP increases from $\psi(r\rightarrow \infty)=0$, to its maximum value at the inner surface of the nanopores, i.e., $\psi(R)=\psi_{\scriptscriptstyle{0}}$, to then decrease to $\psi(r)=\psi_{\scriptscriptstyle{d}}$. Increasing $\sigma_{\scriptscriptstyle{0}}$ produces an increase of $\psi(r), \ \forall \ r \in (0,\infty)$, and a contraction of the MEP curve at the right hand side of the nanopore, which imply an important compactness of the EDL (see \cref{Fig.PCS.g(r).sigmaMR3.5}). The planar nanopore shows the higher MEP increase, while the lowest is for the spherical nanopore. The accompanying increase of the surface charge density on the nanopores (\cref{Fig.PCS_sigma[r]_R3.5-18.8}) with the augment of the MEP  will not have an effect in the calculated capacitance, since in this linear theory the MEP increases linearly with the surface charge. Hence, \textit{within this theory the specific capacitance, $c_{\scriptscriptstyle_{d}}$ is equal to the specific integral capacitance, $c$}. Thence, a larger potential difference in a given region should produce a lower local capacitance, if this potential difference is grater than one. Otherwise, the opposite may occur. We will come back to this point later in \cref{Numerical results capacitance}.

\begin{figure}[!htb]
	\begin{subfigure}{.5\textwidth}
		\centering
		\includegraphics[width=1.05\linewidth]{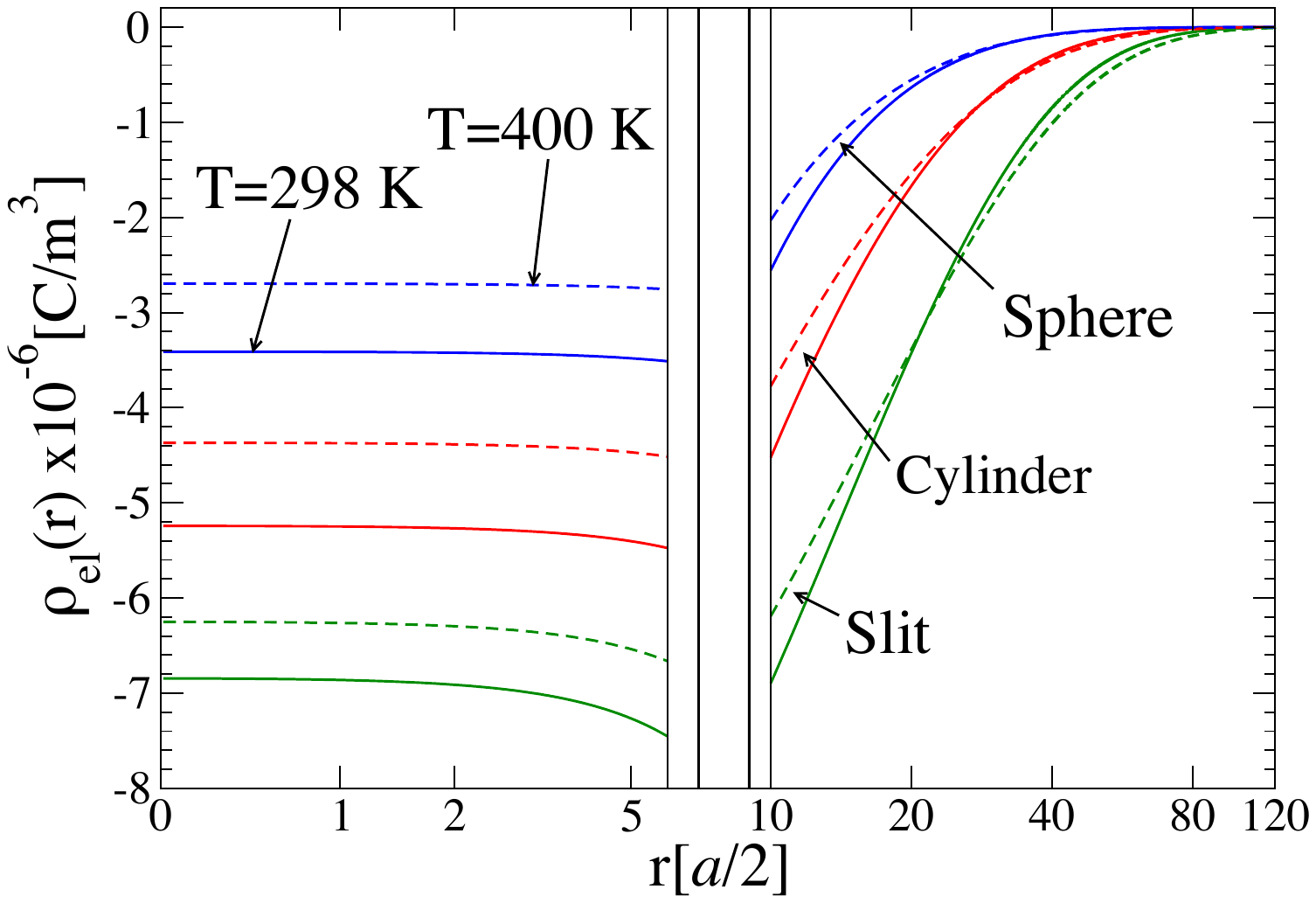}
		\caption{$\rho_{\scriptscriptstyle{el}}(r)$ for $R=\SI{3,5}{\textit{a}}$.}
		\label{PCS_rhoel[r]_R3p5_s0p015_rho0p01}
	\end{subfigure}%
	\begin{subfigure}{.5\textwidth}
		\centering
		\includegraphics[width=1.05\linewidth]{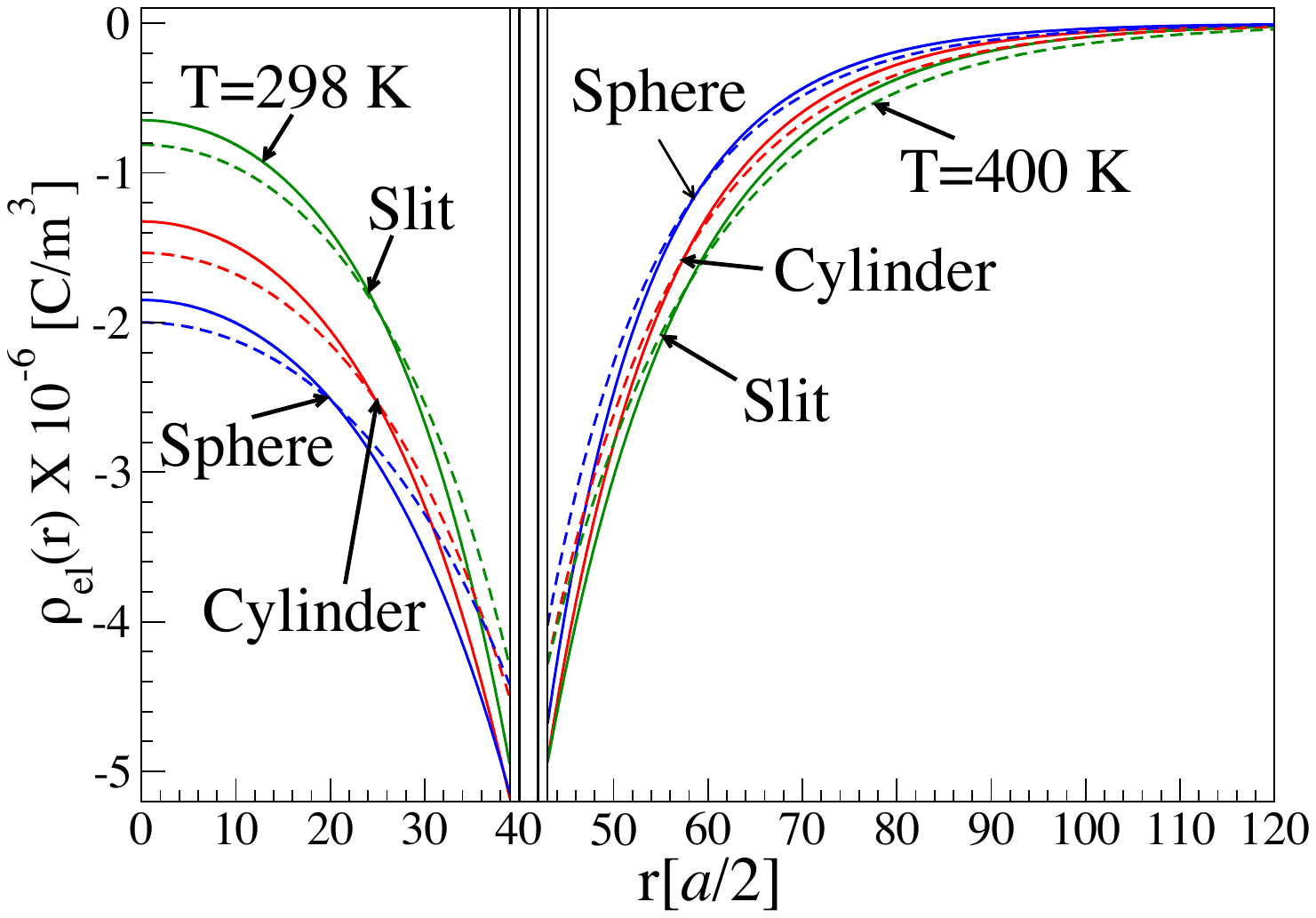}
		\caption{$\rho_{\scriptscriptstyle{el}}(r)$ for $R=\SI{20}{\textit{a}}$.}
		\label{Fig.PCS_rhoel[r]_R20_s0p015_rho0p01}
	\end{subfigure}\\
	\caption{Induced concentration profile, $\rho_{\scriptscriptstyle{el}}(r)$, inside and outside of the nanopores, as a function of the distance to their geometrical center, $r$. Two nanopore radius are considered. In  $(a)$ $R=3.5a$ and $sinh^-1(r)$ scale is used. Whereas in $(b)$ $R=20a$ and the scale is linear. Two system temperatures are investigated, $T=298 K$ (solid lines) and $T=400 K$ (broken lines). In all cases, $\sigma_{\scriptscriptstyle{I}}=\sigma_{\scriptscriptstyle{II}}=\sigma_{\scriptscriptstyle{0}}=0.015 C/m^2$, and $\rho_{\scriptscriptstyle{0}}=0.01M$.}
	\label{Fig.PCS_rhoel[r]_R3p5_R20_s0p015_rho0p01}
\end{figure}

Within this linear theory, from \cref{Ec.Poisson,Ec.ALS_Poisson}, $\rho_{\scriptscriptstyle{el}}(r)=-\varepsilon_{\scriptscriptstyle{0}}\varepsilon \kappa ^2\psi(r)$. Then $\rho^{in}_{\scriptscriptstyle{el}}(r)=-\varepsilon_{\scriptscriptstyle{0}}\varepsilon \kappa ^2\psi_{\scriptscriptstyle{1}}(r)$ and $\rho^{out}_{\scriptscriptstyle{el}}(r)=-\varepsilon_{\scriptscriptstyle{0}}\varepsilon \kappa ^2\psi_{\scriptscriptstyle{5}}(r)$ are the charge concentration profiles inside and outside the nanopores, for the three geometries. A direct calculation with the formulas of $\psi_{\scriptscriptstyle{1}}(r)$ and $\psi_{\scriptscriptstyle{5}}(r)$ given in \cref{Theory}, produces \cref{Fig.PCS_rhoel[r]_R3p5_R20_s0p015_rho0p01}, where we have chosen two values of of the nanopores radius, $R=3.5a$ and $R=20a$ and two different temperatures, $T=298 K$ and $T=400 K$. The surface charge density $\sigma_{\scriptscriptstyle{I}}=\sigma_{\scriptscriptstyle{II}}=\sigma_{\scriptscriptstyle{0}}=0.015 C/m^2$, and salt concentration $\rho_{\scriptscriptstyle{0}}=0.01M$,  for all cases. These concentration profiles are inversely proportional to their radius, since larger radius implies a less violation of the local electroneutrality condition (see \cref{Fig.PCS_sigma[r]_R18.8,Fig.PCS_sigma[r]_R18.8_d20}), thereby decreasing the effective surface charge density, $\sigma_{\scriptscriptstyle{Ho}}$, on the other side of the nanopores walls. On the other hand, in general, they decrease and increase with decreasing or increasing electrolyte concentration and temperature, respectively. This is because they are strongly dependent of $\kappa\propto\sqrt[2]{\rho_{\scriptscriptstyle{0}}/T}$.

Next to the nanopore's walls (internal and external), the electrolyte charge is strongly adsorbed, to then decrease away from the walls (becomes   less negative). For very narrow pores, inside the pore (see \cref{PCS_rhoel[r]_R3p5_s0p015_rho0p01}), $\rho^{in}_{\scriptscriptstyle{el}}(r)$ is nearly constant. However, it is not; they have a slight charge increase (more negative charge), as $r\rightarrow (R-a/2)$ (see \cref{Fig.PCS.psi(r).sigmaMR3.5}). The adsorbed negative charge is larger for the slit-pore, followed by the cylindrical, and spherical pores. The larger the adsorbed negative charge, the larger the induced, negative surface charge, $\sigma_{\scriptscriptstyle{Hi}}$. Thus, the net positive charge, $\sigma_{\scriptscriptstyle{Ho}}$, and therefore $\rho^{out}_{\scriptscriptstyle{el}}(r)$, for the slit nanopore should be above those for the cylindrical and spherical pores. However, the opposite is observed in \cref{PCS_rhoel[r]_R3p5_s0p015_rho0p01} (see also \cref{Fig.PCS_sigma[r]_R3.5}). This is a geometrical effect, since the electrical field $E_{\scriptscriptstyle_{5}}(r)$ decreases faster for the spherical pore, followed by the cylindrical and slit pores (see \cref{Plates-E(r),Cylinder-E(r),Sphere-E(r)}).

For wider pores, for example for $R=20a$, inside the pores the order of the adsorbed electrolyte charge is inverted, i.e., more charge is adsorbed in the spherical pore, followed by the cylindrical and slit pores, in this order (see \cref{Fig.PCS_rhoel[r]_R20_s0p015_rho0p01}). Again, this seems to be meanly a geometrical effect, i.e., for equal dimensions, the volume of a sphere is larger than that for a cylinder, followed by that of a slit, since their corresponding electrical fields at $E_{\scriptscriptstyle_{1}}(r=R-a/2)=\sigma_{\scriptscriptstyle{Hi}}$ have the inverted order (see \cref{Fig.PCS_sigma[r]_R3.5,Fig.PCS_sigma[r]_R18.8}). For this larger radius, the order of $\rho^{out}_{\scriptscriptstyle{el}}(r)$, in relation to the nanopores geometry, is as that for $R=3.5a$. However, $\rho^{in}_{\scriptscriptstyle{el}}(r)\rightarrow\rho^{out}_{\scriptscriptstyle{el}}(r)$, as $R$ increases to balance the local electroneutrality at both sides of the nanopores walls.

There is an interesting effect exhibited by $\rho_{\scriptscriptstyle{el}}(r)$ $\forall\, r \  \cdot \ni \cdot\ (0 \le r \le R-a/2) \ \bigcup \ (R+d+a/2 \le r \le \infty)$, as a function of the temperature $T$; next to the nanopore walls, for the three geometries,  $\rho_{\scriptscriptstyle{el}}(r,T=298 K)$ is larger (more negative) than that for  $\rho_{\scriptscriptstyle{el}}(r,T=400K)$, whereas for large $r$ occurs the opposite. As indicated above, higher temperature, implies lower values of $\kappa$, thus $\rho_{\scriptscriptstyle{el}}(r,T=400K) < \rho_{\scriptscriptstyle{el}}(r,T=298K)$. However, sufficiently away from the walls the opposite is observed, since for the two temperatures $\rho_{\scriptscriptstyle{el}}(r)$ must satisfy overall electroneutrality. In spite of this clear argument, it is interesting that the crossing point of these two curves occurs nearly for the same $r\equiv r_{\scriptscriptstyle_{c}}$, in the three geometries, and that outside the pores $r^{out}_{\scriptscriptstyle_{c}}$ is numerically very close to $\lambda_{\scriptscriptstyle_{D}}$ and to the Bjerrum length, $\lambda_{\scriptscriptstyle_{B}}$, i.e., $r^{out}_{\scriptscriptstyle_{c}}\approx \lambda_{\scriptscriptstyle_{D}} \approx \lambda_{\scriptscriptstyle_{B}}$, for the three geometries. $\lambda_{\scriptscriptstyle_{B}}$ is calculated with the equation  $\psi_{\scriptscriptstyle_{5}}(r=\lambda_{\scriptscriptstyle_{B}})-K_{\scriptscriptstyle_{B}}T/e=0$. Inside the pore, $r^{in}_{\scriptscriptstyle_{c}}$, for sufficiently wide pores, is in general larger than $\lambda_{\scriptscriptstyle_{B}}$, calculated with the previous equation, now with $\psi_{\scriptscriptstyle_{1}}(r=\lambda_{\scriptscriptstyle_{B}})$.

Here, let us point out that although no direct comparison is given here of the concentration and potential of mean force profiles presented in \cref{Fig.PCS.g(r)_s0.002,Fig.PCS.g(r).sigmaMR3.5,Fig.PCS.psi(r).sigmaMR3.5}, there is a qualitative agreement of these results with those obtained from the PB equation for point ions, the hypernetted chain/mean-spherical approximation (HNC/MSA) integral equation and computer simulations for the RPM~\cite{McQuarrie_StatMech} for the three geometries, inside and outside of the nanopores~\cite{Gonzalez_1985,Lozada_1990-I,Degreve_1993,Yeomans1993,Yu_1997}. These is of course to be expected since it has been shown that the HNC/MSA equation reduces to the PB equation in the limit of point ions~\cite{Lozada_1984,Lozada_1983,Gonzalez_1989}, and the PB equation reduces to the LPB equation at low mean electrostatic potentials. Our present equations were analytically derived from the PB equation, with no other approximation.

%%%%%%%%%%%%%%%%%%%%%%%%%%%%%%%%%%%%%%%%%%%%%%%%%%%%%%%%%%%%%%%%%%%%%%%%%%%%%
%This phenomena indicates a higher energy, $\omega_j=W_{j}/A_j(r_i)$, stored in %the planar nanopore, since in general
%\begin{equation}
%	W_i=\frac{1}{2}\varepsilon_{\scriptscriptstyle{0}}\varepsilon %\int_{\Omega}E_i^2(r)\,\dd^3{r}\label{Ec.esf.energy_i},
%\end{equation}
%\noindent where $i$ and $\Omega$ refer to the different nanopore topologies %and total volume around the nanopore. In the case of the slit nanopore, only %the right hand side of the nanopore should be considered. We will come back %to this point in the next subsection.

%%%%%%%%%%%%%%%%%%%%%%%%%%%%%%%%%%%%%%%%%%%%%%%%%%%%%%%%%%%%%%%%%%%%%%%%%%%%%%%
\subsection{Capacitances: numerical results}\label{Numerical results capacitance}

\begin{figure}[!htb]
	\begin{subfigure}{.5\textwidth}
		\centering
		\includegraphics[width=0.95\linewidth]{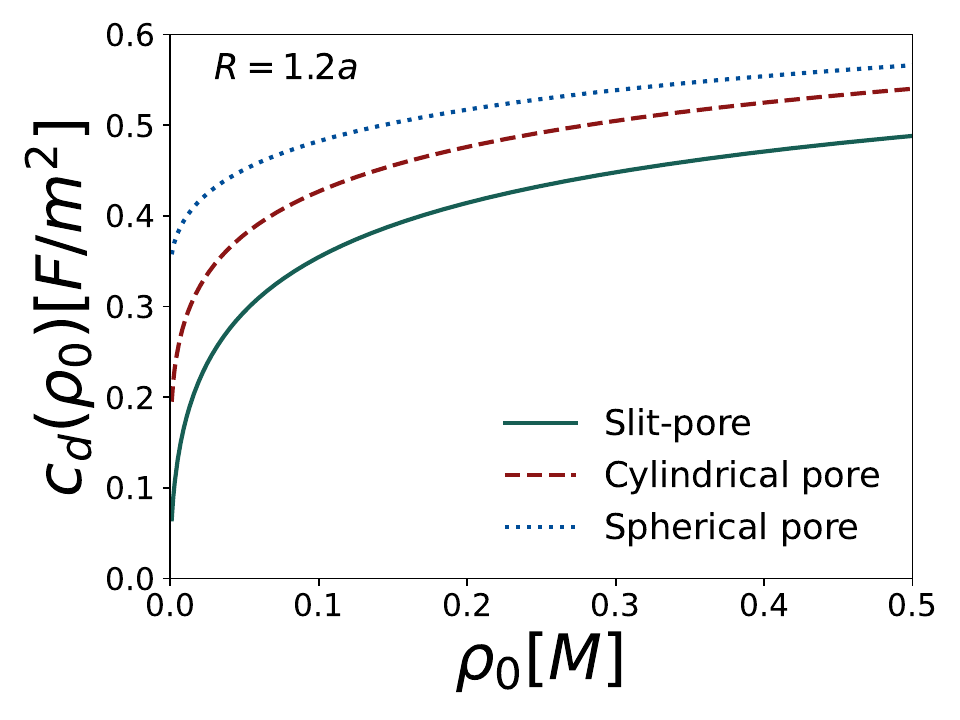}
		\caption{$c_{\scriptscriptstyle{d}}$ for $R=\SI{1,2}{\textit{a}}$.}
		\label{Fig.PCS.Cd(rho).R1.2}
	\end{subfigure}%
	\begin{subfigure}{.5\textwidth}
		\centering
		\includegraphics[width=0.95\linewidth]{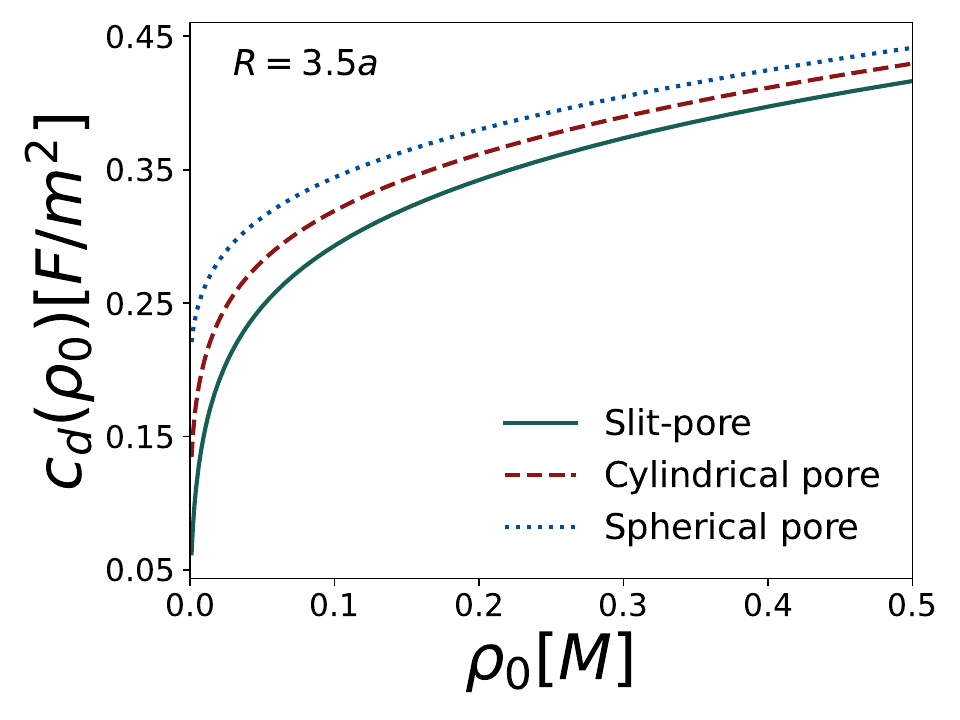}
		\caption{$c_{\scriptscriptstyle{d}}$ for $R=\SI{3,5}{\textit{a}}$.}
		\label{Fig.PCS.Cd(rho).R3.5}
	\end{subfigure}\\
	\begin{subfigure}{.5\textwidth}
		\centering
		\includegraphics[width=0.95\linewidth]{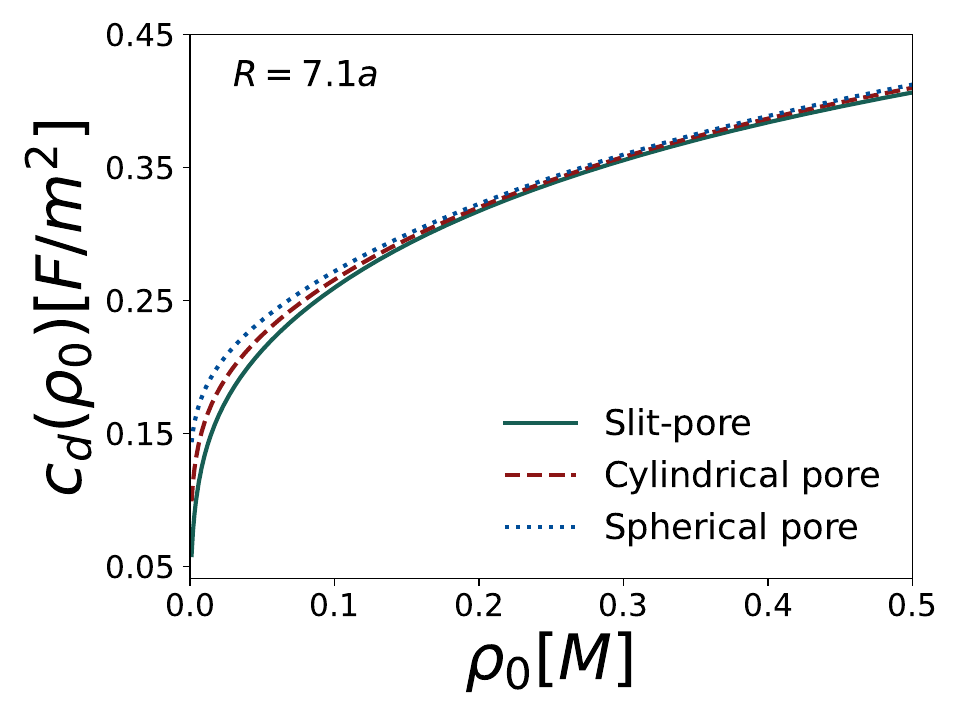}
		\caption{$c_{\scriptscriptstyle{d}}$ for $R=\SI{7,1}{\textit{a}}$.}
		\label{Fig.PCS.Cd(rho).R7.1}
	\end{subfigure}%
	\begin{subfigure}{.5\textwidth}
		\centering
		\includegraphics[width=0.95\linewidth]{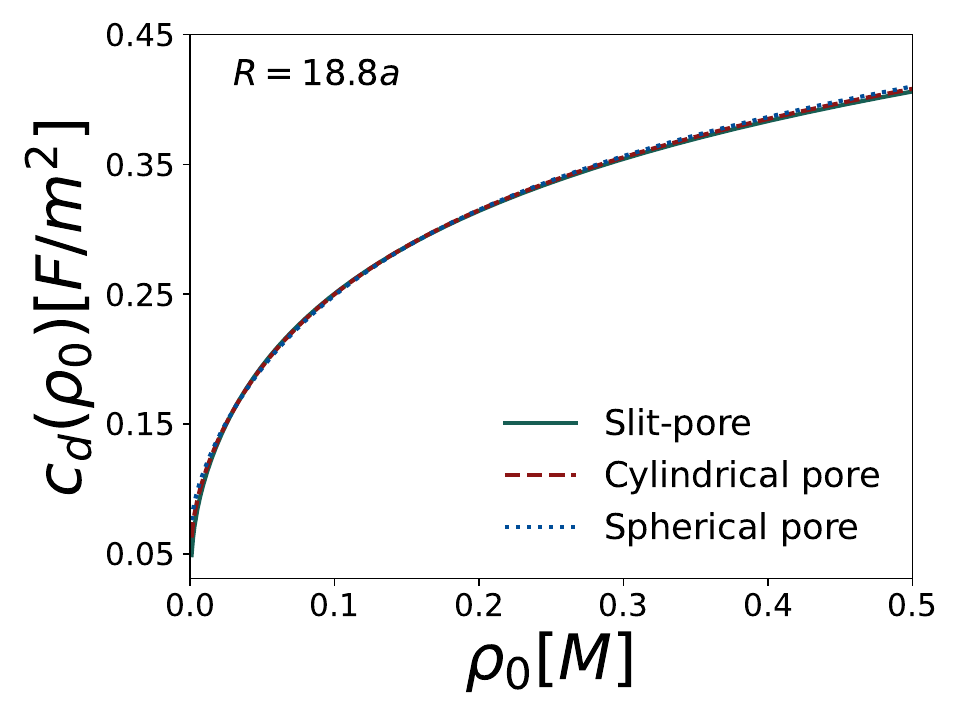}
		\caption{$c_{\scriptscriptstyle{d}}$ for $R=\SI{18,8}{\textit{a}}$.}
		\label{Fig.PCS.Cd(rho).R18.8}
	\end{subfigure}
	\caption{Capacitance of nanopore electrodes, $c_{\scriptscriptstyle{d}}\,[\si[per-mode=symbol]{\farad\per\square\metre}]$, as a function of the molar concentration, $\rho_{\scriptscriptstyle{0}}$, for four radii, of the slit, cylindrical and spherical nanopores. A temperature $T=298K$ is considered.}
	\label{Fig.PCS.Cd(rho).R}
\end{figure}

With the formulas given in \cref{Nanopores-capacitances}, the capacitance of the different electrode's topologies can be analytically calculated. The nanopores are taken to be immersed into a $1:1$ electrolyte with a dielectric constant of $78.5$, and,  unless otherwise indicated, a molar concentration of \SI{0.01}{\Molar}, a temperature of \SI{298}{\kelvin}, and a pore's width of $d= {\textit{a}}$.
The  total capacitances of the three topologies are calculated from \cref{Ec.cT}.
In \cref{Fig.PCS.Cd(rho).R,Tab.SNCs.Cd.rho}, the capacitances of the different nanopore electrodes are obtained for four radii, \SIlist{1,2;3,5;7,1;18,8}{\textit{a}}, while varying the electrolyte's molar concentration.
It is observed that the highest, \textit{specific}, capacitances are obtained for the most packed nanopores, where the spherical pores achieves the highest capacitance followed by the cylindrical pore and lastly the slit nanopore.
As the pore' radius increases the spherical and cylindrical pores tend quickly to the values of the slit pore.
Such that, for a radius of \SI{7,1}{\textit{a}} the three nanopores share almost the same capacitance for any value of the molar concentration, with only a slight difference for $\rho_{\scriptscriptstyle{0}}<0.15M$.
Whereas, with a radius of \SI{18,8}{\textit{a}} the capacitance of all the nanopores almost completely overlap. However, the capacitances of the three nanopores will become equal, only in the limit of $R\rightarrow \infty$, i.e., from the equations given above, straightforwardly one finds 

\begin{equation}
	 \lim_{R\to\infty} c_{\scriptscriptstyle d}(R,d,a,\rho_{\scriptscriptstyle{0}})  =\frac{\varepsilon_{\scriptscriptstyle{0}}\varepsilon \kappa}{2+\kappa (a+d)} \label{Ec.limit_c}.
\end{equation}

\noindent Thereby, in this limit 

\begin{equation}
	 \lim_{R\to\infty}\left(\frac{1}{c_{\scriptscriptstyle d}} \right)=\frac{(a+d)}{\varepsilon_{\scriptscriptstyle{0}}\varepsilon} +\frac{2}{\varepsilon_{\scriptscriptstyle{0}}\varepsilon \kappa} \label{Ec.limit_c_inverse},
\end{equation}

\noindent which is the capacitance of an isolated charge plate of thickness $d$. If only one side of the plate is considered, and the thickness of the plate is neglected, \cref{Ec.limit_c_inverse} reduces to the well known formula for the capacitance of an infinitely wide, single plate~\cite{Henderson2015}.

\begin{figure}[!htb]
	\begin{subfigure}{.5\textwidth}
		\centering
		\includegraphics[width=0.95\linewidth]{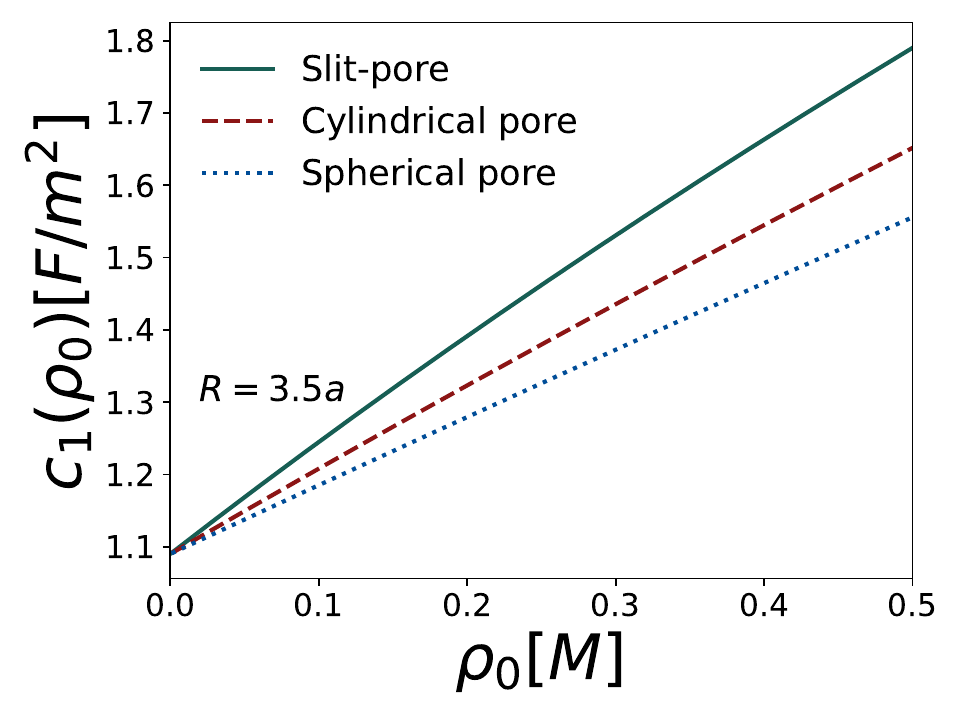}
		\caption{Differential capacitance, $c_{\scriptscriptstyle{1}}$, in region I.}
		\label{Fig.PCS.C1R3.5}
	\end{subfigure}%
	\begin{subfigure}{.5\textwidth}
		\centering
		\includegraphics[width=0.95\linewidth]{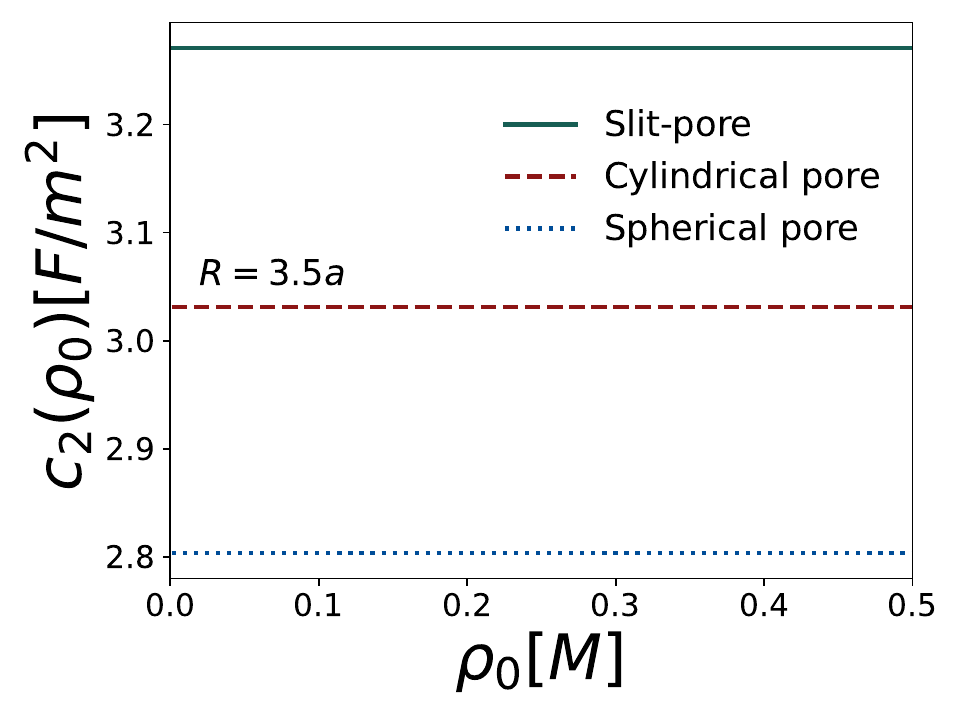}
		\caption{Differential capacitance, $c_{\scriptscriptstyle{2}}$, in region II.}
		\label{Fig.PCS.C2R3.5}
	\end{subfigure}\\
	\begin{subfigure}{.5\textwidth}
		\centering
		\includegraphics[width=0.95\linewidth]{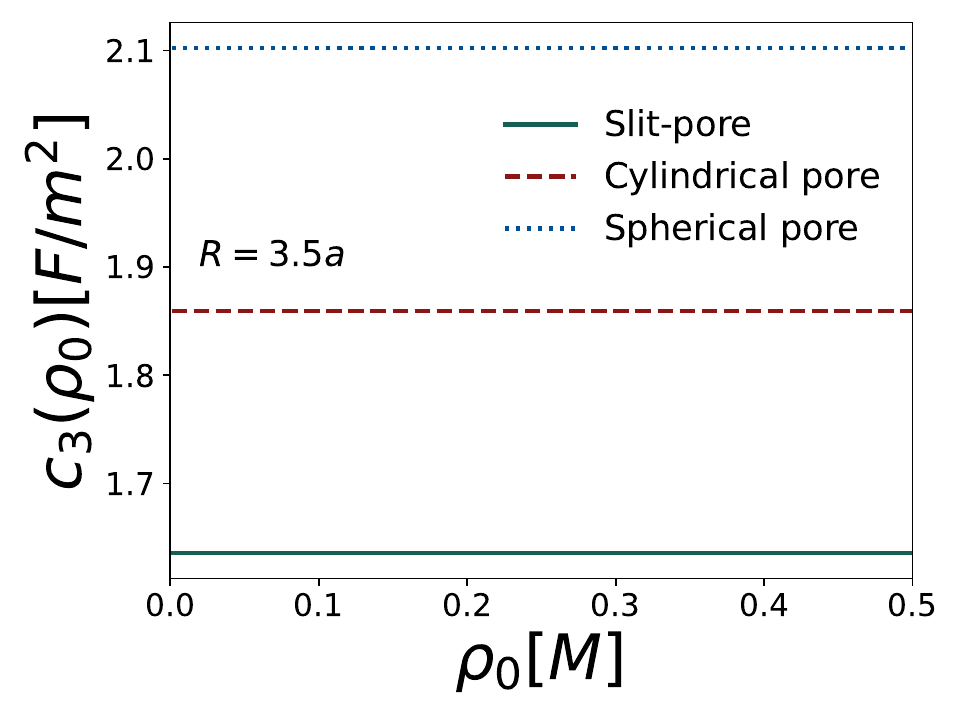}
		\caption{Differential capacitance, $c_{\scriptscriptstyle{3}}$, in region III.}
		\label{Fig.PCS.C3R3.5}
	\end{subfigure}%
	\begin{subfigure}{.5\textwidth}
		\centering
		\includegraphics[width=0.95\linewidth]{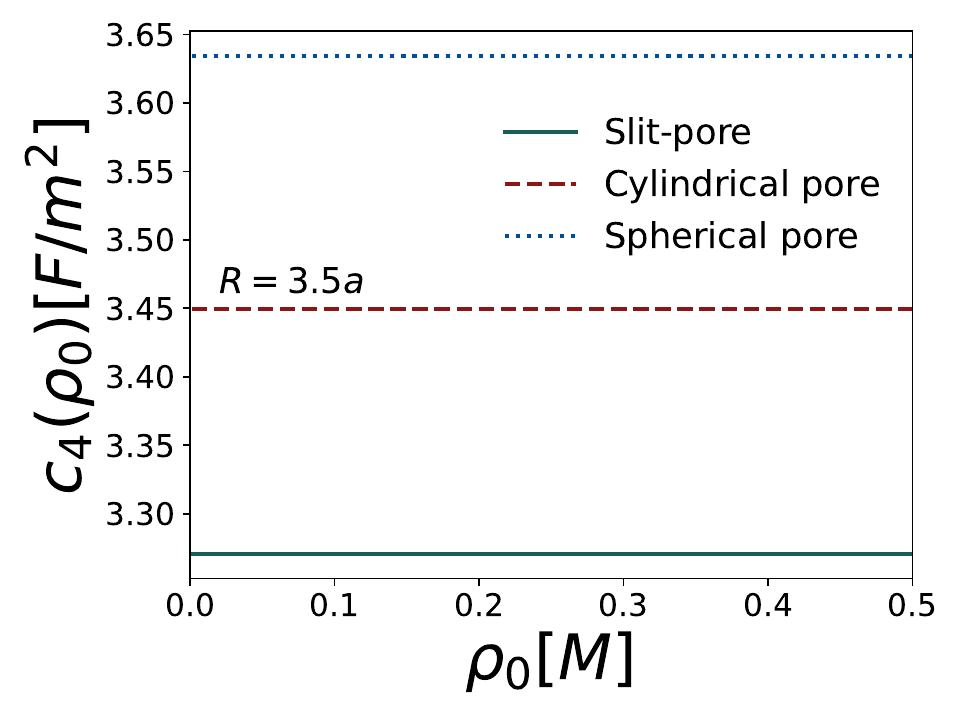}
		\caption{Differential capacitance, $c_{\scriptscriptstyle{4}}$, in region IV.}
		\label{Fig.PCS.C4R3.5}
	\end{subfigure}
\begin{subfigure}{.5\textwidth}
	\centering
	\includegraphics[width=0.95\linewidth]{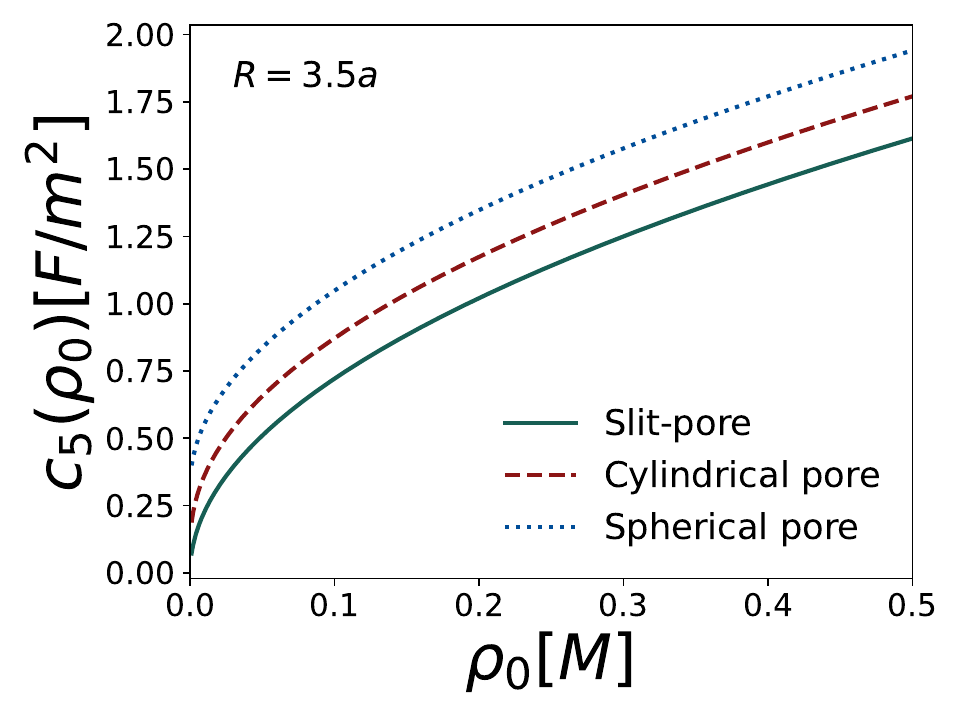}
	\caption{Differential capacitance, $c_{\scriptscriptstyle{5}}$, in region V.}
	\label{Fig.PCS.C5R3.5}
\end{subfigure}
	\caption{Capacitances, $c_{\scriptscriptstyle{i}}\,[\si[per-mode=symbol]{\farad\per\square\metre}]$, corresponding to the different charge distributions in a nanopore (see~\cref{Geometry_electrodes}), as a function of the molar concentration, $\rho_{\scriptscriptstyle{0}}$. The inner radius of the nanopores is $R=3.5a$, and the thickness of their walls is $d=a$.}
	\label{Fig.PCS.Ci(rho).R3.5}
\end{figure}

The increase of the capacitance, with increasing electrolyte concentration is physically appealing, since for any given charge density on the plates, a higher concentration implies a more compact and dense EDL, and thus, more charge is stored, per unit of volume. In \cref{Fig.PCS.Ci(rho).R3.5}, the specific capacitances corresponding to the five regions of the three nanopores geometries are portrayed as a function of $\rho_{\scriptscriptstyle{0}}$. Their nanopores width and thickness are $R=3.5a$ and $d=a$, respectively. As in the case of the integral specific capacitance, $c_{\scriptscriptstyle_{d}}$, the capacitances of regions I, $c_1$, and V, $c_5$, also augment with increasing $\rho_{\scriptscriptstyle{0}}$, while $c_2$, $c_3$ and $c_4$ are constant in their respective regions, since by construction the electrolyte can not penetrate the nanopores walls. In these regions, the capacitances correspond to standard capacitors filled with a material of dielectric permittivity $\varepsilon_{\scriptscriptstyle{0}}\varepsilon$, thus depending only of geometrical factors. In fact, $c_1$, and $c_5$ also depend of geometrical factors, filled with an electrolyte material.

In \cref{Fig.PCS.Cd(rho).R,Fig.PCS.Ci(rho).R3.5} it is observed that as $R$ increases, $c_{\scriptscriptstyle_{d}}$, $c_1$ and $c_5$ decrease for all nanopore geometries. On the hand these quantities increase with increasing bulk electrolyte concentration. A direct comparison with the density functional calculations of O. Pizio, et al.~\cite{Pizio2012} of the specific capacitance of a RPM electrolyte, confined between two plates, is not possible because they used constant potential boundary conditions. However, it is interesting that for low potentials and short distance between the plates or at large distance between the plates their calculated $c_{\scriptscriptstyle_{d}}$ shows the same qualitative behavior, and since they did not consider explicitly the electrical double layer outside the nanopore, their $c_{\scriptscriptstyle_{d}}$ should be compared with our $c_1$ (see Fig. 5 in reference~\cite{Pizio2012}).

\begin{table}[!htb]
	\centering
	\resizebox{\columnwidth}{!}{
		\begin{tabular}{ccccccccccccc}
			\toprule 
			&\multicolumn{3}{c}{$R=\SI{1,2}{\textit{a}}$}&\multicolumn{3}{c}{$R=\SI{3,5}{\textit{a}}$}&\multicolumn{3}{c}{$R=\SI{7,1}{\textit{a}}$}&\multicolumn{3}{c}{$R=\SI{18,8}{\textit{a}}$}\\ \cmidrule{2-13}
			$\rho_{\scriptscriptstyle{0}}\,[\si{\Molar}]$&P&C&S&P&C&S&P&C&S&P&C&S\\ \midrule
			0.001&0.065&0.195&0.358&0.063&0.135&0.221&0.059&0.099&0.143&0.049&0.063&0.076\\
			0.01&0.172&0.285&0.397&0.154&0.208&0.262&0.133&0.158&0.183&0.107&0.111&0.115\\
			0.1&0.355&0.427&0.482&0.293&0.319&0.344&0.260&0.266&0.272&0.250&0.250&0.249\\
			0.5&0.488&0.540&0.566&0.416&0.430&0.441&0.406&0.410&0.412&0.406&0.408&0.410\\
			\bottomrule
	\end{tabular}}
	\caption{The capacitances, $c_{\scriptscriptstyle{d}}\,[\si[per-mode=symbol]{\farad\per\square\metre}]$, of the slit nanopore (P), and cylindrical (C) and spherical (S) pores, for different radii, while the molar concentration, $\rho_{\scriptscriptstyle{0}}$, is varied.}\label{Tab.SNCs.Cd.rho} 
\end{table}

\begin{figure}[!htb]
	\begin{subfigure}{.5\textwidth}
		\centering
		\includegraphics[width=0.95\linewidth]{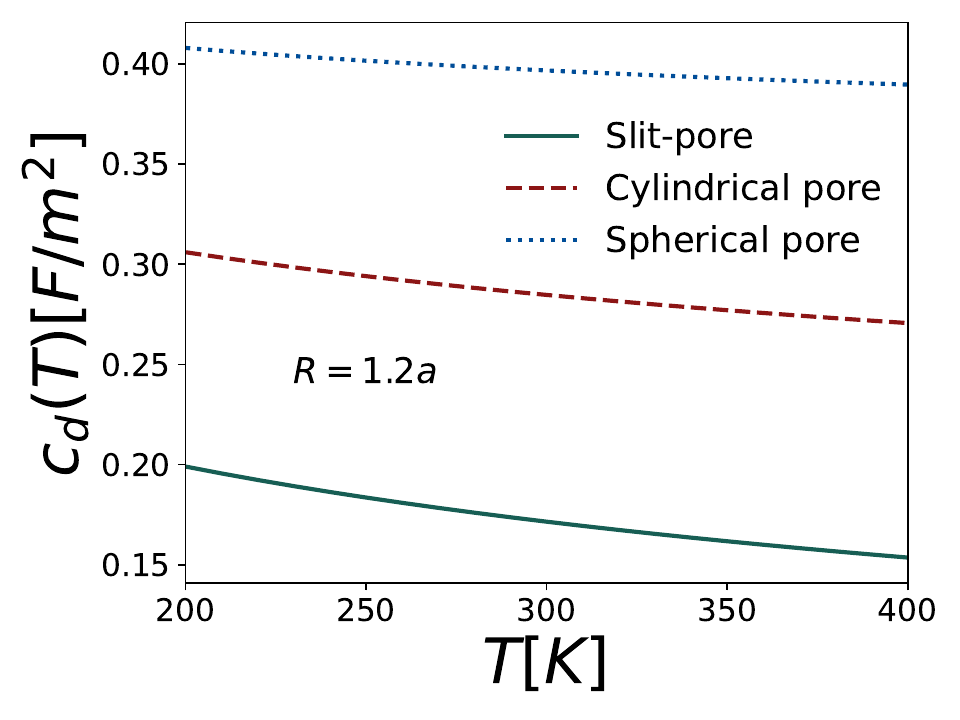}
		\caption{$c_{\scriptscriptstyle{d}}$ for $R=\SI{1,2}{\textit{a}}$.}
		\label{Fig.PCS.Cd(T).R1.2}
	\end{subfigure}%
	\begin{subfigure}{.5\textwidth}
		\centering
		\includegraphics[width=0.95\linewidth]{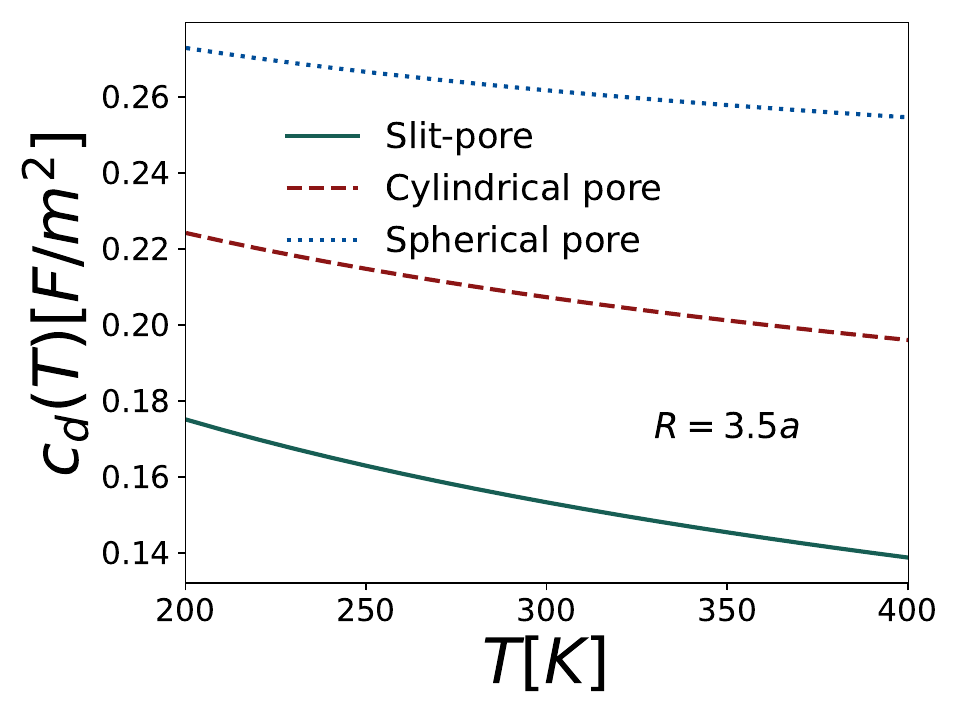}
		\caption{$c_{\scriptscriptstyle{d}}$ for $R=\SI{3,5}{\textit{a}}$.}
		\label{Fig.PCS.Cd(T).R3.5}
	\end{subfigure}\\
	\begin{subfigure}{.5\textwidth}
		\centering
		\includegraphics[width=0.95\linewidth]{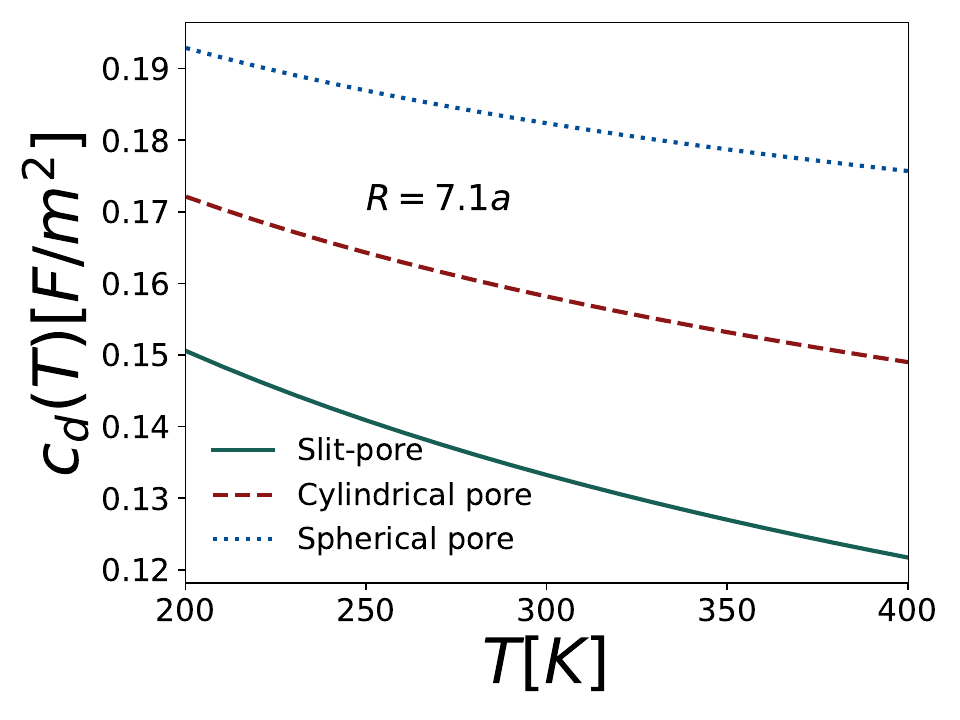}
		\caption{$c_{\scriptscriptstyle{d}}$ for $R=\SI{7,1}{\textit{a}}$.}
		\label{Fig.PCS.Cd(T).R7.1}
	\end{subfigure}%
	\begin{subfigure}{.5\textwidth}
		\centering
		\includegraphics[width=0.95\linewidth]{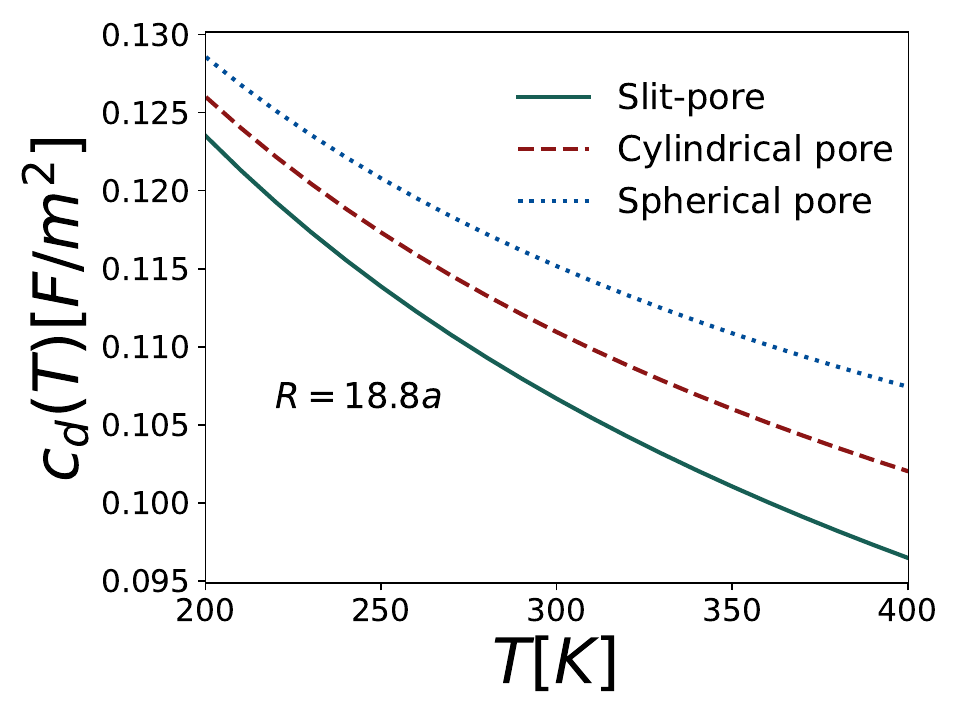}
		\caption{$c_{\scriptscriptstyle{d}}$ for $R=\SI{18,8}{\textit{a}}$.}
		\label{Fig.PCS.Cd(T).R18.8}
	\end{subfigure}
	\caption{Capacitance of nanopore electrodes, $c_{\scriptscriptstyle{d}}\,[\si[per-mode=symbol]{\farad\per\square\metre}]$, as a function of the temperature, $T$, for different radii.}
	\label{Fig.PCS.Cd(T).R}
\end{figure}

In \cref{Fig.PCS.Cd(T).R,Tab.SNCs.Cd.T} the linear capacitance of different topology electrodes are obtained for four radii, \SIlist{1,2;3,5;7,1;18,8}{\textit{a}}, while the electrolyte's temperature is varied.
Conversely to the MEP, the capacitance is inversely proportional to the electrolyte's temperature, i.e., when the temperature increases, the capacitance drops. This effect is to be expected, since as a higher temperature, implies a larger Debye length $\lambda_{\scriptscriptstyle_{D}}$, thereby generating a thicker, less dense EDL on the outside of the nanopores, and, as a consequence less charge per unit of volume. The temperature dependence of $c_1$ is negligible. Therefore, the decrease of $c_{\scriptscriptstyle_{d}}$, with the system's temperature is due almost entirely to $c_5$. Thence, this decrease of $c_{\scriptscriptstyle_{d}}$, with the temperature are in qualitative agreement with those obtained for the capacitance of a planar or slit electrode in contact with a restricted primitive model electrolyte reported in the past~\cite{Mier_1988,Henderson2005,Pizio2012}.

In summary, from \cref{Fig.PCS.Cd(rho).R,Fig.PCS.Cd(T).R,Tab.SNCs.Cd.rho,Tab.SNCs.Cd.T} the \textit{specific} capacitance is enhanced when a small spherical nanopore is used at a higher electrolyte's molar concentration, and a lower temperature, and this qualitative behavior seems to be in agreement with that found with much more sofisticated theories 

\subsection{Capacitance of solid nano-electrodes}

If we now choose to ignore the internal structure of the nanopores for $r<R+d$, the electrical field $E_4(R+d)=\sigma(r=R+d)/\varepsilon_{\scriptscriptstyle{0}}\varepsilon$, will now become that of a nano-electrode, with an effective surface charge density of $\sigma(R+d)$. Hence, the analytical results of regions IV and V, for the three nanopore geometries will be valid for their corresponding solid nano-electrodes. For the slit, cylindrical and spherical nanopores, $\sigma(r=R+d)$ is equal to $\sigma_{\scriptscriptstyle{Ho}}$, $[R_{\scriptscriptstyle_{H}}/(R+d)]\sigma_{\scriptscriptstyle{Ho}}$, and $[R_{\scriptscriptstyle_{H}}^2/(R+d)^2]\sigma_{\scriptscriptstyle{Ho}}$, respectively.

In the past, it was proposed a so-called capacitive compactness of the EDL induced by a spherical, charged electrode, i.e., it was shown that one can assume that the center of the charge induced on the surrounding electrolyte is located at a certain distance $\tau_{\scriptscriptstyle_{c}}$ from the geometrical center of the spherical electrode~\cite{Gonzalez-Tovar-2004}. Hence, one can think this system as a capacitor with charge density $\sigma(r=R+d)$ located on the surface of the electrode, and a charge density $-\sigma(r=R+d)$, located at $r=\tau_{\scriptscriptstyle_{c}}$. A straightforward application of electrostatic theory gives analytical expressions of  $\tau_{\scriptscriptstyle_{c}}$ for a \textit{solid} planar, cylindrical and spherical electrodes, in terms of their surface potential and charge~\cite{Gonzalez-Tovar-2021}. Let us refer to $\tau_{\scriptscriptstyle_{c}}$  as the capacitive compactness~\cite{Gonzalez-Tovar-2018,Gonzalez-Tovar-2021}. The analytical formulas for $\tau_{\scriptscriptstyle_{c}}$ for the planar, cylindrical and spherical electrodes are:

 \begin{equation}
 	\tau_{\scriptscriptstyle_{cp}}=\frac{\varepsilon_{\scriptscriptstyle{0}}\varepsilon \varphi_{\scriptscriptstyle_{0}}}{\sigma(R+d)} \label{Ec.tau_Plate},
 \end{equation}

\begin{equation}
	\tau_{\scriptscriptstyle_{cc}}=(R+d)exp\left(\frac{\varepsilon_{\scriptscriptstyle{0}}\varepsilon \varphi_{\scriptscriptstyle_{0}}}{(R+d)\sigma(R+d)}\right) \label{Ec.tau_Cyl},
\end{equation}

\noindent and

\begin{equation}
	\tau_{\scriptscriptstyle_{cs}}=(R+d)\left[1-\frac{\varepsilon_{\scriptscriptstyle{0}}\varepsilon \varphi_{\scriptscriptstyle_{0}}}{(R+d)\sigma(R+d)}\right]^{-1} \label{Ec.tau_Sphere},
\end{equation}

\noindent respectively. Let us draw attention to the fact that these analytical expressions for $\tau_{\scriptscriptstyle_{c}}$ are exact electrostatic results. With \cref{Ec.tau_Plate,Ec.tau_Cyl,Ec.tau_Sphere} and their corresponding equations for $\varphi_{\scriptscriptstyle_{0}}$ given in \cref{Slit-pore,Cylindrical-pore,Spherical-pore}, it can be straightforwardly show that $\tau_{\scriptscriptstyle_{c}}$ for the planar, cylindrical and spherical electrodes are

	\begin{gather}
	\tau_{\scriptscriptstyle_{cp}}=R_{\scriptscriptstyle_{H}} + \frac{1}{\kappa} \label{tau-Plate},\\
	\tau_{\scriptscriptstyle_{cc}}=R_{\scriptscriptstyle_{H}} exp\left\{\frac{K_{\scriptscriptstyle_{0}}[\kappa R_{\scriptscriptstyle_{H}} ]}{\kappa R_{\scriptscriptstyle_{H}} K_{\scriptscriptstyle_{1}}[\kappa R_{\scriptscriptstyle_{H}} ]}\right\} \label{tau-Cylinder},\\	\tau_{\scriptscriptstyle_{cs}}=R_{\scriptscriptstyle_{H}} + \frac{1}{\kappa}\label{tau-Sphere},	
\end{gather}
\label{Tau-all}
 
\noindent respectively. Clearly,  $\tau_{\scriptscriptstyle_{cp}}$ and $\tau_{\scriptscriptstyle_{cs}}$ are Debye lengths, $\lambda_{\scriptscriptstyle_{D}}$, measured from the geometrical center of the electrode. Although, for the case of the planar electrode, due to the fact that for this geometry the surface charge density at $r=R+d$ and at $r=R+d+a/2$ are the same, $\tau_{\scriptscriptstyle_{cp}}$ can be indistinctly measured from the center of the electrode or its surface. For the cylindrical electrode, its MEP in region V is $\psi_{\scriptscriptstyle_{5}}(r) =\varphi_{\scriptscriptstyle_{H}}K_0[\kappa r]/K_0[\kappa R_{\scriptscriptstyle_{H}}]$ (see \cref{Cylinder-MEP(r)}). A numerical calculation of $\tau_{\scriptscriptstyle_{cs}}$ with this MEP shows that $\tau_{\scriptscriptstyle_{cs}}=\lambda_{\scriptscriptstyle_{D}}=R_{\scriptscriptstyle_{H}}+1/\kappa$. So $\tau_{\scriptscriptstyle_{cc}}$ is also a Debye length.

Thence, considering the capacitances of the nano-electrodes as that of two capacitors connected in series, i.e., $1/c=1/c_{\scriptscriptstyle_{H}}+1/c_{\scriptscriptstyle_{d}}$, corresponding to the Helmholtz and diffuse layers, respectively. The formulas for planar, cylindrical and spherical capacitors, with surface charges $\sigma(R+d)$ and $-\sigma(R+d)$, located at $R+d$ and $\tau_{\scriptscriptstyle_{c}}$, respectively, are:

	\begin{gather}
	1/c_{\scriptscriptstyle_{P}}=1/c_{\scriptscriptstyle_{H}}+1/c_{\scriptscriptstyle_{d}}= \frac{1}{\varepsilon_{\scriptscriptstyle{0}}\varepsilon}(a/2+1/\kappa) \label{c-tau-Plate},\\
	1/c_{\scriptscriptstyle_{C}}=1/c_{\scriptscriptstyle_{H}}+1/c_{\scriptscriptstyle_{d}}= \frac{1}{\varepsilon_{\scriptscriptstyle{0}}\varepsilon}\left\{(R+d)ln\left[\frac{R_{\scriptscriptstyle_{H}}}{R+d}\right]+\frac{K_{\scriptscriptstyle_{0}}[\kappa(R+a/2)]}{\kappa K_{\scriptscriptstyle_{1}}[\kappa(R+a/2)]}\right\}\label{c-tau-Cylinder},\\	
	1/c_{\scriptscriptstyle_{S}}=1/c_{\scriptscriptstyle_{H}}+1/c_{\scriptscriptstyle_{d}}= \frac{1}{\varepsilon_{\scriptscriptstyle{0}}\varepsilon}\left\{\frac{(R+d)a/2}{R_{\scriptscriptstyle_{H}}}+\frac{R_{\scriptscriptstyle_{H}}}{1+\kappa R_{\scriptscriptstyle_{H}}}\right\}\label{c-tau-Sphere},
\end{gather}

\noindent where all the capacitances are \textit{specific}, i.e., capacitance per area, and where we have used the standard formulas for planar, cylindrical and spherical capacitors, as given in elementary physics books~\cite{Resnick-2001}, i.e., $c_{\scriptscriptstyle_{P}}=\varepsilon_{\scriptscriptstyle{0}}\varepsilon/d_{\scriptscriptstyle_{P}}$, $c_{\scriptscriptstyle_{C}}=\varepsilon_{\scriptscriptstyle{0}}\varepsilon/[aln[b/a]]$, and $c_{\scriptscriptstyle_{S}}=\varepsilon_{\scriptscriptstyle{0}}\varepsilon/[a(1-a/b)]$, respectively, and where $d_{\scriptscriptstyle_{P}}$ is the distance between the plates in the planar capacitor, and $a$ and $b$ are inner and outer radius in the cylindrical and spherical capacitors.  \cref{c-tau-Plate,c-tau-Cylinder,c-tau-Sphere} are in complete accordance with those obtained from $1/c=1/c_4 +1/c_5$, with $c_4$ and $c_5$ given by their expressions in \cref{Nanopores-capacitances} for the three nanopore geometries. Our results, validates the concept of an effective center of charge in induced electrical double layers~\cite{Gonzalez-Tovar-2004,Gonzalez-Tovar-2021}. An effective center of charge can also be obtained for the internal EDL of the nanopores, however, we will not present this approach in this publication.

\begin{table}[!htb]
	\centering
	\resizebox{\columnwidth}{!}{
		\begin{tabular}{ccccccccccccc}
			\toprule 
			&\multicolumn{3}{c}{$R=\SI{1,2}{\textit{a}}$}&\multicolumn{3}{c}{$R=\SI{3,5}{\textit{a}}$}&\multicolumn{3}{c}{$R=\SI{7,1}{\textit{a}}$}&\multicolumn{3}{c}{$R=\SI{18,8}{\textit{a}}$}\\ \cmidrule{2-13}
			$T\,[\si{\kelvin}]$&P&C&S&P&C&S&P&C&S&P&C&S\\ \midrule
			278.15&0.176&0.288&0.397&0.157&0.210&0.262&0.136&0.161&0.183&0.110&0.114&0.115\\
			298.15&0.172&0.285&0.397&0.154&0.208&0.262&0.133&0.158&0.183&0.107&0.111&0.115\\
			308.15&0.170&0.283&0.396&0.152&0.206&0.261&0.132&0.157&0.182&0.106&0.100&0.114\\
			328.15&0.166&0.280&0.394&0.149&0.204&0.260&0.130&0.155&0.180&0.103&0.108&0.113\\
			\bottomrule 
	\end{tabular}}
	\caption{The capacitances, $c_{\scriptscriptstyle{d}}\,[\si[per-mode=symbol]{\farad\per\square\metre}]$, of different nanopores for four radii, while the temperature, $T$, is varied.
		The initials (P, C, S) used have the same meaning as in \cref{Tab.SNCs.Cd.rho}.}\label{Tab.SNCs.Cd.T}
\end{table}

\section{Conclusions}\label{Conclusions}

By analytically solving the linear Poisson-Boltzmann equation, analytical expressions are derived for the electrical properties of nano-pores and nano-electrodes. In particular closed formulas are given for the specific capacitance of planar, cylindrical and spherical geometries. Numerical results are presented for some sets of the systems parameters. We find that they qualitatively agree with numerical solutions for the non-linear Poisson-Boltzmann, HNC/MSA, density functional equations and computer simulations existing in the literature. Given the same circumstances, the nanopore electrodes achieve the highest capacitance, compared to the solid nano-electrodes; and the spherical, and cylindrical topologies favor the capacitance, being of course the spherical nanopore topology the most effective.
Furthermore, the capacitance depends on the mean electrostatic potential, which in turns highly depends on the nanopore surface charge and on the ionic concentration.
In consequence, the higher the molar concentration, the lower the mean electrostatic potential and the higher the capacitance, i.e., the larger amount of charge captured by the nanopores.
Lastly, even though the temperature has the least impact on the capacitance (the higher the temperature, the lower the capacitance), in a large array of nanopores, closely packed, as components of a  supercapacitor device, its temperature during operation will increase, hence decreasing the capacitance. Therefore, the temperature must be taken into account to achieve a high capacitance.

As general results, we find that: (a) narrower spherical nanopores, at higher electrolyte concentration, and lower temperature produce the higher capacitance; (b)  In the limit of infinitely wide nanopores, the capacitances of the three geometries become that of an isolated, single, charged plate, i.e., $1/c(r\rightarrow\infty)=(a+d)/\varepsilon_{\scriptscriptstyle{0}}\varepsilon +2/(\varepsilon_{\scriptscriptstyle{0}}\varepsilon \kappa)$; (c) All of our results reduce to those for solid nano-electrodes, assuming the surface charge density on the external nanopores walls, as that of a solid electrode. As a corollary of these results, it is shown that the capacitances of these solid-electrodes can also be obtained by considering an oppositely charged layer, located at the center of charge, $\tau_{\scriptscriptstyle_{c}}$, of the induced electrical double layer outside the nanopores. This outcome validates the concept of capacitive compactness, proposed in the past~\cite{Gonzalez-Tovar-2004,Gonzalez-Tovar-2018,Gonzalez-Tovar-2021}.

We are aware that our results are limited to low surface potentials on the capacitors, and low electrolyte concentration, among many other model limitations, such as ionic sizes and/or dielectric constant on the nanopores~\cite{Ala-Nisila_2011}. However, within the limitations of the linear Poisson-Boltzmann equation, our results are analytical and exact, for these parameters conditions. Hence, our results might be useful not only to understand some basic physics of nanopores in bulk, but also as bench marking of more sophisticated theories, computer simulations, and experimental results.

\section*{Acknowledgement}

The support of UNAM (PAPIIT Clave: IN108023) and CONACYT (Grant No. 169125) is acknowledged.

%\begin{appendix}
\appendix
\section{\\Cylindrical Nanopore}\label{A}
\subsection{\textbf{Region V }}

\begin{gather*}
	\nabla^2 \psi(r) = \kappa^2 \psi(r) \implies \\
	\frac{1}{r} \frac{\partial}{\partial r} \left[ r \frac{\partial \psi(r)}{\partial r} \right] + \frac{1}{r^2} \frac{\partial^2\psi(r)}{\partial \varphi^2} + \frac{\partial^2\psi(r)}{\partial z^2} = \kappa^2\psi(r) \implies \\
	\frac{1}{r} \frac{\partial}{\partial r} \left[ r \frac{\partial \psi(r)}{\partial r} \right] = \kappa^2\psi(r) \implies \\
	\frac{\partial^2\psi(r)}{\partial r^2} + \frac{1}{r}\frac{\partial \psi(r)}{\partial r} = \kappa^2 \psi(r) \implies  \\
	\psi_5(r) = A I_0\left[ \kappa r\right] + B K_0 \left[ \kappa r \right] \quad \textrm{and} \\
	\frac{d\psi(r)}{dr} = A\kappa I_1\left[\kappa r\right] - B\kappa K_1 \left[ \kappa r \right],
\end{gather*}
where $I_0 \left[ \kappa r\right]$, $K_0 \left[ \kappa r \right]$, $I_1\left[ \kappa r\right]$ and $K_1\left[ \kappa r \right]$ are the modified Bessel functions of the first and second kind of order 0 and 1, respectively, and where 
\begin{eqnarray*}
	I_0\left[ \kappa r \right] = \frac{1}{\pi} \int_{0}^{\pi} e^{\kappa r cos\theta}d\theta \\
	K_0 \left[ \kappa r \right] = \int_{0}^{\infty} e^{-\kappa r cosh\theta}d\theta,
\end{eqnarray*}
hence
\begin{eqnarray*}
	\frac{dI_0\left[\kappa r\right]}{dr} = \kappa I_1\left[ \kappa r \right] = \frac{\kappa}{\pi} \int_{0}^{\pi} e^{\kappa r cos\theta}cos\theta d\theta \\
	\frac{dK_0\left[\kappa r\right]}{dr} = -\kappa K_1 \left[ \kappa r \right] = -\kappa \int_{0}^{\infty} e^{-\kappa r cosh\theta}cosh\theta d\theta.
\end{eqnarray*}
\subsubsection{Boundary conditions}
\begin{gather*}
	\psi(\infty) =0, \ \ E(\infty) = - \nabla \psi(r) = 0 \\
	\therefore \psi(\infty) = A I_0 \left[ \infty \right] + B K_0 \left[ \infty \right] = 0 \implies A =0 \\
	E_5(\infty) = B \kappa K_1\left[ \infty \right] \to 0 \\
	\therefore \psi_5(r) = BK_0 \left[ \kappa r \right] \\
	E_5(r) = B \kappa K_1 \left[\kappa r \right]  \\
	\varphi_H = \psi_5 \left( R+d+\frac{a}{2} \right) = B K_0 \left[ \kappa \left(R+d+\frac{a}{2}\right) \right] \\
	\therefore B= \frac{\varphi_H}{K_0\left[\kappa R_{\scriptscriptstyle_{H}} \right]}; \quad R_{\scriptscriptstyle_{H}} \equiv R+d+\frac{a}{2} \\
	\therefore \psi_5(r) = \frac{\varphi_H}{K_0\left[\kappa R_{\scriptscriptstyle_{H}} \right]}K_0 \left[\kappa r\right].
\end{gather*}
On the other hand,
\begin{gather*}
	E_5 (R_{\scriptscriptstyle_{H}}) = \frac{\sigma_{\scriptscriptstyle{Ho}}}{\varepsilon_{\scriptscriptstyle{0}} \varepsilon} \implies E_5(R_{\scriptscriptstyle_{H}}) = \frac{\varphi_H \kappa }{K_0\left[ \kappa R_{\scriptscriptstyle_{H}} \right]} K_1 \left[ \kappa R_{\scriptscriptstyle_{H}} \right] = \frac{\sigma_{\scriptscriptstyle{Ho}}}{\varepsilon_{\scriptscriptstyle{0}} \varepsilon}, \\
	\therefore E_5(r) = \frac{\varphi_H \kappa}{K_0 \left[ \kappa R_{\scriptscriptstyle_{H}} \right]} K_1\left[ \kappa r \right]
\end{gather*}
Hence
\begin{eqnarray*}
	\varphi_H = \frac{K_0 \left[ \kappa R_{\scriptscriptstyle_{H}} \right]}{\kappa K_1 \left[ \kappa R_{\scriptscriptstyle_{H}}\right]} \frac{\sigma_{\scriptscriptstyle{Ho}}}{\varepsilon_{\scriptscriptstyle{0}} \varepsilon}.
\end{eqnarray*}
%%%%%%%%%%%%%%%%%%%%%%%%%%%%%%%%
\subsection{Region IV}

\begin{gather*}
	\nabla^2 \psi(r) = 0 \implies \frac{\partial^2 \psi(r)}{\partial r^2} + \frac{1}{r} \frac{\partial \psi(r)}{\partial r} = 0 \implies \\
	\psi_4(r) = D+Clnr, \ \ E_4(r) = -\nabla \psi_4(r) = -\frac{C}{r} \\
	\textrm{but} \ \ E_4(R+d) = - \frac{C}{R+d} = \frac{\sigma_4 (R+d)}{\varepsilon_{\scriptscriptstyle{0}} \varepsilon} = \frac{Q_{Ho}}{\varepsilon_{\scriptscriptstyle{0}} \varepsilon \left[ 2\pi (R+d)L_0 \right]}  \\
	\textrm{but} \ \ Q_{Ho} = 2\pi (R+d)L_0 \sigma_4(R+d) = 2\pi \left(R+d+\frac{a}{2} \right)L_0 \sigma_{\scriptscriptstyle{Ho}} \implies \\
	E_4(R+d) = -\frac{C}{R+d} = \frac{2\pi R_{\scriptscriptstyle_{H}}L_0 \sigma_{\scriptscriptstyle{Ho}}}{\varepsilon_{\scriptscriptstyle{0}} \varepsilon \left[ 2\pi (R+d)L_0 \right]} = \frac{R_{\scriptscriptstyle_{H}} \sigma_{\scriptscriptstyle{Ho}}}{(R+d)\varepsilon_{\scriptscriptstyle{0}} \varepsilon} \implies \\
	C = -R_{\scriptscriptstyle_{H}} \frac{\sigma_{\scriptscriptstyle{Ho}}}{\varepsilon_{\scriptscriptstyle{0}} \varepsilon} \implies E_4(r) = \frac{R_{\scriptscriptstyle_{H}} \sigma_{\scriptscriptstyle{Ho}}}{\varepsilon_{\scriptscriptstyle{0}} \varepsilon r} \\
	\textrm{and} \ \ \varphi_0 \equiv \psi_4(R+d) = -\frac{R_{\scriptscriptstyle_{H}} \sigma_{\scriptscriptstyle{Ho}}}{\varepsilon_{\scriptscriptstyle{0}} \varepsilon} ln(R+d) + D \implies \\
	D = \varphi_0 + \frac{R_{\scriptscriptstyle_{H}} \sigma_{\scriptscriptstyle{Ho}}}{\varepsilon_{\scriptscriptstyle{0}} \varepsilon} ln(R+d),
\end{gather*}
\noindent 	where $L_0$ is the cylinder axial length.
\begin{gather*}
	\psi_4(r) = - \frac{R_{\scriptscriptstyle_{H}} \sigma_{\scriptscriptstyle{Ho}}}{\varepsilon_{\scriptscriptstyle{0}} \varepsilon} ln r + \varphi_0 + \frac{R_{\scriptscriptstyle_{H}} \sigma_{\scriptscriptstyle{Ho}}}{\varepsilon_{\scriptscriptstyle{0}} \varepsilon} ln(R+d) \implies \\
	\psi_4(r) = \varphi_0 + \frac{R_{\scriptscriptstyle_{H}} \sigma_{\scriptscriptstyle{Ho}}}{\varepsilon_{\scriptscriptstyle{0}} \varepsilon}ln\left[\frac{R+d}{r}\right] \\
	\psi_4(R_{\scriptscriptstyle_{H}}) \equiv \varphi_H = \varphi_0 + \frac{R_{\scriptscriptstyle_{H}} \sigma_{\scriptscriptstyle{Ho}}}{\varepsilon_{\scriptscriptstyle{0}} \varepsilon}ln\left[ \frac{R+d}{R_{\scriptscriptstyle_{H}}} \right] \implies \\
	\varphi_0 - \varphi_H = \frac{R_{\scriptscriptstyle_{H}} \sigma_{\scriptscriptstyle{Ho}}}{\varepsilon_{\scriptscriptstyle{0}} \varepsilon} ln \left[ \frac{R_{\scriptscriptstyle_{H}}}{R+d} \right]
\end{gather*}
%%%%%%%%%%%%%%%%%%%%%%%%%%%%%%%%%%%%%%%%%%%%%%%%%%%%%
\subsection{Region III}

\begin{gather*}
	\varepsilon_{\scriptscriptstyle{0}} \varepsilon E_4(R+d) - \varepsilon_{\scriptscriptstyle{0}} \varepsilon E_3(R+d) = \sigma_0 \\
	\psi_3(r) = Flnr + G \\
	E_3(r) = -\nabla \psi_3(r) = -\frac{F}{r} \\
	\textrm{but} \ \ E_3(R+d) = E_4(R+d) - \frac{\sigma_0}{\varepsilon_{\scriptscriptstyle{0}} \varepsilon}
\end{gather*}
and since $E_4(R+d) = \frac{R_{\scriptscriptstyle_{H}}}{R+d} \frac{\sigma_{\scriptscriptstyle{Ho}}}{\varepsilon_{\scriptscriptstyle{0}} \varepsilon} \implies$
\begin{gather*}
	E_3(R+d) = \frac{R_{\scriptscriptstyle_{H}}}{R+d} \frac{\sigma_{\scriptscriptstyle{Ho}}}{\varepsilon_{\scriptscriptstyle{0}} \varepsilon} - \frac{\sigma_0}{\varepsilon_{\scriptscriptstyle{0}} \varepsilon} = -\frac{F}{R+d} \implies \\
	-F = R_{\scriptscriptstyle_{H}} \frac{\sigma_{\scriptscriptstyle{Ho}}}{\varepsilon_{\scriptscriptstyle{0}} \varepsilon} - \frac{(R+d)\sigma_0}{\varepsilon_{\scriptscriptstyle{0}} \varepsilon} \implies \\
	E_3(r) = \left[ \frac{R_{\scriptscriptstyle_{H}}}{\varepsilon_{\scriptscriptstyle{0}} \varepsilon}\sigma_{\scriptscriptstyle{Ho}} - \frac{(R+d)\sigma_0}{\varepsilon_{\scriptscriptstyle{0}} \varepsilon} \right] \frac{1}{r} \\
	\psi_3(r) = -  \left[ \frac{R_{\scriptscriptstyle_{H}}}{\varepsilon_{\scriptscriptstyle{0}} \varepsilon}\sigma_{\scriptscriptstyle{Ho}} - \frac{(R+d)\sigma_0}{\varepsilon_{\scriptscriptstyle{0}} \varepsilon} \right] ln(r) + G \\
	\psi_0 = \psi_3(R) = Fln(R)+G \\
	\varphi_0 = \psi_3(R+d) = Fln(R+d) + G \\
	\psi_0 - \varphi_0 =  \left[ \frac{R_{\scriptscriptstyle_{H}}}{\varepsilon_{\scriptscriptstyle{0}} \varepsilon}\sigma_{\scriptscriptstyle{Ho}} - \frac{(R+d)\sigma_0}{\varepsilon_{\scriptscriptstyle{0}} \varepsilon} \right] ln \left[\frac{R+d}{R}\right].
\end{gather*}
%%%%%%%%%%%%%%%%%%%%%%%%%%%%%%%%%%%%%%%%%%%%%%%%%%%%%%%%
\subsection{Region II}

\begin{gather*}
	\varepsilon_{\scriptscriptstyle{0}} \varepsilon E_3(R) - \varepsilon_{\scriptscriptstyle{0}} \varepsilon E_2(R) = \sigma_0 \\
	\psi_2(r) = Hlnr + I \\
	E_2(r) = -\nabla \psi_2(r) = -\frac{H}{r} \\
	E_2(R) = E_3(R) - \frac{\sigma_0}{\varepsilon_{\scriptscriptstyle{0}} \varepsilon} = \left[ \frac{R_{\scriptscriptstyle_{H}}}{\varepsilon_{\scriptscriptstyle{0}} \varepsilon}\sigma_{\scriptscriptstyle{Ho}} - \frac{(R+d)\sigma_0}{\varepsilon_{\scriptscriptstyle{0}} \varepsilon} \right] \frac{1}{R} - \frac{\sigma_0}{\varepsilon_{\scriptscriptstyle{0}} \varepsilon} = -\frac{H}{R}
\end{gather*}
\begin{gather*}
	\implies -H = \left[ \frac{R_{\scriptscriptstyle_{H}}}{\varepsilon_{\scriptscriptstyle{0}} \varepsilon}\sigma_{\scriptscriptstyle{Ho}} - \frac{(R+d)\sigma_0}{\varepsilon_{\scriptscriptstyle{0}} \varepsilon} \right] - \frac{R\sigma_0}{\varepsilon_{\scriptscriptstyle{0}} \varepsilon} \implies \\
	E_2(r) = \frac{1}{\varepsilon_{\scriptscriptstyle{0}} \varepsilon} \left[ R_{\scriptscriptstyle_{H}} \sigma_{\scriptscriptstyle{Ho}} - (2R+d)\sigma_0 \right] \frac{1}{r} \\
	\psi_2(r) = -   \frac{1}{\varepsilon_{\scriptscriptstyle{0}} \varepsilon} \left[ R_{\scriptscriptstyle_{H}} \sigma_{\scriptscriptstyle{Ho}} - (2R+d)\sigma_0 \right] ln(r) + I \\
	\psi_0 = \psi_2(R) = Hln(R)+I \implies I = \psi_0 - Hln(R) \\
	\psi_2 (r) = \psi_0 + \frac{1}{\varepsilon_{\scriptscriptstyle{0}} \varepsilon}\left[ R_{\scriptscriptstyle_{H}} \sigma_{\scriptscriptstyle{Ho}} - (2R+d)\sigma_0 \right] ln \left[ \frac{R}{r} \right] \\
	\psi_2\left( R-\frac{a}{2} \right) = \psi_H = \psi_0 + \frac{1}{\varepsilon_{\scriptscriptstyle{0}} \varepsilon} \left[ R_{\scriptscriptstyle_{H}} \sigma_{\scriptscriptstyle{Ho}} - (2R+d)\sigma_0 \right] ln \left[ \frac{R}{R-\frac{a}{2}} \right] \\
	\psi_H - \psi_0 = \frac{1}{\varepsilon_{\scriptscriptstyle{0}} \varepsilon} \left[ R_{\scriptscriptstyle_{H}} \sigma_{\scriptscriptstyle{Ho}} - (2R+d)\sigma_0 \right] ln \left[ \frac{R}{R-\frac{a}{2}} \right]
\end{gather*}
%%%%%%%%%%%%%%%%%%%%%%%%%%%%%%%%%%%%%
\subsection{Region I}

\begin{gather*}
	\psi_1(r) = LI_0\left[ \kappa r \right] + MK_0\left[ \kappa r \right] \\
	E_1(r) = -\nabla \psi_1(r) = -L \kappa I_1\left[ \kappa r \right] + M \kappa K_1\left[ \kappa r \right]
\end{gather*}
such that $\psi_{\scriptscriptstyle_{1}}\left(R-\frac{a}{2}\right) = \psi_2\left(R-\frac{a}{2}\right)$, $E_1\left(R-\frac{a}{2} \right)= E_2\left( R-\frac{a}{2}\right)$
\\ and $\lim_{r\to 0} E_1(r) = 0$; and since $I_1\left[0\right]=0$ and $K_1[0] = \infty$ $\implies$ $M=0$ $\implies$
\begin{gather*}
	E_1(r) = -L\kappa I_1\left[ \kappa r\right] \quad \textrm{and} \\
	\psi_1(r) = LI_0\left[\kappa r\right] \\
	E_1\left(R-\frac{a}{2}\right) = -L\kappa I_1\left[\kappa \left(R-\frac{a}{2}\right)\right] = E_2\left(R-\frac{a}{2}\right) \implies \\
	-L\kappa I_1\left[\kappa \left( R-\frac{a}{2} \right)\right] = \frac{1}{\varepsilon_{\scriptscriptstyle{0}} \varepsilon} \left[ R_{\scriptscriptstyle_{H}}\sigma_{\scriptscriptstyle{Ho}} - (2R+d)\sigma_0 \right] \frac{1}{(R-\frac{a}{2})} \\
	\therefore E_1(r) = \frac{1}{\varepsilon_{\scriptscriptstyle{0}} \varepsilon} \left[ R_{\scriptscriptstyle_{H}}\sigma_{\scriptscriptstyle{Ho}} - (2R+d)\sigma_0 \right] \frac{I_1\left[\kappa r\right]}{\left( R-\frac{a}{2}\right)I_1\left[\kappa \left(R-\frac{a}{2}\right)\right]} \\
	\psi_1(r) = -\frac{1}{\varepsilon_{\scriptscriptstyle{0}} \varepsilon} \left[ R_{\scriptscriptstyle_{H}}\sigma_{\scriptscriptstyle{Ho}} - (2R+d)\sigma_0 \right] \frac{I_0\left[\kappa r\right]}{\kappa \left( R-\frac{a}{2}\right)I_1\left[\kappa \left(R-\frac{a}{2}\right)\right]} \\
	\textrm{but} \ \ \psi_1\left(R-\frac{a}{2}\right) = \psi_2\left( R-\frac{a}{2}\right) \implies \\
	-\frac{1}{\varepsilon_{\scriptscriptstyle{0}} \varepsilon}\left[ R_{\scriptscriptstyle_{H}}\sigma_{\scriptscriptstyle{Ho}} - (2R+d)\sigma_0 \right] \frac{I_0\left[\kappa \left(R-\frac{a}{2}\right)\right]}{\kappa \left( R-\frac{a}{2}\right)I_1\left[\kappa \left(R-\frac{a}{2}\right)\right]} = \psi_H \\
	\psi(0) \equiv \psi_d = -\frac{1}{\varepsilon_{\scriptscriptstyle{0}} \varepsilon}\left[ R_{\scriptscriptstyle_{H}}\sigma_{\scriptscriptstyle{Ho}} - (2R+d)\sigma_0 \right] \frac{1}{\kappa \left( R-\frac{a}{2}\right)I_1\left[\kappa \left(R-\frac{a}{2}\right)\right]} \\
	\psi_d - \psi_H = \frac{1}{\varepsilon_{\scriptscriptstyle{0}} \varepsilon} \frac{\left[ R_{\scriptscriptstyle_{H}}\sigma_{\scriptscriptstyle{Ho}} - (2R+d)\sigma_0 \right]}{\kappa \left( R-\frac{a}{2}\right)I_1\left[\kappa \left(R-\frac{a}{2}\right)\right]} \left( I_0\left[\kappa \left(R-\frac{a}{2}\right)\right]-1 \right)
\end{gather*}

Now, to obtain the formulas for $\sigma_{\scriptscriptstyle{Hi}}$ and $\sigma_{\scriptscriptstyle{Ho}}$ in terms of $\sigma_{\scriptscriptstyle{0}}$, we will list the following useful relation, although some of them have presented in the main text:

\begin{gather*}
	\varphi_H = \frac{K_0\left[ \kappa R_{\scriptscriptstyle_{H}} \right]}{\kappa K_1\left[\kappa R_{\scriptscriptstyle_{H}}\right]}\frac{\sigma_{\scriptscriptstyle{Ho}}}{\varepsilon_{\scriptscriptstyle{0}} \varepsilon} \\
	\psi_H = \psi_1\left( R-\frac{a}{2} \right) = - \frac{1}{\varepsilon_{\scriptscriptstyle{0}} \varepsilon} \left[R_{\scriptscriptstyle_{H}}\sigma_{\scriptscriptstyle{Ho}}-(2R+d)\sigma_0\right] \frac{I_0\left[\kappa \left(R-\frac{a}{2}\right)\right]}{\kappa\left( R-\frac{a}{2}\right)I_1\left[\kappa\left(R-\frac{a}{2}\right)\right]} \\
	\varphi_H = \varphi_0 + \frac{R_{\scriptscriptstyle_{H}} \sigma_{\scriptscriptstyle{Ho}}}{\varepsilon_{\scriptscriptstyle{0}} \varepsilon} ln \left[\frac{R+d}{R_{\scriptscriptstyle_{H}}} \right] \\
	\psi_H = \psi_0 + \frac{1}{\varepsilon_{\scriptscriptstyle{0}} \varepsilon} \left[ R_{\scriptscriptstyle_{H}}\sigma_{\scriptscriptstyle{Ho}} -(2R+d)\sigma_0 \right] ln \left[ \frac{R}{R-\frac{a}{2}} \right] \\
	\psi_0 -\varphi_0 = \frac{1}{\varepsilon_{\scriptscriptstyle{0}} \varepsilon} \left[ R_{\scriptscriptstyle_{H}}\sigma_{\scriptscriptstyle{Ho}} -(R+d)\sigma_0 \right] ln \left[ \frac{R+d}{R}\right]
\end{gather*} 
\begin{eqnarray*}
	\psi_H-\varphi_H &=& \psi_0 - \varphi_0 + \frac{1}{\varepsilon_{\scriptscriptstyle{0}} \varepsilon} \left[ R_{\scriptscriptstyle_{H}} \sigma_{\scriptscriptstyle{Ho}} - (2R+d)\sigma_0 \right] ln \left[ \frac{R}{R-\frac{a}{2}} \right] - \frac{R_{\scriptscriptstyle_{H}}}{\varepsilon_{\scriptscriptstyle{0}} \varepsilon}\sigma_0 ln \left[ \frac{R+d}{R_{\scriptscriptstyle_{H}}} \right] \\
	&=& \frac{1}{\varepsilon_{\scriptscriptstyle{0}} \varepsilon} \left[ R_{\scriptscriptstyle_{H}} \sigma_{\scriptscriptstyle{Ho}} - (R+d)\sigma_0 \right] ln \left[  \frac{R+d}{R}\right] \\
	& &+ \frac{1}{\varepsilon_{\scriptscriptstyle{0}} \varepsilon} \left[ R_{\scriptscriptstyle_{H}} \sigma_{\scriptscriptstyle{Ho}} - (2R+d)\sigma_0 \right] ln \left[  \frac{R}{R-\frac{a}{2}}\right] - \frac{R_{\scriptscriptstyle_{H}}}{\varepsilon_{\scriptscriptstyle{0}} \varepsilon}\sigma_{\scriptscriptstyle{Ho}} ln \left[ \frac{R+d}{R_{\scriptscriptstyle_{H}}} \right] 
\end{eqnarray*}
\begin{gather*}
	\psi_d-\psi_H = \frac{\left[ R_{\scriptscriptstyle_{H}}\sigma_{\scriptscriptstyle{Ho}} - (2R+d)\sigma_0 \right]\left[ I_0\left[\kappa\left(R-\frac{a}{2}\right)-1\right] \right]}{\varepsilon_{\scriptscriptstyle{0}} \varepsilon \kappa \left(R-\frac{a}{2}\right)I_1\left[ \kappa \left( R-\frac{a}{2} \right) \right]} \implies
\end{gather*}
\begin{gather*}
	\psi_H = \frac{K_0\left[ \kappa R_{\scriptscriptstyle_{H}} \right]}{\kappa K_1\left[ \kappa R_{\scriptscriptstyle_{H}} \right]} \frac{\sigma_{\scriptscriptstyle{Ho}}}{\varepsilon_{\scriptscriptstyle{0}} \varepsilon} + \frac{1}{\varepsilon_{\scriptscriptstyle{0}} \varepsilon} \left[ R_{\scriptscriptstyle_{H}}\sigma_{\scriptscriptstyle{Ho}} - (R+d)\sigma_0 \right] ln \left[ \frac{R+d}{R} \right] \\[10pt]
	+ \frac{1}{\varepsilon_{\scriptscriptstyle{0}} \varepsilon} \left[ R_{\scriptscriptstyle_{H}}\sigma_{\scriptscriptstyle{Ho}} - (2R+d)\sigma_0 \right] ln \left[ \frac{R}{R-\frac{a}{2}} \right] - \frac{R_{\scriptscriptstyle_{H}}\sigma_{\scriptscriptstyle{Ho}}}{\varepsilon_{\scriptscriptstyle{0}} \varepsilon} ln \left[ \frac{R+d}{R_{\scriptscriptstyle_{H}}} \right] \\[10pt]
	= - \frac{1}{\varepsilon_{\scriptscriptstyle{0}} \varepsilon} \left[ R_{\scriptscriptstyle_{H}}\sigma_{\scriptscriptstyle{Ho}} - (2R+d)\sigma_0 \right] \frac{I_0\left[\kappa \left(R-\frac{a}{2}\right)\right]}{\kappa \left(R-\frac{a}{2}\right)I_1\left[ \kappa \left(R-\frac{a}{2}\right)\right]} \implies \\[10pt]
	\left( \frac{K_0\left[\kappa R_{\scriptscriptstyle_{H}}\right]}{\kappa K_1\left[\kappa R_{\scriptscriptstyle_{H}}\right]} + R_{\scriptscriptstyle_{H}}ln\left[ \frac{R+d}{R} \right] + R_{\scriptscriptstyle_{H}}ln\left[ \frac{R}{R-\frac{a}{2}} \right] - R_{\scriptscriptstyle_{H}}ln\left[ \frac{R+d}{R_{\scriptscriptstyle_{H}}} \right] \right. \\[10pt]
	+ \left. \frac{R_{\scriptscriptstyle_{H}} I_0\left[\kappa \left(R-\frac{a}{2}\right) \right]}{\kappa \left( R-\frac{a}{2} \right)I_1\left[\kappa \left( R-\frac{a}{2} \right)\right]} \right) \sigma_{\scriptscriptstyle{Ho}} = \left( \frac{(2R+d)I_0\left[ \kappa\left(R-\frac{a}{2}\right) \right]}{\kappa \left( R-\frac{a}{2}\right)I_1\left[\kappa\left(R-\frac{a}{2}\right)\right]} + (R+d)ln\left[ \frac{R+d}{R} \right] \right. \\[10pt]
	+ \left. (2R+d)ln\left[ \frac{R}{R-\frac{a}{2}} \right] \right) \sigma_0 \implies \\[10pt]
	%%%%
	\left( \frac{K_0\left[\kappa R_{\scriptscriptstyle_{H}}\right]}{\kappa K_1\left[\kappa R_{\scriptscriptstyle_{H}}\right]} + R_{\scriptscriptstyle_{H}}ln\left[ \frac{R_{\scriptscriptstyle_{H}}}{R-\frac{a}{2}} \right] + \frac{R_{\scriptscriptstyle_{H}} I_0\left[\kappa \left(R-\frac{a}{2}\right) \right]}{\kappa \left( R-\frac{a}{2} \right)I_1\left[\kappa \left( R-\frac{a}{2} \right)\right]} \right) \sigma_{\scriptscriptstyle{Ho}} \\[10pt]
	= \left( \frac{(2R+d)I_0\left[ \kappa\left(R-\frac{a}{2}\right) \right]}{\kappa \left( R-\frac{a}{2}\right)I_1\left[\kappa\left(R-\frac{a}{2}\right)\right]} + (R+d)ln\left[ \frac{R+d}{R} \right] \right. \\[10pt]
	+ \left. (2R+d)ln\left[ \frac{R}{R-\frac{a}{2}} \right] \right) \sigma_0 \implies \\[10pt]
	\sigma_{\scriptscriptstyle{Ho}} = \frac{L_2}{L_1}\sigma_0.
\end{gather*}
On the other hand,
\begin{gather*}
	2\pi RL_0 \sigma_0 + 2\pi(R+d)L_0\sigma_0 = 2\pi R_{\scriptscriptstyle_{H}}L_0\sigma_{\scriptscriptstyle{Ho}} + 2\pi\left(R-\frac{a}{2}\right)L_0\sigma_{\scriptscriptstyle{Hi}} \implies \\
	\sigma_{\scriptscriptstyle{Hi}} = \frac{(2R+d)}{R-\frac{a}{2}}\sigma_0 - \frac{R_{\scriptscriptstyle_{H}}}{R-\frac{a}{2}}\sigma_{\scriptscriptstyle{Ho}} \\
	L_2 = \left( \frac{(2R+d)I_0\left[ \kappa\left(R-\frac{a}{2}\right) \right]}{\kappa \left( R-\frac{a}{2}\right)I_1\left[\kappa\left(R-\frac{a}{2}\right)\right]} + (R+d)ln\left[ \frac{R+d}{R} \right] \right. \\
	+ \left. (2R+d)ln\left[ \frac{R}{R-\frac{a}{2}} \right] \right)\\
	L_1 = 	\left( \frac{K_0\left[\kappa R_{\scriptscriptstyle_{H}}\right]}{\kappa K_1\left[\kappa R_{\scriptscriptstyle_{H}}\right]} + R_{\scriptscriptstyle_{H}}ln\left[ \frac{R_{\scriptscriptstyle_{H}}}{R-\frac{a}{2}} \right] + \frac{R_{\scriptscriptstyle_{H}} I_0\left[\kappa \left(R-\frac{a}{2}\right) \right]}{\kappa \left( R-\frac{a}{2} \right)I_1\left[\kappa \left( R-\frac{a}{2} \right)\right]} \right)
\end{gather*}
%%%%%%%%%%%%%%%%%%%%%%%%%%%%%
\subsection{Capacitance/Area}

\begin{gather*}
	C_5 = \frac{Q_{Ho}}{\varphi_H} = \frac{2\pi L_0 R_{\scriptscriptstyle_{H}} \sigma_{\scriptscriptstyle{Ho}}}{\varphi_H} \implies c_5 = \frac{C_5}{2\pi L_0 R_{\scriptscriptstyle_{H}}} = \frac{\sigma_{\scriptscriptstyle{Ho}}}{\varphi_H} \implies \\
	c_5 = \frac{\sigma_{\scriptscriptstyle{Ho}}}{\frac{K_0\left[\kappa R_{\scriptscriptstyle_{H}}\right]\sigma_{\scriptscriptstyle{Ho}}}{\varepsilon_{\scriptscriptstyle{0}} \varepsilon \kappa K_1\left[\kappa R_{\scriptscriptstyle_{H}}\right]}} = \frac{\varepsilon_{\scriptscriptstyle{0}} \varepsilon \kappa K_1\left[\kappa R_{\scriptscriptstyle_{H}}\right]}{K_0\left[\kappa R_{\scriptscriptstyle_{H}}\right]} \\
	C_4 = \frac{Q_4(R+d)}{\varphi_0-\varphi_H} = \frac{2\pi L_0 (R+d) \sigma_4(R+d)}{\varphi_0-\varphi_H} \implies \\
	c_4 = \frac{C_4}{2\pi L_0(R+d)} = \frac{\sigma_4(R+d)}{\varphi_0-\varphi_H}
\end{gather*}
but $2\pi L_0(R+d)\sigma_4(R+d) = 2\pi L_0 R_{\scriptscriptstyle_{H}}\sigma_{\scriptscriptstyle{Ho}}$, hence
\begin{gather*}
	c_4 = \frac{R_{\scriptscriptstyle_{H}}\sigma_{\scriptscriptstyle{Ho}}}{(R+d)(\varphi_0-\varphi_H)} = \frac{R_{\scriptscriptstyle_{H}} \sigma_{\scriptscriptstyle{Ho}}}{(R+d)\left[\frac{R_{\scriptscriptstyle_{H}}\sigma_{\scriptscriptstyle{Ho}}}{\varepsilon_{\scriptscriptstyle{0}} \varepsilon}ln\left[\frac{R_{\scriptscriptstyle_{H}}}{R+d}\right]\right]}\\
	\therefore c_4 = \frac{\varepsilon_{\scriptscriptstyle{0}} \varepsilon}{(R+d)ln\left[ \frac{R_{\scriptscriptstyle_{H}}}{R+d} \right]} \\
	C_3 = \frac{Q_3(R)}{\psi_0-\varphi_0} = \frac{2\pi L_0 R \sigma_3 (R)}{\psi_0-\varphi_0} \implies c_3 = \frac{C_3}{2\pi L_0R} = \frac{\sigma_3(R)}{\psi_0-\varphi_0} \\
	\textrm{but} \ \ E_3(R) = \frac{1}{\varepsilon_{\scriptscriptstyle{0}} \varepsilon} \left[ R_{\scriptscriptstyle_{H}}\sigma_{\scriptscriptstyle{Ho}} - (R+d) \sigma_0 \right] \frac{1}{R} = \frac{\sigma_3(R)}{\varepsilon_{\scriptscriptstyle{0}} \varepsilon} \implies \\
	\sigma_3(R) = \frac{R_{\scriptscriptstyle_{H}}\sigma_{\scriptscriptstyle{Ho}}-(R+d)\sigma_0}{R}, \ \ \textrm{thus}\\
	c_3 = \frac{R_{\scriptscriptstyle_{H}}\sigma_{\scriptscriptstyle{Ho}}-(R+d)\sigma_0}{R(\psi_0-\varphi_0)} \\
	c_3 = \frac{R_{\scriptscriptstyle_{H}}\sigma_{\scriptscriptstyle{Ho}}-(R+d)\sigma_0}{R\left[\frac{1}{\varepsilon_{\scriptscriptstyle{0}} \varepsilon}(R_{\scriptscriptstyle_{H}}\sigma_{\scriptscriptstyle{Ho}}-(R+d)\sigma_0)\right]ln\left[\frac{R+d}{R}\right]} \\
	\therefore c_3 = \frac{\varepsilon_{\scriptscriptstyle{0}} \varepsilon}{Rln\left[\frac{R+d}{R}\right]} 
\end{gather*}
\begin{gather*}
	C_2 = \frac{Q_2(R)}{\psi_H-\psi_0} = \frac{2\pi L_0 R \sigma_2(R)}{\psi_H-\psi_0} \\
	\textrm{but} \ \ c_2 = \frac{C_2}{2\pi L_0 R} = \frac{\sigma_2(R)}{\psi_H-\psi_0}, \ \ \textrm{and} \ \ \textrm{since} \\
	E_2(R) = \frac{1}{\varepsilon_{\scriptscriptstyle{0}} \varepsilon} \left[R_{\scriptscriptstyle_{H}} \sigma_{\scriptscriptstyle{Ho}} - (2R+d)\sigma_0\right]\frac{1}{R} = \frac{\sigma_2(R)}{\varepsilon_{\scriptscriptstyle{0}} \varepsilon} \implies \\
	c_2 = \frac{R_{\scriptscriptstyle_{H}}\sigma_{\scriptscriptstyle{Ho}}-(2R+d)\sigma_0}{R[\psi_H-\psi_0]} = \frac{R_{\scriptscriptstyle_{H}}\sigma_{\scriptscriptstyle{Ho}}-(2R+d)\sigma_0}{R\frac{1}{\varepsilon_{\scriptscriptstyle{0}} \varepsilon}\left[ R_{\scriptscriptstyle_{H}} \sigma_{\scriptscriptstyle{Ho}} - (2R+d)\sigma_0 \right]ln\left[ \frac{R}{R-\frac{a}{2}} \right]} \\
	c_2 = \frac{\varepsilon_{\scriptscriptstyle{0}} \varepsilon}{Rln\left[\frac{R}{R-\frac{a}{2}}\right]}
\end{gather*}
\begin{gather*}
	C_1 = \frac{Q_1 \left(R-\frac{a}{2} \right)}{\psi_d-\psi_H} = \frac{2\pi L_0 \left(R-\frac{a}{2}\right) \sigma_1\left(R-\frac{a}{2}\right)}{\psi_d-\psi_H} \implies \\
	c_1 = \frac{C_1}{2\pi L_0 \left( R-\frac{a}{2}\right)} = \frac{\sigma_1\left(R-\frac{a}{2}\right)}{\psi_d-\psi_H}\\
	\textrm{but} \ \ E_1\left(R-\frac{a}{2}\right) = \frac{1}{\varepsilon_{\scriptscriptstyle{0}} \varepsilon} \frac{R_{\scriptscriptstyle_{H}}\sigma_{\scriptscriptstyle{Ho}}-(2R+d)\sigma_0}{R-\frac{a}{2}} = \frac{\sigma_1\left(R-\frac{a}{2}\right)}{\varepsilon_{\scriptscriptstyle{0}} \varepsilon} \implies \\
	\sigma_1 \left(R-\frac{a}{2}\right) = \frac{R_{\scriptscriptstyle_{H}}\sigma_{\scriptscriptstyle{Ho}}-(2R+d)\sigma_0}{R-\frac{a}{2}} \\
	\therefore c_1 = \frac{R_{\scriptscriptstyle_{H}}\sigma_{\scriptscriptstyle{Ho}}-(2R+d)\sigma_0}{\left(R-\frac{a}{2}\right)\frac{1}{\varepsilon_{\scriptscriptstyle{0}} \varepsilon}\frac{(R_{\scriptscriptstyle_{H}}\sigma_{\scriptscriptstyle{Ho}}-(2R+d)\sigma_0)\left(I_0\left[\kappa \left( R-\frac{a}{2} \right)\right] - 1 \right)}{\kappa \left( R-\frac{a}{2}\right)I_1\left[\kappa \left(R-\frac{a}{2}\right)\right]}} \\
	c_1 = \frac{\varepsilon_{\scriptscriptstyle{0}} \varepsilon \kappa I_1 \left[ \kappa \left(R-\frac{a}{2}\right)\right]}{I_0\left[\kappa \left(R-\frac{a}{2}\right) \right]-1}
\end{gather*}
%%%%%%%%%%%%%%%%%%%%%%%%%%%%%%%%%%%%%%%%%%
\newpage
\section{\\Spherical Nanopore}\label{B}
\subsection{Region V}

\begin{gather*}
	\nabla^2 \psi(r) = \kappa^2 \psi(r) \implies \\
	\frac{1}{r^2}\frac{d}{dr}\left[ r^2 \frac{d\psi(r)}{dr} \right] = \kappa^2 \psi(r) \implies \\
	r^2\frac{d^2\psi (r)}{dr^2} + 2r\frac{d\psi(r)}{dr} = \kappa^2 r^2 \psi(r) \implies \\
	\frac{d^2\psi (r)}{dr^2} + \frac{2}{r} \frac{d\psi(r)}{dr} = \kappa^2 \psi(r); \ \ \psi(\infty)=0, \ \  \frac{d\psi(\infty)}{dr}=0 \\
	\implies \psi(r) = A\frac{e^{-\kappa r}}{\kappa r} \ \ \textrm{such} \ \ \textrm{that} \ \ \psi\left(r=R+d+\frac{a}{2}\right) = \varphi_H \implies \\
	\varphi_H = A \frac{e^{-\kappa \left( R+d+\frac{a}{2}\right)}}{\kappa \left(R+d+\frac{a}{2} \right)}  \implies A = \kappa \left(R+d+\frac{a}{2} \right) \varphi_H e^{\kappa \left(R+d+\frac{a}{2}\right)}  \\
	R_{\scriptscriptstyle_{H}} = R+d+\frac{a}{2} \implies \\
	A = \kappa R_{\scriptscriptstyle_{H}} \varphi_H e^{\kappa R_{\scriptscriptstyle_{H}}} \\
	\psi_5(r)= R_{\scriptscriptstyle_{H}} \varphi_H \frac{e^{-\kappa\left[r-R_{\scriptscriptstyle_{H}}\right]}}{r} \\
	E_5(r) = -\nabla\psi(r) = -R_{\scriptscriptstyle_{H}}\varphi_H \left[ \frac{-\kappa r e^{-\kappa \left(r-R_{\scriptscriptstyle_{H}}\right)}-e^{-\kappa\left(r-R_{\scriptscriptstyle_{H}} \right)}}{r^2} \right] \implies \\
	E_5(r) =  R_{\scriptscriptstyle_{H}} \varphi_H (1+\kappa r) \frac{e^{-\kappa(r-R_{\scriptscriptstyle_{H}})}}{r^2} \\
	E_5(R_{\scriptscriptstyle_{H}}) = \frac{R_{\scriptscriptstyle_{H}} \varphi_H (1+\kappa R_{\scriptscriptstyle_{H}})}{R_{\scriptscriptstyle_{H}}^2} = \frac{\sigma_{\scriptscriptstyle{Ho}}}{\varepsilon_{\scriptscriptstyle{0}} \varepsilon} \implies \\
	\varphi_H = \frac{R_{\scriptscriptstyle_{H}}}{1+\kappa R_{\scriptscriptstyle_{H}}} \frac{\sigma_{\scriptscriptstyle{Ho}}}{\varepsilon_{\scriptscriptstyle{0}} \varepsilon} \implies \psi_5(r) = \frac{R_{\scriptscriptstyle_{H}}^2\sigma_{\scriptscriptstyle{Ho}}}{1+\kappa R_{\scriptscriptstyle_{H}}}\frac{e^{-\kappa[r-R_{\scriptscriptstyle_{H}}]}}{r\varepsilon_{\scriptscriptstyle{0}} \varepsilon}
\end{gather*}
%%%%%%%%%%%%%%%%%%%%%%%

\subsection{Region IV}

\begin{gather*}
	\nabla^2 \psi(r) = 0 \implies r^2\psi''(r)+2r\psi'(r)= 0 \implies \\
	\frac{d\psi'}{dr} = - \frac{2\psi'}{r} \implies \frac{d\psi'}{\psi'}=-\frac{2}{r}dr \implies \psi'(r) = -\frac{C}{r^2} \implies \\
	\psi_4(r)=-\frac{C}{r}+D; \ \ E_4(r) = -\nabla\psi_4(r) = - \frac{C}{r^2} \\
	\textrm{but} \ \ E(R+d) = -\frac{C}{(R+d)^2} = \frac{\sigma_4(R+d)}{\varepsilon_{\scriptscriptstyle{0}} \varepsilon} = \frac{Q_{Ho}}{\varepsilon_{\scriptscriptstyle{0}} \varepsilon[4\pi (R+d)^2]}\\
	\textrm{Notice}: \ \ \sigma_{\scriptscriptstyle{Ho}} = \frac{Q_{Ho}}{4\pi R_{\scriptscriptstyle_{H}}^2} \\
	\textrm{Hence}: \ \ E(R+d) = - \frac{C}{(R+d)^2} = \frac{R_{\scriptscriptstyle_{H}}^2\sigma_{\scriptscriptstyle{Ho}}}{(R+d)^2\varepsilon_{\scriptscriptstyle{0}} \varepsilon} \implies \\
	-C = R_{\scriptscriptstyle_{H}}^2 \frac{\sigma_{\scriptscriptstyle{Ho}}}{\varepsilon_{\scriptscriptstyle{0}} \varepsilon} \implies \\
	E_4(r) = \frac{R_{\scriptscriptstyle_{H}}^2\sigma_{\scriptscriptstyle{Ho}}}{r^2 \varepsilon_{\scriptscriptstyle{0}} \varepsilon} \\
	\textrm{but} \ \ \psi_4(R+d) \equiv \varphi_0 = \frac{R_{\scriptscriptstyle_{H}}^2\sigma_{\scriptscriptstyle{Ho}}}{\varepsilon_{\scriptscriptstyle{0}} \varepsilon(R+d)}+D \implies \\
	D = \varphi_0 - \frac{R_{\scriptscriptstyle_{H}}^2\sigma_{\scriptscriptstyle{Ho}}}{\varepsilon_{\scriptscriptstyle{0}} \varepsilon(R+d)} \implies \\
	\psi_4(r) = R_{\scriptscriptstyle_{H}}^2 \frac{\sigma_{\scriptscriptstyle{Ho}}}{\varepsilon_{\scriptscriptstyle{0}} \varepsilon r} + \varphi_0 - R_{\scriptscriptstyle_{H}}^2 \frac{\sigma_{\scriptscriptstyle{Ho}}}{\varepsilon_{\scriptscriptstyle{0}} \varepsilon (R+d)} \implies \\
	\psi_4(r) = \varphi_0 + \frac{ R_{\scriptscriptstyle_{H}}^2 \sigma_{\scriptscriptstyle{Ho}}}{\varepsilon_{\scriptscriptstyle{0}} \varepsilon} \left[ \frac{1}{r} - \frac{1}{R+d} \right] \\
	\psi_4(r) = \varphi_0 + \frac{ R_{\scriptscriptstyle_{H}}^2 \sigma_{\scriptscriptstyle{Ho}}}{\varepsilon_{\scriptscriptstyle{0}} \varepsilon (R+d)} \left[ \frac{R+d}{r} - 1 \right] \\
	\therefore \psi_4(R_{\scriptscriptstyle_{H}}) \equiv \varphi_H = \varphi_0 + \frac{R_{\scriptscriptstyle_{H}}^2\sigma_{\scriptscriptstyle{Ho}}}{\varepsilon_{\scriptscriptstyle{0}} \varepsilon(R+d)} \left[\frac{R+d}{R_{\scriptscriptstyle_{H}}}-1\right] \\
	= \varphi_0 + \frac{R_{\scriptscriptstyle_{H}}^2\sigma_{\scriptscriptstyle{Ho}}}{\varepsilon_{\scriptscriptstyle{0}} \varepsilon(R+d)} \left[-\frac{\frac{a}{2}}{R_{\scriptscriptstyle_{H}}}\right] \implies \\
\end{gather*}
\begin{gather*}
		\varphi_0 = \varphi_H + \frac{R_{\scriptscriptstyle_{H}}\sigma_{\scriptscriptstyle{Ho}}}{\varepsilon_{\scriptscriptstyle{0}} \varepsilon(R+d)} \left[\frac{a}{2}\right] \implies \\
	\varphi_0 - \varphi_H = \frac{R_{\scriptscriptstyle_{H}}\sigma_{\scriptscriptstyle{Ho}}}{\varepsilon_{\scriptscriptstyle{0}} \varepsilon(R+d)} \left[\frac{a}{2}\right]
\end{gather*}
%%%%%%%%%%%%%%%%%%%%%%%%%%%%%%%%%%%%%
\subsection{Region III}

\begin{gather*}
	\varepsilon_{\scriptscriptstyle{0}} \varepsilon E_4(R+d) - \varepsilon_{\scriptscriptstyle{0}} \varepsilon E_3(R+d) = \sigma_0 \\
	\psi_3(r) = - \frac{F}{r}+G \\
	E_3(r) = - \nabla\psi_3(r)=-\frac{F}{r^2} \\
	\textrm{but} \ \ E_3(R+d) = E_4(R+d) - \frac{\sigma_0}{\varepsilon_{\scriptscriptstyle{0}} \varepsilon} \quad \textrm{and} \\
	E_4(R+d) = \frac{R_{\scriptscriptstyle_{H}}^2}{(R+d)^2} \frac{\sigma_{\scriptscriptstyle{Ho}}}{\varepsilon_{\scriptscriptstyle{0}} \varepsilon} \implies \\
	E_3(R+d) = \frac{R_{\scriptscriptstyle_{H}}^2}{(R+d)^2} \frac{\sigma_{\scriptscriptstyle{Ho}}}{\varepsilon_{\scriptscriptstyle{0}} \varepsilon} - \frac{\sigma_0}{\varepsilon_{\scriptscriptstyle{0}} \varepsilon} = -\frac{F}{(R+d)^2} \implies \\
	-F = R_{\scriptscriptstyle_{H}}^2\frac{\sigma_{\scriptscriptstyle{Ho}}}{\varepsilon_{\scriptscriptstyle{0}} \varepsilon} - (R+d)^2\frac{\sigma_{\scriptscriptstyle{0}}}{\varepsilon_{\scriptscriptstyle{0}} \varepsilon} \implies \\
	E_3(r) = \frac{1}{\varepsilon_{\scriptscriptstyle{0}} \varepsilon} \left[ R_{\scriptscriptstyle_{H}}^2\sigma_{\scriptscriptstyle{Ho}} - (R+d)^2\sigma_0 \right] \frac{1}{r^2} \\
	\psi_3(r) = \frac{1}{\varepsilon_{\scriptscriptstyle{0}} \varepsilon}\left[ R_{\scriptscriptstyle_{H}}^2\sigma_{\scriptscriptstyle{Ho}} - (R+d)^2\sigma_0 \right] \frac{1}{r} + G \\
	\psi_0 \equiv \psi_3(R) = \frac{1}{\varepsilon_{\scriptscriptstyle{0}} \varepsilon}\left[ R_{\scriptscriptstyle_{H}}^2\sigma_{\scriptscriptstyle{Ho}} - (R+d)^2\sigma_0 \right] \frac{1}{R} + G \quad \textrm{and} \\
	\varphi_0 \equiv \psi_3(R+d) = \frac{1}{\varepsilon_{\scriptscriptstyle{0}} \varepsilon}\left[ R_{\scriptscriptstyle_{H}}^2\sigma_{\scriptscriptstyle{Ho}} - (R+d)^2\sigma_0 \right] \frac{1}{R+d} + G \implies \\
	\psi_0 - \varphi_0 = \frac{1}{\varepsilon_{\scriptscriptstyle{0}} \varepsilon} \left[ R_{\scriptscriptstyle_{H}}^2\sigma_{\scriptscriptstyle{Ho}} - (R+d)^2\sigma_0 \right] \left[\frac{1}{R} - \frac{1}{R+d}\right] \implies \\
	\psi_0 - \varphi_0 = \frac{1}{\varepsilon_{\scriptscriptstyle{0}} \varepsilon} \left[ R_{\scriptscriptstyle_{H}}^2\sigma_{\scriptscriptstyle{Ho}} - (R+d)^2\sigma_0 \right] \left[\frac{d}{R(R+d)}\right]
\end{gather*}

To obtain $\psi_{\scriptscriptstyle_{3}}(r)$, since the MEP must be equal at $r=R+d$, we find that $G=\varphi_{\scriptscriptstyle_{0}}-\frac{1}{\varepsilon_{\scriptscriptstyle{0}}\varepsilon}[R_{\scriptscriptstyle_{H}}^2\sigma_{\scriptscriptstyle{Ho}}-(R+d)^2\sigma_{\scriptscriptstyle{0}}]/(R+d)$. Thus, 

\begin{equation*}
	\psi_{\scriptscriptstyle_{3}}(r)=\varphi_{\scriptscriptstyle_{0}}+\frac{1}{\varepsilon_{\scriptscriptstyle{0}}\varepsilon}[R_{\scriptscriptstyle_{H}}^2\sigma_{\scriptscriptstyle{Ho}}-(R+d)^2\sigma_{\scriptscriptstyle{0}}]\left[\frac{1}{r}-\frac{1}{(R+d)}\right]
\end{equation*}
\noindent or using the above relations between $\psi_{\scriptscriptstyle_{0}}$ and $\varphi_{\scriptscriptstyle_{0}}$,
\begin{equation*}
		\psi_{\scriptscriptstyle_{3}}(r)=\psi_{\scriptscriptstyle_{0}}+\frac{1}{\varepsilon_{\scriptscriptstyle{0}}\varepsilon}[R_{\scriptscriptstyle_{H}}^2\sigma_{\scriptscriptstyle{Ho}}-(R+d)^2\sigma_{\scriptscriptstyle{0}}]\left[\frac{1}{r}-\frac{1}{R}\right]
\end{equation*}
%%%%%%%%%%%%%%%%%%%%%%%%%%%%%%%%%%%%%
\subsection{Region II}

\begin{gather*}
	\varepsilon_{\scriptscriptstyle{0}} \varepsilon E_3(R) - \varepsilon_{\scriptscriptstyle{0}} \varepsilon E_2(R) = \sigma_0 \\
	\psi_2(r) = -\frac{H}{r} + I \\
	E_2(r) = -\nabla \psi_2(r) = -\frac{H}{r^2} \implies \\
	E_2(R) = -\frac{H}{R^2} = \frac{1}{\varepsilon_{\scriptscriptstyle{0}} \varepsilon} \left[R_{\scriptscriptstyle_{H}}^2\sigma_{\scriptscriptstyle{Ho}} - (R+d)^2\sigma_0 \right]\frac{1}{R^2} - \frac{\sigma_0}{\varepsilon_{\scriptscriptstyle{0}} \varepsilon} \implies \\
	-H = \frac{1}{\varepsilon_{\scriptscriptstyle{0}} \varepsilon} \left[ R_{\scriptscriptstyle_{H}}^2\sigma_{\scriptscriptstyle{Ho}} - (R+d)^2\sigma_0 \right] - \frac{1}{\varepsilon_{\scriptscriptstyle{0}} \varepsilon}R^2\sigma_0 \implies \\
	E_2(r) = \frac{1}{\varepsilon_{\scriptscriptstyle{0}} \varepsilon} \left( R_{\scriptscriptstyle_{H}}^2\sigma_{\scriptscriptstyle{Ho}} - \left[ R^2+(R+d)^2 \right]\sigma_0 \right) \frac{1}{r^2} \\
	\psi_2(R) \equiv \psi_0 = \frac{1}{\varepsilon_{\scriptscriptstyle{0}} \varepsilon} \left( R_{\scriptscriptstyle_{H}}^2\sigma_{\scriptscriptstyle{Ho}} - \left[ R^2+(R+d)^2 \right] \sigma_0 \right) \frac{1}{R} + I \quad \textrm{and} \\
	\psi_2\left(R-\frac{a}{2}\right) \equiv \psi_H = \frac{1}{\varepsilon_{\scriptscriptstyle{0}} \varepsilon} \left( R_{\scriptscriptstyle_{H}}^2\sigma_{\scriptscriptstyle{Ho}} - \left[ R^2+(R+d)^2 \right] \sigma_0 \right) \frac{1}{R-\frac{a}{2}} + I \\
	\psi_0 - \psi_H = \frac{-1}{\varepsilon_{\scriptscriptstyle{0}} \varepsilon} \left( R_{\scriptscriptstyle_{H}}^2\sigma_{\scriptscriptstyle{Ho}} - \left[R^2+(R+d)^2\right]\sigma_0 \right) \frac{\frac{a}{2}}{R\left(R-\frac{a}{2}\right)}
\end{gather*}

Now, since the MEP must be equal at $r=R$, i.e., $\psi_{\scriptscriptstyle_{2}}(R)=\psi_{\scriptscriptstyle_{3}}(R)$, we find that $I=\psi_{\scriptscriptstyle_{0}}-\frac{1}{\varepsilon_{\scriptscriptstyle{0}}\varepsilon}[R_{\scriptscriptstyle_{H}}^2\sigma_{\scriptscriptstyle{Ho}}-[R^2+(R+d)^2]\sigma_{\scriptscriptstyle{0}}]1/R$. Thus,

\begin{equation*}
	\psi_{\scriptscriptstyle{2}}(r)=\psi_{\scriptscriptstyle_{0}}+\frac{1}{\varepsilon_{\scriptscriptstyle{0}}\varepsilon}[R_{\scriptscriptstyle_{H}}^2\sigma_{\scriptscriptstyle{Ho}}-[R^2+(R+d)^2]\sigma_{\scriptscriptstyle{0}}]\left[\frac{1}{r}-\frac{1}{R}\right],
\end{equation*}
\noindent or from the above equation for $\psi_{\scriptscriptstyle{0}}$,

\begin{equation*}
	\psi_{\scriptscriptstyle{2}}(r)=\psi_{\scriptscriptstyle_{H}}+\left[\frac{[R_{\scriptscriptstyle_{H}}^2\sigma_{\scriptscriptstyle{Ho}}-[R^2+(R+d)^2]\sigma_{\scriptscriptstyle{0}}]}{\varepsilon_{\scriptscriptstyle{0}}\varepsilon(R-a/2)}\right]\left[\frac{(R-a/2)}{r}-1\right].
\end{equation*}

%%%%%%%%%%%%%%%%%%%%%%%%%%%%%%%%%%%%%
\subsection{Region I}

\begin{gather*}
	\psi_1(r) = L \frac{e^{-\kappa r}}{r} + M \frac{e^{\kappa r}}{r} \implies \\
	E_1(r) = -\nabla\psi_1(r) = L \left[ \frac{1+\kappa r}{r^2} \right] e^{-\kappa r} + M \left[ \frac{1-\kappa r}{r^2} \right] e^{\kappa r} \\
	\textrm{but} \ \ \psi_1\left( R-\frac{a}{2} \right) = \psi_H, \ \ E_1\left( R-\frac{a}{2} \right) = E_2\left(R-\frac{a}{2}\right), \ \ \textrm{and} \ \ E_1(r=0) = 0 \\
	\therefore \lim_{r \to 0} E_1(r) = L \lim_{r \to 0} \frac{(1+\kappa r)e^{-\kappa r}}{r^2} + M \lim_{r \to 0}  \frac{(1-\kappa r)e^{\kappa r}}{r^2} \\
	= L \lim_{r \to 0} \left[ \frac{-\kappa(1+\kappa r)e^{-\kappa r}+\kappa e^{-\kappa r}}{2r} \right] + M \lim_{r \to 0} \left[ \frac{\kappa(1-\kappa r)e^{\kappa r}-\kappa e^{\kappa r}}{2r} \right] \\
	= L \lim_{r \to 0} \left[ \frac{-\kappa^2 r e^{-\kappa r}}{2r} \right] +  M \lim_{r \to 0} \left[ \frac{-\kappa^2 r e^{\kappa r}}{2r} \right] \\
	= L \lim_{r \to 0} \left[-\frac{1}{2}\kappa^2e^{-\kappa r}\right] + M \lim_{r \to 0} \left[-\frac{1}{2}\kappa^2e^{\kappa r}\right] \\
	= -\frac{1}{2} \kappa^2 \left[ L + M \right] = 0 \implies L=-M \\
	\psi_d \equiv \psi_1(r=0) = \lim_{r \to 0} \left[ L\frac{e^{-\kappa r}}{r} + M\frac{e^{\kappa r}}{r} \right] \\
	= M \lim_{r \to 0} \left[ \frac{e^{\kappa r}}{r} - \frac{e^{-\kappa r}}{r} \right] = \lim_{r \to 0} \left[ \frac{\kappa e^{\kappa r}+\kappa e^{-\kappa r}}{1} \right] \\
	\implies \psi_d = 2\kappa M
\end{gather*}
\begin{gather*}
	E_1(r) = \frac{M}{r^2} \left( [1-\kappa r]e^{\kappa r} - [1+\kappa r]e^{-\kappa r} \right) \\
	\psi_1(r) = \frac{M}{r} \left( e^{\kappa r} - e^{-\kappa r} \right) = \frac{2sinh[\kappa r]M}{r} \\
	E_1 \left( R-\frac{a}{2} \right) = E_2 \left( R-\frac{a}{2} \right) \implies \\
	\frac{M}{\left(R-\frac{a}{2}\right)^2} \left( \left[ 1-\kappa\left( R-\frac{a}{2} \right) \right]e^{\kappa \left(R-\frac{a}{2}\right)} - \left[ 1+\kappa\left( R-\frac{a}{2} \right) \right]e^{-\kappa \left(R-\frac{a}{2}\right)} \right) \\
	= \frac{1}{\varepsilon_{\scriptscriptstyle{0}} \varepsilon \left(R-\frac{a}{2}\right)^2} \left( R_{\scriptscriptstyle_{H}}^2\sigma_{\scriptscriptstyle{Ho}}-[R^2+(R+d)^2]\sigma_0 \right) \implies \\
	M = \frac{R_{\scriptscriptstyle_{H}}^2\sigma_{\scriptscriptstyle{Ho}} - [R^2+(R+d)^2]\sigma_0}{2\varepsilon_{\scriptscriptstyle{0}} \varepsilon\left[ sinh\left[\kappa \left(R-\frac{a}{2}\right)\right] - \kappa \left(R-\frac{a}{2}\right)cosh\left[ \kappa  \left(R-\frac{a}{2}\right) \right] \right]} \\
	\therefore \psi_1(r) = \frac{1}{\varepsilon_{\scriptscriptstyle{0}} \varepsilon} \left(  \frac{\left[R_{\scriptscriptstyle_{H}}^2\sigma_{\scriptscriptstyle{Ho}} - [R^2+(R+d)^2]\sigma_0\right]sinh[\kappa r]}{sinh\left[\kappa \left(R-\frac{a}{2}\right)\right] - \kappa \left(R-\frac{a}{2}\right)cosh\left[ \kappa  \left(R-\frac{a}{2}\right) \right]} \right) \frac{1}{r} \implies \\
	\psi_1\left(R-\frac{a}{2}\right) \equiv \psi_H = \frac{1}{\varepsilon_{\scriptscriptstyle{0}} \varepsilon} \left(  \frac{\left[R_{\scriptscriptstyle_{H}}^2\sigma_{\scriptscriptstyle{Ho}} - [R^2+(R+d)^2]\sigma_0\right]sinh\left[\kappa \left(R-\frac{a}{2}\right) \right]}{sinh\left[\kappa \left(R-\frac{a}{2}\right)\right] - \kappa \left(R-\frac{a}{2}\right)cosh\left[ \kappa  \left(R-\frac{a}{2}\right) \right]} \right) \frac{1}{R-\frac{a}{2}}
\end{gather*}

\begin{eqnarray*}
	\psi_d = \psi_1(r=0) &=& 2M\kappa \\
	&=& \frac{\kappa}{\varepsilon_{\scriptscriptstyle{0}} \varepsilon} \left( \frac{R_{\scriptscriptstyle_{H}}^2\sigma_{\scriptscriptstyle{Ho}} - [R^2+(R+d)^2]\sigma_0}{sinh\left[\kappa \left(R-\frac{a}{2}\right)\right] - \kappa \left(R-\frac{a}{2}\right)cosh\left[ \kappa  \left(R-\frac{a}{2}\right) \right]} \right)
\end{eqnarray*}

Now, to find the relation of $\sigma_{\scriptscriptstyle{Hi}}$ and $\sigma_{\scriptscriptstyle{Ho}}$ with $\sigma_{\scriptscriptstyle{0}}$, from our previous equations we know that,
\begin{equation*}
	\varphi_H = \frac{R_{\scriptscriptstyle_{H}}}{1+\kappa R_{\scriptscriptstyle_{H}}} \frac{\sigma_{\scriptscriptstyle{Ho}}}{\varepsilon_{\scriptscriptstyle{0}} \varepsilon}
\end{equation*}
\begin{equation*}
	\psi_0-\varphi_0 = \frac{1}{\varepsilon_{\scriptscriptstyle{0}} \varepsilon} \left[R_{\scriptscriptstyle_{H}}^2\sigma_{\scriptscriptstyle{Ho}}-(R+d)^2\sigma_0\right] \frac{d}{R(R+d)}
\end{equation*}
\begin{equation*}
	\psi_H = \psi_0 + \frac{1}{\varepsilon_{\scriptscriptstyle{0}} \varepsilon} \left[ R_{\scriptscriptstyle_{H}}^2\sigma_{\scriptscriptstyle{Ho}} - [R^2+(R+d)^2]\sigma_0 \right] \frac{\frac{a}{2}}{R\left(R-\frac{a}{2}\right)}
\end{equation*}
\begin{equation*}
	\varphi_H = \varphi_0 - \frac{R_{\scriptscriptstyle_{H}}\sigma_{\scriptscriptstyle{Ho}}}{\varepsilon_{\scriptscriptstyle{0}} \varepsilon(R+d)}\frac{a}{2}
\end{equation*}
\begin{equation*}
	\psi_H-\varphi_H = \psi_0 - \varphi_0 + \frac{R_{\scriptscriptstyle_{H}} \sigma_{\scriptscriptstyle{Ho}}}{\varepsilon_{\scriptscriptstyle{0}} \varepsilon (R+d)} \frac{a}{2} + \frac{1}{\varepsilon_{\scriptscriptstyle{0}} \varepsilon}\left[ R_{\scriptscriptstyle_{H}}^2\sigma_{\scriptscriptstyle{Ho}} - [R^2+(R+d)^2]\sigma_0 \right] \frac{\frac{a}{2}}{R\left(R-\frac{a}{2}\right)}
\end{equation*}
\begin{eqnarray*}
	\psi_d - \psi_H &=& \frac{1}{\varepsilon_{\scriptscriptstyle{0}} \varepsilon\left( R-\frac{a}{2} \right)} \left[ \frac{R_{\scriptscriptstyle_{H}}^2\sigma_{\scriptscriptstyle{Ho}}-[R^2-(R+d)^2]\sigma_0}{sinh\left[\kappa \left(R-\frac{a}{2}\right)\right]-\kappa \left(R-\frac{a}{2}\right)cosh\left[\kappa \left(R-\frac{a}{2}\right)\right]} \right] \\
	& & \times \left( \kappa \left(R-\frac{a}{2}\right) - sinh\left[\kappa \left(R-\frac{a}{2}\right)\right] \right)
\end{eqnarray*}

\noindent Therefore, we find the following equality,

\begin{gather*}
	\psi_H = \frac{R_{\scriptscriptstyle_{H}}}{(1+\kappa R_{\scriptscriptstyle_{H}})} \frac{\sigma_{\scriptscriptstyle{Ho}}}{\varepsilon_{\scriptscriptstyle{0}} \varepsilon} +  \frac{1}{\varepsilon_{\scriptscriptstyle{0}} \varepsilon} \left[R_{\scriptscriptstyle_{H}}^2\sigma_{\scriptscriptstyle{Ho}}-(R+d)^2\sigma_0\right] \frac{d}{R(R+d)} \\
	+ \frac{R_{\scriptscriptstyle_{H}} \sigma_{\scriptscriptstyle{Ho}}}{\varepsilon_{\scriptscriptstyle{0}} \varepsilon(R+d)}\frac{a}{2} + \frac{1}{\varepsilon_{\scriptscriptstyle{0}} \varepsilon}\left[ R_{\scriptscriptstyle_{H}}^2\sigma_{\scriptscriptstyle{Ho}} - [R^2+(R+d)^2]\sigma_0 \right] \frac{\frac{a}{2}}{R\left(R-\frac{a}{2}\right)} \\
	= \left(  \frac{R_{\scriptscriptstyle_{H}}^2\sigma_{\scriptscriptstyle{Ho}} - [R^2+(R+d)^2]\sigma_0}{sinh\left[\kappa \left(R-\frac{a}{2}\right)\right] - \kappa \left(R-\frac{a}{2}\right)cosh\left[ \kappa  \left(R-\frac{a}{2}\right) \right]} \right) \frac{sinh\left[\kappa \left(R-\frac{a}{2}\right) \right]}{\varepsilon_{\scriptscriptstyle{0}} \varepsilon \left(R-\frac{a}{2}\right)} \implies \\
	\left( \frac{R_{\scriptscriptstyle_{H}}}{1+\kappa R_{\scriptscriptstyle_{H}}} +\frac{R_{\scriptscriptstyle_{H}}^2d}{R(R+d)} + \frac{R_{\scriptscriptstyle_{H}} a}{2(R+d)} + \frac{R_{\scriptscriptstyle_{H}}^2 a}{2R\left(R-\frac{a}{2}\right)} \right. \\
	\left. -\frac{R_{\scriptscriptstyle_{H}}^2}{R-\frac{a}{2}}\left[ \frac{sinh\left[\kappa \left(R-\frac{a}{2}\right)\right]}{sinh\left[\kappa \left(R-\frac{a}{2}\right)\right] - \kappa \left(R-\frac{a}{2}\right)cosh\left[ \kappa  \left(R-\frac{a}{2}\right) \right]} \right] \right) \sigma_{\scriptscriptstyle{Ho}} \\
	= \left( \frac{(R+d)^2d}{R(R+d)} + \frac{[R^2+(R+d)^2]a}{2R\left( R-\frac{a}{2} \right)} \right. \\
	\left. -\frac{\left[R^2+(R+d)^2\right]sinh\left[\kappa \left(R-\frac{a}{2}\right)\right]}{\left(R-\frac{a}{2}\right)\left(sinh\left[\kappa \left(R-\frac{a}{2}\right)\right] - \kappa \left(R-\frac{a}{2}\right)cosh\left[ \kappa  \left(R-\frac{a}{2}\right) \right] \right)} \right) \sigma_0 \\
	\implies \sigma_{\scriptscriptstyle{Ho}} = \frac{L_2}{L_1}\sigma_0 
\end{gather*}
where
\begin{gather*}
	L_2 \equiv  \left( \frac{(R+d)d}{R} + \frac{[R^2+(R+d)^2]a}{2R\left( R-\frac{a}{2} \right)} \right. \\
	\left. -\frac{\left[R^2+(R+d)^2\right]sinh\left[\kappa \left(R-\frac{a}{2}\right)\right]}{\left(R-\frac{a}{2}\right)\left(sinh\left[\kappa \left(R-\frac{a}{2}\right)\right] - \kappa \left(R-\frac{a}{2}\right)cosh\left[ \kappa  \left(R-\frac{a}{2}\right) \right] \right)} \right)\\
	L_1 \equiv \left( \frac{R_{\scriptscriptstyle_{H}}}{1+\kappa R_{\scriptscriptstyle_{H}}} +\frac{R_{\scriptscriptstyle_{H}}^2d}{R(R+d)} + \frac{R_{\scriptscriptstyle_{H}} a}{2(R+d)} + \frac{R_{\scriptscriptstyle_{H}}^2 a}{2R\left(R-\frac{a}{2}\right)} \right. \\
	\left. -\frac{R_{\scriptscriptstyle_{H}}^2}{R-\frac{a}{2}}\left[ \frac{sinh\left[\kappa \left(R-\frac{a}{2}\right)\right]}{sinh\left[\kappa \left(R-\frac{a}{2}\right)\right] - \kappa \left(R-\frac{a}{2}\right)cosh\left[ \kappa  \left(R-\frac{a}{2}\right) \right]} \right] \right)
\end{gather*}
On the other hand, by electroneutrality 
\begin{gather*}
	\sigma_0 [R^2+(R+d)^2]=\left(R-a/2\right)^2\sigma_{\scriptscriptstyle{Hi}} + R_{\scriptscriptstyle_{H}}^2\sigma_{\scriptscriptstyle{Ho}} \implies \\
	\sigma_{\scriptscriptstyle{Hi}} = \left[ \frac{R^2+(R+d)^2}{\left(R-\frac{a}{2}\right)^2} \right]\sigma_0 - \left[\frac{R_{\scriptscriptstyle_{H}}^2}{\left(R-\frac{a}{2}\right)^2}\right] \sigma_{\scriptscriptstyle{Ho}}
\end{gather*}
%%%%%%%%%%%%%%%%%%%%%%%%%%%%%%%%%%%%%%%%%%%%%%
\subsection{Capacitance/Area}

\begin{gather*}
	C_5 = \frac{Q_{Ho}}{\varphi_H} = \frac{4\pi R_{\scriptscriptstyle_{H}}^2 \sigma_{\scriptscriptstyle{Ho}}}{\varphi_H} \implies c_5 = \frac{C_5}{4\pi R_{\scriptscriptstyle_{H}}^2} = \frac{\sigma_{\scriptscriptstyle{Ho}}}{\varphi_H} \\
	\therefore c_5 = \frac{\sigma_{\scriptscriptstyle{Ho}}}{\frac{R_{\scriptscriptstyle_{H}} \sigma_{\scriptscriptstyle{Ho}}}{\varepsilon_{\scriptscriptstyle{0}} \varepsilon (1+\kappa R_{\scriptscriptstyle_{H}})}} = \frac{\varepsilon_{\scriptscriptstyle{0}} \varepsilon (1+\kappa R_{\scriptscriptstyle_{H}})}{R_{\scriptscriptstyle_{H}}} \implies \\
	c_5 = \frac{\varepsilon_{\scriptscriptstyle{0}} \varepsilon(1+\kappa R_{\scriptscriptstyle_{H}})}{R_{\scriptscriptstyle_{H}}} \\
	C_4 = \frac{Q_4(R+d)}{\varphi_0-\varphi_H} = \frac{4\pi (R+d)^2\sigma_4(R+d)}{\varphi_0-\varphi_H} \implies \\
	c_4 = \frac{C_4}{4\pi (R+d)^2} = \frac{\sigma_4(R+d)}{\varphi_0-\varphi_H} \\
	\textrm{but} \ \ 4\pi(R+d)^2\sigma_4(R+d) = 4\pi R_{\scriptscriptstyle_{H}}^2\sigma_{\scriptscriptstyle{Ho}} \implies \\
	c_4 = \frac{R_{\scriptscriptstyle_{H}}^2 \sigma_{\scriptscriptstyle{Ho}}}{(R+d)^2[\varphi_0-\varphi_H]} = \frac{R_h^2 \sigma_{\scriptscriptstyle{Ho}}}{(R+d)^2\left[\frac{R_{\scriptscriptstyle_{H}} \sigma_{\scriptscriptstyle{Ho}}}{\varepsilon_{\scriptscriptstyle{0}} \varepsilon (R+d)} \frac{a}{2}\right]} \implies \\
	c_4 = \frac{\varepsilon_{\scriptscriptstyle{0}} \varepsilon R_{\scriptscriptstyle_{H}}}{(R+d)\frac{a}{2}} \\
	C_3 = \frac{Q_3(R)}{\psi_0-\varphi_0} = \frac{4 \pi R^2 \sigma_3(R)}{\psi_0-\varphi_0} \implies c_3 = \frac{C_3}{4\pi R^2} = \frac{\sigma_3(R)}{\psi_0-\varphi_0} \\
	\textrm{but} \ \ E_3(R) = \frac{R_{\scriptscriptstyle_{H}}^2\sigma_{\scriptscriptstyle{Ho}} - (R+d)^2\sigma_0}{\varepsilon_{\scriptscriptstyle{0}} \varepsilon R^2} = \frac{\sigma_3(R)}{\varepsilon_{\scriptscriptstyle{0}} \varepsilon} \implies \\
	\sigma_3(R) = \frac{R_{\scriptscriptstyle_{H}}^2\sigma_{\scriptscriptstyle{Ho}} - (R+d)^2\sigma_0}{R^2} \implies \\
	c_3 = \frac{R_{\scriptscriptstyle_{H}}^2\sigma_{\scriptscriptstyle{Ho}} - (R+d)^2\sigma_0}{R^2[\psi_0-\varphi_0]} \\
	= \frac{R_{\scriptscriptstyle_{H}}^2\sigma_{\scriptscriptstyle{Ho}} - (R+d)^2\sigma_0}{R^2\left( \frac{1}{\varepsilon_{\scriptscriptstyle{0}} \varepsilon} \left[ R_{\scriptscriptstyle_{H}}^2\sigma_{\scriptscriptstyle{Ho}} - (R+d)^2\sigma_0 \right] \right) \frac{d}{R(R+d)}} \\
	c_3 = \frac{\varepsilon_{\scriptscriptstyle{0}} \varepsilon (R+d)}{Rd}
\end{gather*}
\begin{gather*}
	C_2 = \frac{Q_2(R)}{\psi_H-\psi_0} = \frac{4\pi R^2\sigma_2(R)}{\psi_H-\psi_0} \implies \\
	c_2 = \frac{C_2}{4\pi R^2} = \frac{\sigma_2(R)}{\psi_H-\psi_0}, \ \ \textrm{and} \ \ \textrm{since} \\
	E_2(R) = \frac{R_{\scriptscriptstyle_{H}}^2\sigma_{\scriptscriptstyle{Ho}} - [R^2+(R+d)^2]\sigma_0}{\varepsilon_{\scriptscriptstyle{0}} \varepsilon R^2} = \frac{\sigma_2(R)}{\varepsilon_{\scriptscriptstyle{0}} \varepsilon} \implies \\
	\sigma_2(R) = \frac{R_{\scriptscriptstyle_{H}}^2\sigma_{\scriptscriptstyle{Ho}} - [R^2+(R+d)^2]\sigma_0}{R^2} \\
	\therefore c_2 = \frac{R_{\scriptscriptstyle_{H}}^2\sigma_{\scriptscriptstyle{Ho}} - [R^2+(R+d)^2]\sigma_0}{R^2\frac{1}{\varepsilon_{\scriptscriptstyle{0}} \varepsilon} \left[ R_{\scriptscriptstyle_{H}}^2\sigma_{\scriptscriptstyle{Ho}} - [R^2+(R+d)^2]\sigma_0 \right]\frac{\frac{a}{2}}{R\left(R-\frac{a}{2}\right)}} \implies \\
	c_2 = \frac{\varepsilon_{\scriptscriptstyle{0}} \varepsilon \left(R-\frac{a}{2}\right)}{R\frac{a}{2}}
\end{gather*}
\begin{gather*}
	C_1 = \frac{Q_1\left(R-\frac{a}{2}\right)}{\psi_d-\psi_H} = \frac{4 \pi \left(R-\frac{a}{2}\right)^2\sigma_1\left(R-\frac{a}{2}\right)}{\psi_d-\psi_H} \\
	\therefore c_1 = \frac{C_1}{4\pi \left(R-\frac{a}{2}\right)^2} = \frac{\sigma_1\left(R-\frac{a}{2}\right)}{\psi_d-\psi_H} \\
	\textrm{but} \ \ E_1\left(R-\frac{a}{2}\right) = \frac{2M}{\left(R-\frac{a}{2}\right)^2} \left[ sinh\left[\kappa r \right] - \kappa r cosh\left[\kappa r \right]\right] \quad \textrm{and} \\
	M = \frac{R_{\scriptscriptstyle_{H}}^2\sigma_{\scriptscriptstyle{Ho}} - [R^2+(R+d)^2]\sigma_0}{2\varepsilon_{\scriptscriptstyle{0}} \varepsilon \left[ sinh \left[\kappa \left(R-\frac{a}{2}\right)\right] - \kappa \left(R-\frac{a}{2}\right) cosh\left[\kappa \left(R-\frac{a}{2}\right)\right] \right]} \\
	\therefore E_1\left(R-\frac{a}{2}\right) = \frac{R_{\scriptscriptstyle_{H}}^2\sigma_{\scriptscriptstyle{Ho}} - [R^2+(R+d)^2]\sigma_0}{\varepsilon_{\scriptscriptstyle{0}} \varepsilon \left(R-\frac{a}{2}\right)^2} = \frac{\sigma_1\left(R-\frac{a}{2}\right)}{\varepsilon_{\scriptscriptstyle{0}} \varepsilon} \implies \\
	\sigma_1\left(R-\frac{a}{2}\right) =  \frac{R_{\scriptscriptstyle_{H}}^2\sigma_{\scriptscriptstyle{Ho}} - [R^2+(R+d)^2]\sigma_0}{\left(R-\frac{a}{2}\right)^2} \implies \\
	c_1 = \frac{R_{\scriptscriptstyle_{H}}^2\sigma_{\scriptscriptstyle{Ho}} - [R^2+(R+d)^2]\sigma_0}{\left(R-\frac{a}{2}\right)^2 \left( \frac{\left[R_{\scriptscriptstyle_{H}}^2\sigma_{\scriptscriptstyle{Ho}} - [R^2+(R+d)^2]\sigma_0\right]\left[\kappa \left(R-\frac{a}{2}\right)-sinh\left[\kappa \left(R-\frac{a}{2}\right)\right]\right]}{\varepsilon_{\scriptscriptstyle{0}} \varepsilon \left(R-\frac{a}{2}\right)\left[sinh\left[\kappa \left(R-\frac{a}{2}\right)\right]-\kappa \left(R-\frac{a}{2}\right)cosh\left[\kappa \left(R-\frac{a}{2}\right)\right]\right]} \right) } \implies\\
	c_1 = \frac{\varepsilon_{\scriptscriptstyle{0}} \varepsilon \left(sinh\left[\kappa \left(R-\frac{a}{2}\right)\right]-\kappa \left(R-\frac{a}{2}\right)cosh\left[\kappa \left(R-\frac{a}{2}\right)\right] \right)}{\left(R-\frac{a}{2}\right)\left(\kappa \left(R-\frac{a}{2}\right)-sinh\left[\kappa \left(R-\frac{a}{2}\right)\right]\right)}
\end{gather*}

%\end{appendix}
\newpage
%\bibliography{Adrian-Alejandra-nano}

\begin{thebibliography}{10}
	\expandafter\ifx\csname url\endcsname\relax
	\def\url#1{\texttt{#1}}\fi
	\expandafter\ifx\csname urlprefix\endcsname\relax\def\urlprefix{URL }\fi
	\expandafter\ifx\csname href\endcsname\relax
	\def\href#1#2{#2} \def\path#1{#1}\fi
	
	\bibitem{Chmiola_2006}
	J.~Chmiola, G.~Yushin, Y.~Gogotsi, C.~Portet, P.~Simon, P.~L. Taberna,
	Anomalous {I}ncrease in {C}arbon {C}apacitance at {P}ore {S}izes {L}ess
	{T}han 1 {N}anometer, Science 313~(5794) (2006) 1760--1763.
	
	\bibitem{Kim_2013}
	T.~Y. Kim, G.~Jung, S.~Yoo, K.~S. Suh, R.~S. Ruoff, Activated
	{G}raphene-{B}ased {C}arbons as {S}upercapacitor {E}lectrodes with {M}acro-
	and {M}esopores, ACS Nano 7~(8) (2013) 6899--6905.
	
	\bibitem{Supercapacitors-Book-2013}
	F.~Beguin, E.~Frackowiak (Eds.), Supercapacitors: Materials, Systems, and
	Applications, Materials for Sustainable Energy and Development, Wiley-VCH,
	2013.
	
	\bibitem{Beguin_2014}
	F.~B{\'e}guin, V.~Presser, A.~Balducci, E.~Frackowiak, Carbons and
	{E}lectrolytes for {A}dvanced {S}upercapacitors carbons and electrolytes for
	advanced supercapacitors, Adv. Mater. 26~(14) (2014) 2219--2251.
	
	\bibitem{ElKady_2014}
	M.~F. El-Kady, M.~Ihns, M.~Li, J.~Y. Hwang, M.~F. Mousavi, L.~Chaney, A.~T.
	Lech, R.~B. Kaner, Engineering three-dimensional hybrid supercapacitors and
	microsupercapacitors for high-performance integrated energy storage, PNAS
	112~(14) (2015) 4233--4238.
	
	\bibitem{Ke_2016}
	Q.~Ke, J.~Wang, Graphene-based materials for supercapacitor electrodes -- {A}
	review, J. Materiomics 2~(1) (2016) 37--54.
	
	\bibitem{Lozada_1984}
	M.~Lozada-Cassou, The force between two planar electrical double layers, J.
	Chem. Phys. 80~(7) (1984) 3344--3349.
	
	\bibitem{Mier_1988}
	L.~Mier~y Teran, E.~Diaz-Herrera, M.~Lozada-Cassou, D.~Henderson,
	\href{https://doi.org/10.1021/j100333a044}{Temperature dependence of the
		primitive-model double-layer differential capacitance: a hypernetted
		chain/mean spherical approximation calculation}, The Journal of Physical
	Chemistry 92~(22) (1988) 6408--6413.
	\newblock \href {http://arxiv.org/abs/https://doi.org/10.1021/j100333a044}
	{\path{arXiv:https://doi.org/10.1021/j100333a044}}, \href
	{https://doi.org/10.1021/j100333a044} {\path{doi:10.1021/j100333a044}}.
	\newline\urlprefix\url{https://doi.org/10.1021/j100333a044}
	
	\bibitem{Vlachy1989}
	V.~Vlachy, A.~D.~J. Haymet,
	\href{https://doi.org/10.1021/ja00184a012}{Electrolytes in charged
		micropores}, Journal of the American Chemical Society 111~(2) (1989)
	477--481.
	\newblock \href {http://arxiv.org/abs/https://doi.org/10.1021/ja00184a012}
	{\path{arXiv:https://doi.org/10.1021/ja00184a012}}, \href
	{https://doi.org/10.1021/ja00184a012} {\path{doi:10.1021/ja00184a012}}.
	\newline\urlprefix\url{https://doi.org/10.1021/ja00184a012}
	
	\bibitem{Yeomans1993}
	L.~Yeomans, S.~E. Feller, E.~Sánchez, M.~Lozada‐Cassou,
	\href{https://doi.org/10.1063/1.464308}{The structure of electrolytes in
		cylindrical pores}, The Journal of Chemical Physics 98~(2) (1993) 1436--1450.
	\newblock \href {http://arxiv.org/abs/https://doi.org/10.1063/1.464308}
	{\path{arXiv:https://doi.org/10.1063/1.464308}}, \href
	{https://doi.org/10.1063/1.464308} {\path{doi:10.1063/1.464308}}.
	\newline\urlprefix\url{https://doi.org/10.1063/1.464308}
	
	\bibitem{Yu_1997}
	J.~Yu, L.~Degr{\`e}ve, M.~Lozada-Cassou, Charge {S}eparation in {C}onfined
	{C}harge {F}luids, Phys. Rev. Lett. 79~(19) (1997) 3656--3659.
	
	\bibitem{Vlachy2001}
	V.~Vlachy, \href{https://doi.org/10.1021/la000826e}{Ion-partitioning between
		charged capillaries and bulk electrolyte solution: An axample of negative
		``rejection''}, Langmuir 17~(2) (2001) 399--402.
	\newblock \href {http://arxiv.org/abs/https://doi.org/10.1021/la000826e}
	{\path{arXiv:https://doi.org/10.1021/la000826e}}, \href
	{https://doi.org/10.1021/la000826e} {\path{doi:10.1021/la000826e}}.
	\newline\urlprefix\url{https://doi.org/10.1021/la000826e}
	
	\bibitem{Grosse-2002}
	J.~J. L{\'o}pez-Garc{\'i}a, J.~Horno, C.~Grosse, Numerical solution of the
	poisson-boltzmann equation for a spherical cavity, Journal of colloid and
	interface science 251~(1) (2002) 85--93.
	\newblock \href {https://doi.org/10.1006/jcis.2002.8396}
	{\path{doi:10.1006/jcis.2002.8396}}.
	
	\bibitem{Henderson2005}
	J.~Reszko-Zygmunt, S.~Sokołowski, D.~Henderson, D.~Boda,
	\href{https://doi.org/10.1063/1.1850453}{{Temperature dependence of the
			double layer capacitance for the restricted primitive model of an electrolyte
			solution from a density functional approach}}, The Journal of Chemical
	Physics 122~(8), 084504 (02 2005).
	\newblock \href
	{http://arxiv.org/abs/https://pubs.aip.org/aip/jcp/article-pdf/doi/10.1063/1.1850453/15364981/084504\_1\_online.pdf}
	{\path{arXiv:https://pubs.aip.org/aip/jcp/article-pdf/doi/10.1063/1.1850453/15364981/084504\_1\_online.pdf}},
	\href {https://doi.org/10.1063/1.1850453} {\path{doi:10.1063/1.1850453}}.
	\newline\urlprefix\url{https://doi.org/10.1063/1.1850453}
	
	\bibitem{Aguilar_2007}
	G.~E. Aguilar-Pineda, F.~Jim{\'e}nez-{\'A}ngeles, J.~Yu, M.~Lozada-Cassou, Van
	der {W}aals-{L}ike {I}sotherms in a {C}onfined {E}lectrolyte by {S}pherical
	and {C}ylindrical {N}anopores, J. Phys. Chem. B 111~(8) (2007) 2033--2044.
	
	\bibitem{Peng2009}
	B.~Peng, Y.-X. Yu, \href{https://doi.org/10.1063/1.3243873}{{Ion distributions,
			exclusion coefficients, and separation factors of electrolytes in a charged
			cylindrical nanopore: A partially perturbative density functional theory
			study}}, J. Chem. Phys. 131~(13) (2009) 134703.
	\newblock \href {https://doi.org/10.1063/1.3243873}
	{\path{doi:10.1063/1.3243873}}.
	\newline\urlprefix\url{https://doi.org/10.1063/1.3243873}
	
	\bibitem{Henderson2012}
	D.~Henderson,
	\href{https://www.sciencedirect.com/science/article/pii/S0021979712000835}{Oscillations
		in the capacitance of a nanopore containing an electrolyte due to pore width
		and nonzero size ions}, Journal of Colloid and Interface Science 374~(1)
	(2012) 345--347.
	\newblock \href {https://doi.org/https://doi.org/10.1016/j.jcis.2012.01.050}
	{\path{doi:https://doi.org/10.1016/j.jcis.2012.01.050}}.
	\newline\urlprefix\url{https://www.sciencedirect.com/science/article/pii/S0021979712000835}
	
	\bibitem{Pizio2012}
	O.~Pizio, S.~Sokołowski, Z.~Sokołowska,
	\href{https://doi.org/10.1063/1.4771919}{{Electric double layer capacitance
			of restricted primitive model for an ionic fluid in slit-like nanopores: A
			density functional approach}}, The Journal of Chemical Physics 137~(23),
	234705 (12 2012).
	\newblock \href
	{http://arxiv.org/abs/https://pubs.aip.org/aip/jcp/article-pdf/doi/10.1063/1.4771919/15458286/234705\_1\_online.pdf}
	{\path{arXiv:https://pubs.aip.org/aip/jcp/article-pdf/doi/10.1063/1.4771919/15458286/234705\_1\_online.pdf}},
	\href {https://doi.org/10.1063/1.4771919} {\path{doi:10.1063/1.4771919}}.
	\newline\urlprefix\url{https://doi.org/10.1063/1.4771919}
	
	\bibitem{Lamperski-2014}
	S.~Lamperski, J.~Sosnowska, L.~B. Bhuiyan, D.~Henderson,
	\href{https://doi.org/10.1063/1.4851456}{{Size asymmetric hard spheres as a
			convenient model for the capacitance of the electrical double layer of an
			ionic liquid}}, The Journal of Chemical Physics 140~(1), 014704 (01 2014).
	\newblock \href
	{http://arxiv.org/abs/https://pubs.aip.org/aip/jcp/article-pdf/doi/10.1063/1.4851456/15470290/014704\_1\_online.pdf}
	{\path{arXiv:https://pubs.aip.org/aip/jcp/article-pdf/doi/10.1063/1.4851456/15470290/014704\_1\_online.pdf}},
	\href {https://doi.org/10.1063/1.4851456} {\path{doi:10.1063/1.4851456}}.
	\newline\urlprefix\url{https://doi.org/10.1063/1.4851456}
	
	\bibitem{Henderson2015}
	D.~Henderson, {SOME ANALYTIC EXPRESSIONS FOR THE CAPACITANCE AND PROFILES OF
		THE ELECTRIC DOUBLE LAYER FORMED BY IONS NEAR AN ELECTRODE}, HUNGARIAN
	JOURNAL OF INDUSTRY AND CHEMISTRY 43~(2) (2015) 55--66.
	\newblock \href {https://doi.org/10.1515/hjic-2015-0010}
	{\path{doi:10.1515/hjic-2015-0010}}.
	
	\bibitem{Yang2019}
	F.~Yang,
	\href{https://www.sciencedirect.com/science/article/pii/S037596011930369X}{Size
		effect on electric-double-layer capacitances of conducting structures},
	Physics Letters A 383~(20) (2019) 2353--2360.
	\newblock \href
	{https://doi.org/https://doi.org/10.1016/j.physleta.2019.04.051}
	{\path{doi:https://doi.org/10.1016/j.physleta.2019.04.051}}.
	\newline\urlprefix\url{https://www.sciencedirect.com/science/article/pii/S037596011930369X}
	
	\bibitem{biagooi-Nature2020}
	M.~Biagooi, S.~Nedaaee~Oskoee,
	\href{https://doi.org/10.1038/s41598-020-62943-7}{The effects of slit-pore
		geometry on capacitive properties: a molecular dynamics study}, Scientific
	Reports 10~(1) (2020) 6533.
	\newblock \href {https://doi.org/10.1038/s41598-020-62943-7}
	{\path{doi:10.1038/s41598-020-62943-7}}.
	\newline\urlprefix\url{https://doi.org/10.1038/s41598-020-62943-7}
	
	\bibitem{Enrique-Henry2021}
	E.~Gonz{\'a}lez-Tovar, J.~A. Mart{\'i}nez-Gonz{\'a}lez, C.~G. Galv{\'a}n~Peña,
	G.~I. Guerrero-Garc{\'i}a, \href{https://doi.org/10.1063/5.0043028}{{On the
			expected value of the electrostatic potential produced by a charged electrode
			neutralized by a Coulombic fluid: The capacitive compactness}}, The Journal
	of Chemical Physics 154~(9), 096101 (03 2021).
	\newblock \href
	{http://arxiv.org/abs/https://pubs.aip.org/aip/jcp/article-pdf/doi/10.1063/5.0043028/15591400/096101\_1\_online.pdf}
	{\path{arXiv:https://pubs.aip.org/aip/jcp/article-pdf/doi/10.1063/5.0043028/15591400/096101\_1\_online.pdf}},
	\href {https://doi.org/10.1063/5.0043028} {\path{doi:10.1063/5.0043028}}.
	\newline\urlprefix\url{https://doi.org/10.1063/5.0043028}
	
	\bibitem{Keshavarzi_2022}
	E.~Keshavarzi, M.~Abareghi,
	\href{https://dx.doi.org/10.1149/1945-7111/ac52fe}{The effect of stern layer
		thickness on the diffuse capacitance for size asymmetric electrolyte inside
		the charged spherical cavities by density functional theory}, Journal of The
	Electrochemical Society 169~(2) (2022) 020547.
	\newblock \href {https://doi.org/10.1149/1945-7111/ac52fe}
	{\path{doi:10.1149/1945-7111/ac52fe}}.
	\newline\urlprefix\url{https://dx.doi.org/10.1149/1945-7111/ac52fe}
	
	\bibitem{Feng-nanopores-topology-2023}
	T.~Mo, Z.~Wang, L.~Zeng, M.~Chen, A.~A. Kornyshev, M.~Zhang, Y.~Zhao, G.~Feng,
	\href{https://onlinelibrary.wiley.com/doi/abs/10.1002/adma.202301118}{Energy
		storage mechanism in supercapacitors with porous graphdiynes: Effects of pore
		topology and electrode metallicity}, Advanced Materials n/a~(n/a) (2023)
	2301118.
	\newblock \href
	{http://arxiv.org/abs/https://onlinelibrary.wiley.com/doi/pdf/10.1002/adma.202301118}
	{\path{arXiv:https://onlinelibrary.wiley.com/doi/pdf/10.1002/adma.202301118}},
	\href {https://doi.org/https://doi.org/10.1002/adma.202301118}
	{\path{doi:https://doi.org/10.1002/adma.202301118}}.
	\newline\urlprefix\url{https://onlinelibrary.wiley.com/doi/abs/10.1002/adma.202301118}
	
	\bibitem{Verwey_TheoryStabilityLyophobicColloids_1948}
	E.~J.~W. Verwey, J.~T.~G. Overbeek, Theory of the {S}tability of {L}yophobic
	{C}olloids, Elsevier, Netherlands, 1948.
	
	\bibitem{Israelachvili-book}
	J.~Israelachvili, {Intermolecular forces \& Surface Forces}, 2nd Edition,
	Academic Press, New York, NY, 1991.
	
	\bibitem{Evans-Wennerstrom-1999}
	D.~F. Evans, H.~Wennestr{\"o}m, {The colloidal domain: where physics, chemistry
		and thecnology meet}, 2nd Edition, Wiley-VCH, New York, NY, 1999.
	
	\bibitem{Bohinc_2008}
	K.~Bohinc, T.~Slivnik, A.~Igli{\v c}, V.~Kralj-Igcli{\v c}, {M}embrane
	{E}lectrostatics---{A} {S}tatistical {M}echanical {A}pproach to the
	{F}unctional {D}ensity {T}heory of {E}lectric {D}ouble {L}ayer, in: A.~L. Liu
	(Ed.), Advances in {P}lanar {L}ipid {B}ilayers and {L}iposomes, Vol.~8,
	Academic Press, Inc., 2008, pp. 107--154.
	
	\bibitem{Bohinc-2018}
	M.~Špadina, S.~Gourdin-Bertin, G.~Dražić, A.~Selmani, J.-F. Dufrêche,
	K.~Bohinc, \href{https://doi.org/10.1021/acsami.7b18737}{Charge properties of
		tio2 nanotubes in nano3 aqueous solution}, ACS Applied Materials \&
	Interfaces 10~(15) (2018) 13130--13142, pMID: 29620855.
	\newblock \href {http://arxiv.org/abs/https://doi.org/10.1021/acsami.7b18737}
	{\path{arXiv:https://doi.org/10.1021/acsami.7b18737}}, \href
	{https://doi.org/10.1021/acsami.7b18737} {\path{doi:10.1021/acsami.7b18737}}.
	\newline\urlprefix\url{https://doi.org/10.1021/acsami.7b18737}
	
	\bibitem{Coffey-Biology-2023}
	J.~L.~F. Dennis K.~Jeppesen, Qin~Zhang, R.~J. Coffey,
	\href{https://doi.org/10.1016/j.tcb.2023.01.002}{Extracellular vesicles and
		nanoparticles: emerging complexities}, Trends in Cell Biology 33 (2023)
	667--681.
	\newblock \href {https://doi.org/10.1016/j.tcb.2023.01.002}
	{\path{doi:10.1016/j.tcb.2023.01.002}}.
	\newline\urlprefix\url{https://doi.org/10.1016/j.tcb.2023.01.002}
	
	\bibitem{HUANG-oil-1996}
	J.~S. Huang, R.~Varadaraj,
	\href{https://www.sciencedirect.com/science/article/pii/S1359029496801245}{Colloid
		and interface science in the oil industry}, Current Opinion in Colloid \&
	Interface Science 1~(4) (1996) 535--539.
	\newblock \href {https://doi.org/https://doi.org/10.1016/S1359-0294(96)80124-5}
	{\path{doi:https://doi.org/10.1016/S1359-0294(96)80124-5}}.
	\newline\urlprefix\url{https://www.sciencedirect.com/science/article/pii/S1359029496801245}
	
	\bibitem{Oil-Recovery-book-2019}
	S.~Taylor (Ed.), {Colloids and Interfaces in Oil Recovery}, MDPI, Basel, 2019.
	\newblock \href
	{https://doi.org/https://doi.org/10.3390/books978-3-03921-107-4}
	{\path{doi:https://doi.org/10.3390/books978-3-03921-107-4}}.
	
	\bibitem{Grahame_1947}
	D.~C. Grahame, The {E}lectrical {D}ouble {L}ayer and the {T}heory of
	{E}lectrocapillarity, Chem. Rev. 41~(3) (1947) 441--501.
	
	\bibitem{Helmholtz_1879}
	H.~Helmholtz, Studien {\"u}ber electrische grenzschichten, Ann. d. Phy. u.
	Chem. 243~(7) (1879) 337--382.
	
	\bibitem{Gouy_1910}
	M.~Gouy, Sur la {C}onstitution de la {C}harge {{\'E}}lectrique a la {S}urface
	d'un {{\'E}}lectrolyte, J. Phys. T. Ap. 9~(1) (1910) 457--468.
	
	\bibitem{Chapman_1913}
	D.~L. Chapman, {LI}. {A} contribution to the theory of electrocapillarity,
	Phil. Mag. S. 6 25~(148) (1913) 475--481.
	
	\bibitem{Greberg_1998}
	H.~Greberg, R.~Kjellander, Charge inversion in electric double layers and
	effects of different sizes for counterions and coions, J. Chem. Phys. 108~(7)
	(1998) 2940--2953.
	
	\bibitem{Henderson_1992_FIF}
	D.~Henderson, Integral {E}quation {T}heories for {I}nhomogeneous {F}luids, in:
	D.~Henderson (Ed.), Fundamentals of {I}nhomogeneous {F}luids, Marcel Dekker,
	New York, 1992, Ch.~4, pp. 177--199.
	
	\bibitem{Lozada_1992_FIF}
	M.~Lozada-Cassou, Fluids {B}etween {W}alls and in {P}ores, in: D.~Henderson
	(Ed.), Fundamentals of {I}nhomogeneous {F}luids, Marcel Dekker, New York,
	1992, Ch.~8, pp. 303--361.
	
	\bibitem{Attard_1996}
	P.~Attard, Electrolytes and the {E}lectric {D}ouble {L}ayer, in: I.~Prigogine,
	S.~A. Rice (Eds.), Advances in Chemical Physics, Volume XCII, John Wiley \&
	Sons, Inc., Australia, 1996, pp. 1--159.
	
	\bibitem{Croxton_1981}
	T.~L. Croxton, D.~A. McQuarrie, The electrical double layer in the
	{B}orn-{G}reen-{Y}von equation, Mol. Phys. 42~(1) (1981) 141--151.
	
	\bibitem{Henderson_1982}
	D.~Henderson, A simple theory of the electric double layer including solvent
	effects, J. Electroanal. Chem. 132 (1982) 1--13.
	
	\bibitem{Shi_1996}
	H.~Shi, Activated carbons and double layer capcitance, Electrochim. Acta
	41~(10) (1996) 1633--1639.
	
	\bibitem{Goel_2008}
	T.~Goel, C.~N. Patra, S.~K. Ghosh, T.~Mukherjee, Structure of cylindrical
	electric double layers: {A} systematic study by {M}onte {C}arlo simulations
	and density functional theory, J. Chem. Phys. 129~(15) (2008)
	154906(1)--154906(12).
	
	\bibitem{Huang_2008}
	J.~Huang, B.~G. Sumpter, V.~Meunier, A {U}niversal {M}odel for {N}anoporous
	{C}arbon {S}upercapacitors {A}pplicable to {D}iverse {P}ore {R}egimes,
	{C}arbon {M}aterials, and {E}lectrolytes, Chem.: Eur. J. 14~(22) (2008)
	6614--6626.
	
	\bibitem{Lian_2016}
	C.~Lian, D.~E. Jiang, H.~Liu, J.~Wu, A {G}eneric {M}odel for {E}lectric
	{D}ouble {L}ayers in {P}orous {E}lectrodes, J. Phys. Chem. C 120~(16) (2016)
	8704--8710.
	
	\bibitem{Hartel_2017}
	A.~H{\"a}rtel, Structure of electric double layers in capacitive systems and to
	what extent (classical) density functional theory describes it, J. Phys.:
	Condens. Matter 29~(42) (2017) 423002(1)--423002(24).
	
	\bibitem{Patra_1994}
	C.~N. Patra, S.~K. Ghosh, A nonlocal density functional theory of the electric
	double layer: {S}ymmetric electrolytes, J. Chem. Phys. 100~(7) (1994)
	5219--5229.
	
	\bibitem{Patra_2020}
	C.~N. Patra, Size and charge correlations in spherical electric double layers:
	a case study with fully asymmetric mixed electrolytes within the solvent
	primitive model, RSC Adv. 10~(64) (2020) 39017--39025.
	
	\bibitem{Gillespie_2005}
	D.~Gillespie, M.~Valisk{\'o}, D.~Boda, Density functional theory of the
	electrical double layer: the {RFD} functional, J. Phys.: Condens. Matter
	17~(42) (2005) 6609--6626.
	
	\bibitem{Stevens_1990}
	M.~J. Stevens, M.~O. Robbins, Density {F}unctional {T}heory of {I}onic
	{S}creening:{W}hen {D}o {L}ike {C}harges {A}ttract?, Europhys. Lett. 12~(1)
	(1990) 81--86.
	
	\bibitem{Bhuiyan_1994}
	L.~B. Bhuiyan, C.~W. Outhwaite, The cylindrical electric double layer in the
	modified {P}oisson-{B}oltzmann theory, Phil. Mag. B 69~(5) (1994) 1051--1058.
	
	\bibitem{Outhwaite_1986}
	C.~W. Outhwaite, A {M}odified {P}oisson-{B}oltzmann {E}quation for the {I}onic
	{A}tmosphere around a {C}ylindrical {W}all, J. Chem. Soc. Faraday Trans. 2
	82~(5) (1986) 789--794.
	
	\bibitem{Bhuiyan_1993_CMT}
	L.~B. Bhuiyan, C.~W. Outhwaite, A {M}odified {P}oisson-{B}oltzmann {T}reatment
	of an {I}solated {C}ylindrical {E}lectric {D}ouble {L}ayer, in: L.~Blum,
	F.~Malik (Eds.), Condensed {M}atter {T}heories, Vol.~8, Springer, Boston,
	1993, pp. 551--559.
	
	\bibitem{Stern_1924}
	O.~Stern-Hamburg, Zur theorie der elektrolytischen doppelschicht, Z.
	Elektrochem. 30~(21-22) (1924) 508--516.
	
	\bibitem{Lyklema_FundIntCollSci}
	J.~Lyklema, Fundamentals of {I}nterface and {C}olloid {S}cience, Vol. II:
	{S}olid-{L}iquid {I}nterface, Academic Press, Inc., London, 1995.
	
	\bibitem{Dyachkov_2005}
	L.~G. D'yachkov, Analytical {S}olution of the {P}oisson-{B}oltzmann {E}quation
	in {C}ases of {S}pherical and {A}xial {S}ymmetry, Tech. Phys. Lett. 31~(3)
	(2005) 204--207.
	
	\bibitem{Lozada_1982}
	M.~Lozada-Cassou, R.~Saavedra-Barrera, D.~Henderson, The application of the
	hypernetted chain approximation to the electrical double layer: Comparison
	with monte carlo results for symmetric salts, J. Chem. Phys. 77~(10) (1982)
	5150--5156.
	
	\bibitem{Gonzalez_1985}
	E.~Gonz{\'a}lez-Tovar, M.~Lozada-Cassou, D.~Henderson, Hypernetted chain
	approximation for the distribution of ions around a cylindrical electrode.
	{II}. {N}umerical solution for a model cylindrical polyelectrolyte, J. Chem.
	Phys. 83~(1) (1985) 361--372.
	
	\bibitem{Gonzalez_1989}
	E.~Gonz{\'a}lez-Tovar, M.~Lozada-Cassou, The {S}pherical {D}ouble {L}ayer: {A}
	{H}ypernetted {C}hain {M}ean {S}pherical {A}pproximation {C}alculation for a
	{M}odel {S}pherical {C}olloid {P}article, J. Phys. Chem. 93~(9) (1989)
	3761--3768.
	
	\bibitem{Limin-self-assembly-2012}
	L.~Hu, M.~Chen, X.~Fang, L.~Wu,
	\href{http://dx.doi.org/10.1039/C1CS15189D}{Oil–water interfacial
		self-assembly: a novel strategy for nanofilm and nanodevice fabrication},
	Chem. Soc. Rev. 41 (2012) 1350--1362.
	\newblock \href {https://doi.org/10.1039/C1CS15189D}
	{\path{doi:10.1039/C1CS15189D}}.
	\newline\urlprefix\url{http://dx.doi.org/10.1039/C1CS15189D}
	
	\bibitem{Chai-Advance-Materials-self-assembly-2020}
	Z.~Chai, A.~Korkmaz, C.~Yilmaz, A.~A. Busnaina,
	\href{https://onlinelibrary.wiley.com/doi/abs/10.1002/adma.202000747}{High-rate
		printing of micro/nanoscale patterns using interfacial convective assembly},
	Advanced Materials 32~(22) (2020) 2000747.
	\newblock \href
	{http://arxiv.org/abs/https://onlinelibrary.wiley.com/doi/pdf/10.1002/adma.202000747}
	{\path{arXiv:https://onlinelibrary.wiley.com/doi/pdf/10.1002/adma.202000747}},
	\href {https://doi.org/https://doi.org/10.1002/adma.202000747}
	{\path{doi:https://doi.org/10.1002/adma.202000747}}.
	\newline\urlprefix\url{https://onlinelibrary.wiley.com/doi/abs/10.1002/adma.202000747}
	
	\bibitem{vialetto-JACS-self-assembly-2021}
	J.~Vialetto, S.~Rudiuk, M.~Morel, D.~Baigl,
	\href{https://doi.org/10.1021/jacs.1c04220}{Photothermally reconfigurable
		colloidal crystals at a fluid interface, a generic approach for optically
		tunable lattice properties}, Journal of the American Chemical Society
	143~(30) (2021) 11535--11543, pMID: 34309395.
	\newblock \href {http://arxiv.org/abs/https://doi.org/10.1021/jacs.1c04220}
	{\path{arXiv:https://doi.org/10.1021/jacs.1c04220}}, \href
	{https://doi.org/10.1021/jacs.1c04220} {\path{doi:10.1021/jacs.1c04220}}.
	\newline\urlprefix\url{https://doi.org/10.1021/jacs.1c04220}
	
	\bibitem{jia-RSC-adv-2023}
	H.~Jia, Z.~Huang, M.~Kaynak, M.~S. Sakar,
	\href{http://dx.doi.org/10.1039/D3RA03591C}{Colloidal self-assembly of soft
		neural interfaces from injectable photovoltaic microdevices}, RSC Adv. 13
	(2023) 19888--19897.
	\newblock \href {https://doi.org/10.1039/D3RA03591C}
	{\path{doi:10.1039/D3RA03591C}}.
	\newline\urlprefix\url{http://dx.doi.org/10.1039/D3RA03591C}
	
	\bibitem{Lozada1996}
	M.~Lozada-Cassou, W.~Olivares, B.~Sulbar\'an,
	\href{https://link.aps.org/doi/10.1103/PhysRevE.53.522}{Violation of the
		electroneutrality condition in confined charged fluids}, Phys. Rev. E 53
	(1996) 522--530.
	\newblock \href {https://doi.org/10.1103/PhysRevE.53.522}
	{\path{doi:10.1103/PhysRevE.53.522}}.
	\newline\urlprefix\url{https://link.aps.org/doi/10.1103/PhysRevE.53.522}
	
	\bibitem{Wu-nano-letters2011}
	D.-e. Jiang, Z.~Jin, J.~Wu,
	\href{https://doi.org/10.1021/nl202952d}{Oscillation of capacitance inside
		nanopores}, Nano Letters 11~(12) (2011) 5373--5377, pMID: 22029395.
	\newblock \href {http://arxiv.org/abs/https://doi.org/10.1021/nl202952d}
	{\path{arXiv:https://doi.org/10.1021/nl202952d}}, \href
	{https://doi.org/10.1021/nl202952d} {\path{doi:10.1021/nl202952d}}.
	\newline\urlprefix\url{https://doi.org/10.1021/nl202952d}
	
	\bibitem{Lozada-Cassou-PRL1996}
	M.~Lozada-Cassou, J.~Yu,
	\href{https://link.aps.org/doi/10.1103/PhysRevLett.77.4019}{Correlation of
		charged fluids separated by a wall}, Phys. Rev. Lett. 77 (1996) 4019--4022.
	\newblock \href {https://doi.org/10.1103/PhysRevLett.77.4019}
	{\path{doi:10.1103/PhysRevLett.77.4019}}.
	\newline\urlprefix\url{https://link.aps.org/doi/10.1103/PhysRevLett.77.4019}
	
	\bibitem{Lozada-Cassou-PRE1997}
	M.~Lozada-Cassou, J.~Yu,
	\href{https://eurekamag.com/research/082/731/082731479.php}{Correlation of
		charged fluids separated by a wall of finite thickness: Dependence on the
		charge of the fluid and the wall}, Physical Review E 56~(3) (1997)
	2958--2965.
	\newline\urlprefix\url{https://eurekamag.com/research/082/731/082731479.php}
	
	\bibitem{ninham-JTB-1971}
	B.~W. Ninham, V.~A. Parsegian, Electrostatic potential between surfaces bearing
	ionizable groups in ionic equilibrium with physiologic saline solution, J
	Theor Biol . 1971 Jun;31(3):405-28. doi: 10.1016/0022-5193(71)90019-1 31
	(1971) 405--428.
	\newblock \href {https://doi.org/10.1016/0022-5193(71)90019-1}
	{\path{doi:10.1016/0022-5193(71)90019-1}}.
	
	\bibitem{Jackson_2001}
	J.~D. Jackson, \href{https://cds.cern.ch/record/100964}{{Classical
			electrodynamics; 3rd ed.}}, Wiley, New York, NY, 1999.
	\newline\urlprefix\url{https://cds.cern.ch/record/100964}
	
	\bibitem{McQuarrie_StatMech}
	D.~A. McQuarrie, Statistical {M}echanics, Harper \& Row, New York, 1976.
	
	\bibitem{Henderson-Contact-JCP-1978}
	D.~Henderson, L.~Blum, \href{https://doi.org/10.1063/1.436535}{{Some exact
			results and the application of the mean spherical approximation to charged
			hard spheres near a charged hard wall}}, The Journal of Chemical Physics
	69~(12) (1978) 5441--5449.
	\newblock \href
	{http://arxiv.org/abs/https://pubs.aip.org/aip/jcp/article-pdf/69/12/5441/8133329/5441\_1\_online.pdf}
	{\path{arXiv:https://pubs.aip.org/aip/jcp/article-pdf/69/12/5441/8133329/5441\_1\_online.pdf}},
	\href {https://doi.org/10.1063/1.436535} {\path{doi:10.1063/1.436535}}.
	\newline\urlprefix\url{https://doi.org/10.1063/1.436535}
	
	\bibitem{Bari-Contact-Mol-Phys-2015}
	W.~Silvestre-Alcantara, D.~Henderson, L.~B. Bhuiyan,
	\href{https://doi.org/10.1080/00268976.2015.1026857}{Contact condition for
		the density profiles in spherical and cylindrical double layers}, Molecular
	Physics 113~(22) (2015) 3403--3405.
	\newblock \href
	{http://arxiv.org/abs/https://doi.org/10.1080/00268976.2015.1026857}
	{\path{arXiv:https://doi.org/10.1080/00268976.2015.1026857}}, \href
	{https://doi.org/10.1080/00268976.2015.1026857}
	{\path{doi:10.1080/00268976.2015.1026857}}.
	\newline\urlprefix\url{https://doi.org/10.1080/00268976.2015.1026857}
	
	\bibitem{Holovko-Contact-2023}
	M.~Holovko, V.~Vlachy, D.~{di Caprio},
	\href{https://www.sciencedirect.com/science/article/pii/S016773222202579X}{On
		the contact conditions for the density and charge profiles in the theory of
		electrical double layer: From planar to spherical and cylindrical geometry},
	Journal of Molecular Liquids 371 (2023) 121040.
	\newblock \href {https://doi.org/https://doi.org/10.1016/j.molliq.2022.121040}
	{\path{doi:https://doi.org/10.1016/j.molliq.2022.121040}}.
	\newline\urlprefix\url{https://www.sciencedirect.com/science/article/pii/S016773222202579X}
	
	\bibitem{Lozada_1990-I}
	M.~Lozada‐Cassou, E.~Díaz‐Herrera,
	\href{https://doi.org/10.1063/1.458128}{{Three point extension for
			hypernetted chain and other integral equation theories: Numerical results}},
	The Journal of Chemical Physics 92~(2) (1990) 1194--1210.
	\newblock \href
	{http://arxiv.org/abs/https://pubs.aip.org/aip/jcp/article-pdf/92/2/1194/11167912/1194\_1\_online.pdf}
	{\path{arXiv:https://pubs.aip.org/aip/jcp/article-pdf/92/2/1194/11167912/1194\_1\_online.pdf}},
	\href {https://doi.org/10.1063/1.458128} {\path{doi:10.1063/1.458128}}.
	\newline\urlprefix\url{https://doi.org/10.1063/1.458128}
	
	\bibitem{Odriozola_2017}
	G.~Ordiozola, M.~Lozada-Cassou, Equivalence between particles and fields: {A}
	general statistical mechanics theory for short and long range many-body
	forces, Fortschr. Phys. 65~(6-8) (2017) 1600072(1)--1600072(21).
	
	\bibitem{Degreve_1993}
	L.~Degr{\`e}ve, M.~Lozada-Cassou, E.~S{\'a}nchez, E.~Gonz{\'a}lez-Tovar, Monte
	{C}arlo simulation for a symmetrical electrolyte next to a charged spherical
	colloid particle, J. Chem. Phys. 98~(11) (1993) 8905--8909.
	
	\bibitem{Lozada_1983}
	M.~Lozada-Cassou, Hypernetted {C}hain {T}heory for the {D}istribution of {I}ons
	around a {C}ylindrical {E}lectrode, J. Phys. Chem. 87~(19) (1983) 3729--3732.
	
	\bibitem{Gonzalez-Tovar-2004}
	E.~Gonz\'alez-Tovar, F.~Jiménez-\'Angeles, R.~Messina, M.~Lozada-Cassou,
	\href{https://doi.org/10.1063/1.1710861}{{A new correlation effect in the
			Helmholtz and surface potentials of the electrical double layer}}, The
	Journal of Chemical Physics 120~(20) (2004) 9782--9792.
	\newblock \href
	{http://arxiv.org/abs/https://pubs.aip.org/aip/jcp/article-pdf/120/20/9782/10856673/9782\_1\_online.pdf}
	{\path{arXiv:https://pubs.aip.org/aip/jcp/article-pdf/120/20/9782/10856673/9782\_1\_online.pdf}},
	\href {https://doi.org/10.1063/1.1710861} {\path{doi:10.1063/1.1710861}}.
	\newline\urlprefix\url{https://doi.org/10.1063/1.1710861}
	
	\bibitem{Gonzalez-Tovar-2021}
	E.~Gonz\'alez-Tovar, J.~A. Mart\'inez-Gonz\'alez, C.~G. Galv\'an~Peña, G.~I.
	Guerrero-Garc\'ia, \href{https://doi.org/10.1063/5.0043028}{{On the expected
			value of the electrostatic potential produced by a charged electrode
			neutralized by a Coulombic fluid: The capacitive compactness}}, The Journal
	of Chemical Physics 154~(9), 096101 (03 2021).
	\newblock \href
	{http://arxiv.org/abs/https://pubs.aip.org/aip/jcp/article-pdf/doi/10.1063/5.0043028/15591400/096101\_1\_online.pdf}
	{\path{arXiv:https://pubs.aip.org/aip/jcp/article-pdf/doi/10.1063/5.0043028/15591400/096101\_1\_online.pdf}},
	\href {https://doi.org/10.1063/5.0043028} {\path{doi:10.1063/5.0043028}}.
	\newline\urlprefix\url{https://doi.org/10.1063/5.0043028}
	
	\bibitem{Gonzalez-Tovar-2018}
	C.~L. Moraila-Mart\'inez, G.~I. Guerrero-Garc\'ia, M.~Ch\'avez-P\'aez,
	E.~Gonz\'alez-Tovar, \href{https://doi.org/10.1063/1.5024553}{{An
			experimental/theoretical method to measure the capacitive compactness of an
			aqueous electrolyte surrounding a spherical charged colloid}}, The Journal of
	Chemical Physics 148~(15), 154703 (04 2018).
	\newblock \href
	{http://arxiv.org/abs/https://pubs.aip.org/aip/jcp/article-pdf/doi/10.1063/1.5024553/13855034/154703\_1\_online.pdf}
	{\path{arXiv:https://pubs.aip.org/aip/jcp/article-pdf/doi/10.1063/1.5024553/13855034/154703\_1\_online.pdf}},
	\href {https://doi.org/10.1063/1.5024553} {\path{doi:10.1063/1.5024553}}.
	\newline\urlprefix\url{https://doi.org/10.1063/1.5024553}
	
	\bibitem{Resnick-2001}
	D.~Halliday, R.~Resnick, J.~Walker, {Fundamentas of Physics}, 6th Edition,
	Vol.~2, John Wiley and Sons, Inc., New York, NY, 2001.
	
	\bibitem{Ala-Nisila_2011}
	S.~Buyukdagli, C.~V. Achim, T.~Ala-Nissila,
	\href{https://dx.doi.org/10.1088/1742-5468/2011/05/P05033}{Ion size effects
		upon ionic exclusion from dielectric interfaces and slit nanopores}, Journal
	of Statistical Mechanics: Theory and Experiment 2011~(05) (2011) P05033.
	\newblock \href {https://doi.org/10.1088/1742-5468/2011/05/P05033}
	{\path{doi:10.1088/1742-5468/2011/05/P05033}}.
	\newline\urlprefix\url{https://dx.doi.org/10.1088/1742-5468/2011/05/P05033}
	
\end{thebibliography}

\end{document}